\documentclass[preprint2]{aastex}
\shortauthors{JENKINS}
\shorttitle{FRACTIONAL IONIZATION OF THE WARM NEUTRAL ISM}
\begin{document}
\title{The Fractional Ionization of the Warm Neutral Interstellar Medium  
}
\author{Edward B. Jenkins}
\affil{Princeton University Observatory\\
Princeton, NJ 08544-1001}
\email{ebj@astro.princeton.edu}

\begin{abstract}
When the neutral interstellar medium is exposed to EUV and soft X-ray 
radiation, the argon atoms in it are far more susceptible to being ionized 
than the hydrogen atoms.  We make use of this fact to determine the level of 
ionization in the nearby, warm, neutral medium (WNM). By analyzing {\it 
FUSE\/} observations of ultraviolet spectra of 44 hot subdwarf stars a few 
hundred pc away from the Sun, we can compare column densities of Ar~I to 
those of O~I, where the relative ionization of oxygen can be used as a proxy 
for that of hydrogen.  The measured deficiency $[{\rm Ar~I/O~I}]=-0.427\pm 
0.11\,$dex below the expectation for a fully neutral medium implies that the 
electron density $n(e)\approx 0.04\,{\rm cm}^{-3}$ if $n({\rm H})=0.5\,{\rm 
cm}^{-3}$.  This amount of ionization is considerably larger than what we 
expect from primary photoionizations resulting from cosmic rays, the diffuse 
X-ray background, and X-ray emitting sources within the medium, along with 
the additional ionizations caused by energetic secondary photoelectrons, 
Auger electrons, and photons from helium recombinations.  We favor an 
explanation that bursts of radiation created by previous, nearby supernova 
remnants that have faded by now may have elevated the ionization, and the gas 
has not yet recombined to a quiescent level.  A different alternative is that 
the low energy portion of the soft X-ray background is poorly shielded by the 
H~I because it is frothy and has internal pockets of very hot, X-ray emitting 
gases.
\end{abstract}
\section{INTRODUCTION}\label{intro}

In large part, we are aware of the principal processes that create free 
electrons in otherwise completely neutral parts of the interstellar medium 
(ISM) in the disk of our Galaxy.  However, for some of the contributing 
factors, ones that either enhance or diminish the relative ionization of the 
gas, there is a need to validate our understanding of their strengths.  Most 
of these processes are well understood qualitatively, but one major 
quantitative uncertainty is the effectiveness of extreme 
ultraviolet (EUV) and soft X-ray photons 
in ionizing the gas.  Two factors contribute to this uncertainty: one is the 
difficulty in measuring the fluxes of diffuse photons with energies above the 
ionization potential of hydrogen (13.6~eV) but below about 100~eV, and the 
other is an overall assessment of how well these photons can penetrate the 
neutral regions, which depends on the porosity of the gas structures and the 
distribution of radiation sources. 

Our ultimate goal is not only to understand these processes better, but also 
to obtain an estimate for the fractional amount of the gas that is in an 
ionized state.  This information is relevant to gauging the strength of 
heating of the gas due to the photoelectric effect from dust grains 
irradiated by starlight, since the grain charge, which regulates its rate, 
depends on the electron density (Weingartner \& Draine 2001b).  
While generally considered to be less important than the photoelectric 
effect, other means of creating thermal energy, such as the heating caused by 
secondary electrons from cosmic ray and X-ray ionizations and the dissipation 
of Alfv\'en waves and magnetosonic turbulence (Kulsrud \& Pearce 1969; McIvor
1977; Spangler 1991; Minter \& Spangler 1997; Lerche et al. 2007), depend on
the relative fractions of electrons.  Cooling of the gas through the
collisional excitation of the fine-structure levels of C$^+$, Si$^+$ and
Fe$^+$ or, at temperatures approaching $10^4\,$K various metastable levels
and the L$\alpha$ transition of hydrogen, are likewise governed by the degree
of partial ionization  (Dalgarno \& McCray 1972). In the following
paragraphs, we consider three different pathways for creating small amounts
of ionization in the mostly neutral ISM.

Except for the interiors of dense clouds where there is significant 
extinction, all of the space in the disk of our Galaxy is exposed to 
ultraviolet starlight photons that are capable of ionizing atoms that have 
first ionization potentials less than that of hydrogen (13.6~eV).  Since the 
recombination rates of the ions are slow relative to their ionization rates, 
the concentrations of the ionized states of these atoms are strongly 
dominant.  Thus, it is a simple matter to add together the contributions from 
various elements that are able to supply free electrons.  The only 
uncertainty here is an accounting of the strengths of depletions from the gas 
phase caused by these elements condensing into solid form onto dust grains.  
These strengths vary collectively for all of the elements from one region to 
the next.  If we take such variations into account (Jenkins 2009), we 
can state that most of the gas will have free electron contributions that 
should be somewhere in the range $n({\rm M}^+)=0.8-1.7\times 10^{-4}n({\rm 
H}_{\rm tot})$, where $n({\rm M}^+)$ is the number density of heavy elements 
that are capable of being ionized\footnote{This range was computed from the 
solar abundances and representative values of the gas fractions $[{\rm 
X}_{\rm gas}/{\rm H}]_1$ and $[{\rm X}_{\rm gas}/{\rm H}]_0$ listed in 
Jenkins (2009) for elements that can be ionized by 
starlight, except that the gas-phase abundance of C was lowered by a factor 
of 0.43, in accord with the recommendation by Sofia et al. (2011) that
earlier determinations of $N({\rm C~II})$ were systematically too high by a
factor of 1/0.43.} and $n({\rm H}_{\rm tot})$ is the total density of hydrogen
in both neutral and ionized forms.

As cosmic ray particles collide with gas atoms in the Galaxy, they heat and 
ionize the ISM.  We are unable to observe the flux of the lowest energy 
particles because we are shielded from them by the heliospheric magnetic 
field, and extrapolations from the observed higher energy flux distributions 
are uncertain (Spitzer \& Tomasko 1968).  Nevertheless, from 
measurements of the relative abundances of trace molecular species, the 
cosmic ray ionization rates $\zeta_{\rm CR}\approx 0.5-9\times 10^{-16}{\rm 
s}^{-1}$ seems to be the most plausible range for the general ISM 
 (Wagenblast \& Williams 1996; Liszt 2003; Indriolo et al. 2007; Neufeld et
al. 2010; Indriolo \& McCall 2012), although details in the chemical models
may introduce some uncertainty [cf. Le Petit et al. (2004) and Shaw et al.
(2008)].  The chemical models of Bayet et al. (2011) suggest that in some
particularly active regions the ionization rates may increase to $\zeta_{\rm
CR} > 1\times 10^{-14}{\rm s}^{-1}$.

We now consider a third mechanism for ionizing the gas, one that is harder to 
quantify than the other two.  EUV and soft X-ray 
radiation can ionize atoms, through both the action of primary 
photoionizations and by creating a cascade of energetic, secondary 
photoelectrons that can collisionally ionize other atoms.  Estimates of the 
effectiveness of these agents are difficult to synthesize, since there are 
many complicating factors.  At high energies, much of the radiation arises 
from cooling, very hot ($T>10^6\,$K) gas coming from recently shocked regions 
within the disk and halo of our Galaxy.  For energies slightly above about 
100~eV, the photons can survive a journey through the neutral medium up to 
about $N({\rm H}^0)=2\times 10^{19}\,{\rm cm}^{-2}$, and this penetration 
depth progressively increases with energy.  Supplementing this ionizing 
radiation is that coming from several kinds of sources that are embedded 
within the neutral medium.  These include stars over a wide range of spectral 
types on the main sequence, active X-ray binaries, and the plentiful, but 
faint white dwarf (WD) stars.

\section{STRATEGY OF THE INVESTIGATION}\label{strategy}

The objective of the current study is to use specialized observations to help 
resolve the uncertainties that were mentioned above and gain a quantitative 
insight on how effectively the neutral regions are partially photoionized by 
EUV and X-ray radiation.  We do this by repeating a method developed by Sofia 
\& Jenkins (1998, hereafter SJ98), who examined 
interstellar UV absorption features that could be seen in the spectra of 
background stars so that they could compare the abundance of the neutral form 
of argon, which is highly susceptible to being photoionized, to that of 
hydrogen, which has an ionization potential very close to that of argon but 
with a markedly lower ionization cross section.  We propose that it is safe 
to assume that the abundance of argon (both neutral and ionized) relative to 
that of hydrogen should be equal to the solar ratio. For instance, SJ98 
presented arguments that support the principle that argon in the gas phase is 
not likely to be depleted by being incorporated into an atomic matrix within 
interstellar dust grains.  We will reinforce this idea with some indirect 
observational evidence in Section~\ref{reference_abund}.  Thus, we operate on 
the principle that any deficiency of neutral argon (Ar~I) below our 
expectation for the amount of H~I that is present may be considered to arise 
from the conversion of Ar to its ionized form, which is invisible.

The investigation by SJ98 covered an extremely limited number of target stars 
observed with the {\it Interstellar Medium Absorption Profile Spectrograph\/} 
({\it IMAPS\/}) (Jenkins et al. 1996).  Later, Jenkins et al. 
(2000b) and Lehner et al. (2003) 
reported on observations of absorption features of Ar~I observed by the {\it 
Far Ultraviolet Spectroscopic Explorer\/} ({\it FUSE\/}) (Moos et al. 2000;
Sahnow et al. 2000) toward collections of WD stars inside and
slightly beyond the edge of the Local Bubble,\footnote{The Local Bubble is an
irregularly shaped region with an unusually low average density with a radius
of about 80~pc that is approximately centered on the Sun  (Vergely et al.
2010; Welsh et al. 2010; Reis et al. 2011).  It contains small, partly
ionized clouds immersed in a much lower density medium (Redfield 2006;
Redfield \& Linsky 2008).} in order to infer ionization conditions of clouds
subjected to the characteristic radiation field in our immediate
neighborhood.  Our current goal is to extend our reach well beyond the stars
surveyed in these two studies, again by using {\it FUSE\/} spectra, so that
we can sample regions of more typical densities well outside the Local
Bubble.  We do this by downloading from the Mikulski Archive for Space
Telescopes (MAST) at the Space Telescope Science
Institute a large collection of {\it FUSE\/} spectra
of hot subdwarf stars that are situated several hundred pc away from us, well
beyond the boundary of the Local Bubble.

Our ultimate objective is to compare the neutral fractions of Ar and H, as 
had been done in past investigations.  However, a conventional approach of 
simply deriving the two column densities and dividing one by the other is not 
easily achievable with the data in this survey for two reasons.  First, it is 
difficult to measure $N$(Ar~I) because the lines are saturated, but only 
moderately so, and recorded at low resolution.  Second, the amount of H~I on 
a sight line can usually be determined from the damping wings of L$\alpha$, 
but when the interstellar column density is low and there is a significant 
stellar L$\alpha$ absorption, one must know the star's effective temperature 
and surface gravity and then create a model for the stellar feature, against 
which the interstellar feature is superimposed.  Also, as emphasized by Sofia 
et al. (2011) measurements of $N$(H~I) using the 
damping wings of L$\alpha$ can give a misleading outcome if one does not know 
about and correct for the effects of the velocity structures of the gas.  
This problem is probably most severe for low column density cases in the 
present study.

We can overcome the difficulties mentioned above if we replace H with O as 
the comparison element.  In the wavelength coverage of {\it FUSE\/} there are 
a large number of O~I absorption features, and these lines cover a broad 
range of transition probabilities.  Most important, the strengths of the O~I 
features are comparable to the one available feature of Ar~I.\footnote{There 
are two transitions of Ar~I in the {\it FUSE\/} wavelength coverage.  We can 
use only the one at 1048.220$\,$\AA\ because the 1066.660$\,$\AA\ line has 
interference from a pair of strong stellar Si~IV lines at nearly the same 
wavelength.}  By a simple comparison of the strength of this Ar~I line to 
those of O~I, we can determine $N({\rm Ar~I})/N({\rm O~I})$.  We describe 
this process in more detail in Section~\ref{derivation}.

The ionization fraction of O is strongly locked to that of H through a strong 
charge exchange reaction\footnote{We add a caution that deviations from a 
nearly one-to-one relationship for the ionization fractions of O and H can 
occur at low temperatures because the ionization potential of O is slightly 
higher than that of H ($\Delta E/k = 229\,$K) for the lowest fine-structure 
level in the O~I ground state.  However, since the ionization fractions are 
small and most of the gas we are considering is at temperatures much higher 
than $\Delta E/k$, this deviation is generally small enough to ignore.  
Nevertheless, we performed explicit calculations of the O and H ionization 
fractions, as described later in Section~\protect\ref{charge_exchange}.} 
 (Field \& Steigman 1971; Chambaud et al. 1980; Stancil et al. 1999).  Thus
we can use O as a proxy for H.  The only shortcoming of this tactic is that O
can be depleted in the ISM, but the depletion factors are not very large in
the regimes of low densities considered here  (Cartledge et al. 2004, 2008;
Jenkins 2009).

Section~\ref{obs} of this paper describes the selection of archival {\it 
FUSE\/} spectra and how they were processed to yield useful presentations of 
the absorption features for measurements of equivalent widths.
Our method of interpreting the spectra to yield the ionization of Ar relative 
to that of O is presented in Section~\ref{analysis}.  Section~\ref{beyond} 
contains a short digression on how we verify that the target stars are beyond 
the edge of the Local Bubble. In Section~\ref{fundamentals} we outline the 
basic equations that take into account the processes that influence the 
partial ionizations of Ar, relative to those of H and O.  The equations 
presented in this section are virtually identical to those outlined by SJ98, 
but with some new refinements (i.e., a few reactions that were not included 
earlier).  We consider the creation of free electrons from the starlight 
ionization of heavy elements and the effects from cosmic rays as processes 
that are mostly understood and already accounted for, and view the actions 
arising from EUV and X-ray ionizations as the principal unknowns whose 
strengths are to be determined.  In Section~\ref{known_photoionization} we 
make a prediction for the degree of ionization produced by known sources of 
radiation, but find that in order to satisfy the general outcomes for our 
measured ratios of Ar~I to O~I, an extraordinarily low volume density of 
hydrogen $n({\rm H}_{\rm tot})$ is required.  In order to obtain the same 
results for higher densities, we must propose a means of achieving higher 
levels of ionization.  In Section~\ref{additional_photoionization} we propose 
two possibilities: (1) there is a large residual ionization left over from 
effects of radiation emitted by nearby, but now extinct supernova remnants (SNRs)
over the past several Myr or (2) the neutral medium is porous enough to allow 
external, low-energy photons to penetrate the gas with less than the expected 
amount of attenuation.  Section~\ref{discussion} presents an overview of the 
implications of our results on an assortment of physical processes and 
various other kinds of observations that depend on electron fractions in the 
diffuse, neutral medium.  The paper ends with a summary of the main 
conclusions (Section~\ref{summary}).

Appendices to this paper give descriptions of various atomic processes that 
were incorporated into the calculations, but at a level of detail that most 
readers may wish to ignore.  A general section on the ionizations arising 
from secondary electrons (Appendix~\ref{gamma_s}) is broken into two 
subsections: one treats the effects from electrons liberated by the 
ionizations of H and He (Section~\ref{electrons_H_He}), while the second one 
discusses primary and Auger electrons created by the inner shell X-ray 
ionizations of heavy elements (Section~\ref{heavy_elem}).  
Appendix~\ref{gamma_he} gives the equations for evaluating the effects from a 
multitude of different kinds of ionizing photons that arise from the 
recombination of He$^{++}$ and He$^+$ ions with free electrons.  Finally, 
Appendix~\ref{cr_ioniz} describes how we can estimate the rates of cosmic ray 
ionization of H$^0$, He$^0$ and Ar$^0$ from the observed rates that apply to 
molecular hydrogen in dense clouds. 

\section{OBSERVATIONS}\label{obs}

\subsection{Target Selection}\label{tgt selection}

Our objective was to make use of target stars that represented intermediate 
cases between nearby WD stars, whose sight lines are entirely or 
heavily influenced by conditions in the Local Bubble, and the much more 
distant hot main-sequence, giant or supergiant stars that can create their 
own enhanced ionizations in atypical concentrations of gas associated with 
their formation.  Hot subdwarf stars represent a class of objects that fall 
into this intermediate category.  They have distances that are of order a few 
hundred parsecs from us, which reduces the contribution from material in the 
Local Bubble to a very minor level.  Since they are old, they have had 
adequate time to escape from their progenitorial gas clouds, and thus their 
locations are essentially random and should show no preference for dense gas 
complexes.  They have the additional advantage that they are bright enough to 
yield good quality spectra, but they are not so bright that they exceed the 
maximum allowed count-rate levels for {\it FUSE}.

In an initial screening of prospective targets, we examined the quick-look 
plots of all stars classified as sdO and sdB spectral types in the MAST 
archive of {\it FUSE\/} data.  In this step, we rejected all spectra that 
either seemed to show very strong stellar spectral features (8 stars) or that 
had an inadequate signal-to-noise ratio ($S/N$) at wavelengths in the 
vicinity of 920$\,$\AA\ (70 stars), which is where most of the O~I lines are 
situated.  A few further rejections were made after the spectra were 
downloaded and found to have observing anomalies (an extraordinarily large 
number of missing observations caused by channel misalignments: 3 stars), 
strong stellar features that were not evident in the quick-look plots (3 
stars), or molecular hydrogen lines that were strong enough to seriously 
compromise the lines that we wanted to measure (2 stars).  The lack of stars 
with exceptionally strong H$_2$ features helped to eliminate sight lines that 
penetrate dense, cool gas clouds.

\subsection{Creation of the Spectra}\label{creation}

All of the downloads of the calibrated {\it FUSE\/} data from MAST occurred 
well after the final pipeline reductions were performed for the archive with 
CalFUSE version 3.2.3 (Dixon et al. 2007).  For every target, we 
accepted data from all of the available observing sessions (identified by 
unique archive root names) but rejected any subexposures that had an 
extraordinarily low count level caused by a channel misalignment during the 
observation.  We used exposures obtained during both orbital day and night.  
Normally, one must be cautious about observations of features for either O~I 
or N~I because they can be filled in by diffuse telluric emission lines 
during daytime observing.  However the O~I transition strengths considered 
here are so weak that the telluric contributions are insignificant.

All subexposures and spectral channels that passed our initial screening were 
coadded with weight factors based on the inverse squares of their respective 
values of $S/N$ for intensities smoothed over a wavelength interval of 
0.12$\,$\AA\ (or 9 independent spectral elements -- this ensures that weights 
are not strongly influenced by random noise excursions).  Before this 
coaddition took place, we examined some strong interstellar features and 
aligned the individual spectra in wavelength against a preliminary coaddition 
with no wavelength shifts.  This process enabled us to virtually eliminate 
any degradation in resolution caused by drifts of the spectra in the 
wavelength direction from one subexposure to the next.  However, there can 
still be overall small systematic errors in radial velocity of about 
$10\,{\rm km~s}^{-1}$; see Appendix~A of Bowen et al. (2008).  In a few
cases, the spectral $S/N$ values were too low to allow such shifts to be made
with much confidence, even after the intensities were smoothed with a median
filter for viewing.  Such spectra were combined without any shifts.  For
every target, two combined spectra were created: one was made up with shifts
appropriate to the spectral region covering the Ar~I line at 1048$\,$\AA,
while the other had differentials that were optimized for the wavelengths
that covered the weakest O~I lines near 920$\,$\AA.

\section{ANALYSIS}\label{analysis}

\subsection{Equivalent Widths and Their Errors}\label{EW}

We measured equivalent widths of the Ar~I and O~I lines by integrating 
intensity deficits below best-fitting Legendre polynomials for the continua 
defined from intensities at locations somewhat removed from the features. 
Special precautions were made to account for various sources of error, which 
are important for later analysis stages that assign relative weights to 
different measurements and also for the estimates of the ultimate errors in 
the results.  First, we accounted for the direct effect that random 
count-rate variations can have on the equivalent width outcome (Jenkins et
al. 1973).  Next, the weakest lines are subject to uncertainties arising from
improper definitions of the continua.  To construct the probable errors, we
evaluated the expected formal errors in the polynomial coefficients, as
described by Sembach \& Savage (1992), and then we multiplied them by 2 in
order to make an approximate allowance for additional uncertainties caused by
the arbitrariness in selecting the most appropriate polynomial order.  To
find the effects of these continuum errors on our measurements, the
equivalent widths were re-evaluated using the probable excursions of the
continua on either side of the preferred ones.  Errors in the background
subtraction in the {\it FUSE\/} data processing are small compared to the
other errors.

\begin{deluxetable}
{
l   
c   
c
c
c
c
c
c
c
c
c
c
c
c
c
}
\tabletypesize{\scriptsize}
\rotate
\tablecolumns{14}
\tablewidth{0pt}
\tablecaption{Equivalent Widths\tablenotemark{a}\label{EW_table}}
\tablehead{
\colhead{} & \colhead{Ar I\tablenotemark{b}} & 
\multicolumn{12}{c}{O~I\tablenotemark{b,c}}\\
\cline{3-14}
\colhead{} & \colhead{1048.220} & \colhead{919.917} & \colhead{922.200} & 
\colhead{925.446} &
\colhead{916.815} & \colhead{930.257} & \colhead{919.658} & \colhead{921.857} 
&
\colhead{924.950} & \colhead{950.885} & \colhead{976.448} & \colhead{948.686} 
&
\colhead{971.738}\\
\colhead{Star Name} & \colhead{2.440} & \colhead{$-0.788$} & 
\colhead{$-0.645$} &
\colhead{$-0.484$} & \colhead{$-0.362$} & \colhead{$-0.301$} & 
\colhead{$-0.137$} &
\colhead{$-0.001$} & \colhead{0.155} & \colhead{0.176} & \colhead{0.509} &
\colhead{0.778} & \colhead{1.052}\\
\colhead{} & \colhead{\nodata} & \colhead{11.6} & \colhead{15.3} & 
\colhead{20.6} &
\colhead{25.0} & \colhead{27.8} & \colhead{34.5} & \colhead{40.4} & 
\colhead{46.9} &
\colhead{49.0} & \colhead{\nodata} & \colhead{\nodata} & \colhead{\nodata}\\
\colhead{(1)}& \colhead{(2)}& \colhead{(3)}& \colhead{(4)}&
 \colhead{(5)}& \colhead{(6)}& \colhead{(7)}& \colhead{(8)}& \colhead{(9)}&
\colhead{(10)}& \colhead{(11)}& \colhead{(12)}& \colhead{(13)}& 
\colhead{(14)}
}
\startdata
2MASS15265306\\
~~~~~+7941307\dotfill&$  77\pm 18$
&$(  14\pm 16)$&$(  -4\pm 16)$&$  29\pm 12$&       \nodata&$  24\pm 22$&$  
41\pm 22$
&$  39\pm 19$&$  84\pm 11$&$  59\pm 15$&$  97\pm 12$&$ 111\pm 11$&$( 120\pm 
12)$\\
AA Dor\dotfill&$  60\pm 10$
&$  30\pm 18$&       \nodata&$  56\pm 13$&       \nodata&$  75\pm 31$&$  
51\pm 37$
&       \nodata&$  93\pm 13$&$  73\pm 21$&$( 121\pm 19)$&       \nodata&$( 
225\pm  9)$\\
AGK+81 266\dotfill&$  76\pm 19$
&$  22\pm 15$&$  44\pm 15$&$  33\pm 15$&       \nodata&$  60\pm 19$&$  76\pm 
15$
&$  87\pm 18$&$  99\pm 14$&$ 122\pm 14$&$( 116\pm 18)$&$( 102\pm 15)$&$( 
125\pm 14)$\\
BD+18 2647\dotfill&$  43\pm 18$
&$  22\pm 13$&       \nodata&$  37\pm 11$&       \nodata&$  26\pm 16$&$  
39\pm 18$
&       \nodata&$  78\pm 11$&$  68\pm 11$&$(  63\pm 13)$&$( 131\pm 12)$&$( 
159\pm 11)$\\
BD+25 4655\dotfill&$  33\pm 18$
&$   3\pm 12$&       \nodata&$  18\pm 12$&       \nodata&$  22\pm 11$&$  
21\pm 12$
&       \nodata&$  53\pm 11$&$  41\pm 12$&$  64\pm 13$&$  77\pm 12$&$(  75\pm 
11)$\\
BD+28 4211\dotfill&$  31\pm  8$
&$  14\pm  3$&$  27\pm  5$&$  24\pm  3$&$  25\pm  4$&$  37\pm  4$&$  34\pm  
4$
&$  35\pm  4$&$  42\pm  3$&$  51\pm  3$&$(  42\pm  4)$&$(  83\pm  4)$&$(  
62\pm  3)$\\
BD+37 442\dotfill&$ 166\pm  7$
&       \nodata&$  61\pm 28$&$ 107\pm 16$&       \nodata&$  95\pm 24$&       
\nodata
&$ 166\pm 17$&$ 128\pm 16$&       \nodata&       \nodata&       \nodata&       
\nodata\\
BD+39 3226\dotfill&$  54\pm  3$
&$(  21\pm  8)$&$(  81\pm  8)$&$  34\pm  7$&$  40\pm  8$&$  48\pm 10$&$  
47\pm  8$
&$  38\pm  8$&$  52\pm  7$&$  59\pm  7$&$  54\pm  7$&$  69\pm  8$&$(  73\pm  
7)$\\
CPD$-$31 1701\dotfill&$  16\pm  8$
&$  11\pm  5$&       \nodata&$  19\pm  4$&       \nodata&$  20\pm 11$&$  
39\pm  6$
&       \nodata&$  45\pm  4$&$(  28\pm  5)$&$(  45\pm  7)$&$(  68\pm  5)$&$(  
91\pm  4)$\\
CPD$-$71D172\dotfill&$  34\pm 10$
&$  31\pm 13$&$  34\pm 13$&$  52\pm 12$&       \nodata&$  57\pm 13$&$  47\pm 
15$
&$  36\pm 18$&$  67\pm 12$&$  67\pm 13$&$(  61\pm 13)$&$(  93\pm 13)$&$(  
83\pm 14)$\\
EC11481$-$2303\dotfill&$ 132\pm 16$
&$(  39\pm 17)$&       \nodata&$  75\pm 16$&       \nodata&$  92\pm 16$&$  
99\pm 17$
&$  80\pm 34$&$ 145\pm 15$&$ 121\pm 17$&$ 153\pm 17$&$ 146\pm 16$&$ 177\pm 
16$\\
Feige 34\dotfill&$  36\pm 17$
&$(   9\pm 13)$&       \nodata&$  13\pm 13$&       \nodata&$  28\pm 14$&$  
38\pm 13$
&       \nodata&$  42\pm 13$&$  46\pm 14$&$  41\pm 21$&$  86\pm 19$&       
\nodata\\
HD$\,$113001\dotfill&$ 115\pm 20$
&$(  28\pm 10)$&       \nodata&$(  50\pm 10)$&       \nodata&$(  53\pm 15)$&$  
63\pm 10$
&       \nodata&$  96\pm 10$&$  93\pm 12$&$ 100\pm 13$&$ 125\pm 11$&$ 128\pm 
10$\\
JL 119\dotfill&$ 112\pm  5$
&$  52\pm 14$&$  39\pm 27$&$  89\pm  8$&       \nodata&$ 103\pm 15$&$  83\pm 
25$
&$  80\pm 23$&$ 138\pm  7$&$ 111\pm 19$&$( 169\pm  7)$&$( 180\pm  6)$&$( 
174\pm  7)$\\
JL 25\dotfill&$  77\pm 19$
&$  29\pm 17$&       \nodata&$  43\pm 17$&       \nodata&$  72\pm 22$&$  
29\pm 17$
&       \nodata&$  69\pm 17$&$  94\pm 17$&       \nodata&$ 111\pm 18$&$ 
106\pm 17$\\
JL 9\dotfill&$ 100\pm 10$
&$  46\pm 12$&       \nodata&$  67\pm  9$&       \nodata&$  95\pm 16$&$  
87\pm 14$
&       \nodata&$ 106\pm  9$&$ 116\pm  9$&       \nodata&$ 129\pm 11$&$( 
142\pm 12)$\\
LB 1566\dotfill&$  57\pm  6$
&$  32\pm 13$&       \nodata&$  51\pm 11$&       \nodata&$  56\pm 14$&$  
48\pm 15$
&       \nodata&$  81\pm 11$&$  73\pm 19$&$(  98\pm 11)$&$( 108\pm 14)$&$( 
111\pm 11)$\\
LB 1766\dotfill&$  67\pm 16$
&$  -3\pm 13$&       \nodata&$  61\pm 12$&       \nodata&$  58\pm 14$&$  
38\pm 14$
&       \nodata&$  98\pm 11$&       \nodata&$ 120\pm 11$&$( 148\pm 10)$&$( 
168\pm 10)$\\
LB 3241\dotfill&$  56\pm  9$
&$  29\pm  8$&$  25\pm  9$&$  54\pm  8$&       \nodata&$  53\pm  8$&$  52\pm  
8$
&$  97\pm  8$&$  82\pm  7$&$ 106\pm  7$&$( 112\pm  7)$&$( 123\pm  7)$&$( 
128\pm  7)$\\
LS 1275\dotfill&$  62\pm 15$
&$  29\pm 10$&$  42\pm 12$&$  45\pm  9$&$  52\pm 14$&$  65\pm 13$&$  46\pm 
10$
&$(  99\pm 13)$&$  58\pm  9$&$  74\pm 10$&$  80\pm 10$&$  89\pm 11$&$(  89\pm  
9)$\\
LSE 234\dotfill&$ 115\pm 12$
&$(  47\pm 16)$&$  72\pm 17$&$  76\pm 15$&       \nodata&$  94\pm 16$&$  
85\pm 18$
&$ 123\pm 17$&$ 107\pm 15$&$ 122\pm 15$&       \nodata&$ 154\pm 18$&$ 130\pm 
15$\\
LSE 259\dotfill&$ 149\pm  6$
&       \nodata&       \nodata&$  88\pm 14$&       \nodata&$ 134\pm 20$&       
\nodata
&       \nodata&$ 123\pm 13$&$ 156\pm 13$&       \nodata&       \nodata&       
\nodata\\
LSE 263\dotfill&$ 119\pm 17$
&$(  49\pm 14)$&       \nodata&$  73\pm 12$&$ 101\pm 19$&$  80\pm 15$&$  
76\pm 14$
&       \nodata&$  91\pm 13$&$  76\pm 13$&$  95\pm 17$&$ 139\pm 14$&$ 159\pm 
12$\\
LSE 44\dotfill&$  93\pm 12$
&$  56\pm 11$&$  68\pm 12$&$  68\pm 10$&       \nodata&$  78\pm 15$&$  61\pm 
15$
&$  94\pm 12$&$  92\pm  9$&$ 132\pm 10$&       \nodata&$( 140\pm 10)$&$( 
107\pm 11)$\\
LSII+18 9\dotfill&$  91\pm 15$
&       \nodata&       \nodata&$  54\pm 21$&$  66\pm 38$&$  73\pm 23$&$  
72\pm 21$
&       \nodata&$  78\pm 21$&$ 125\pm 21$&       \nodata&       \nodata&$( 
119\pm 21)$\\
LSII+22 21\dotfill&$  45\pm 12$
&$  27\pm 10$&       \nodata&$  35\pm  9$&$  40\pm 15$&$  52\pm 13$&$  44\pm 
11$
&$( 116\pm  8)$&$  63\pm  8$&$  74\pm  9$&$(  91\pm  8)$&$( 100\pm  9)$&$( 
119\pm  8)$\\
LSIV+10 9\dotfill&$ 179\pm 12$
&$(  58\pm 10)$&       \nodata&$  88\pm  9$&       \nodata&$ 106\pm 18$&$  
88\pm 12$
&       \nodata&$ 152\pm  9$&$ 170\pm 10$&$ 231\pm  9$&       \nodata&$( 
192\pm 10)$\\
LSS 1362\dotfill&$  88\pm  3$
&$(  39\pm 11)$&$  52\pm 14$&$  59\pm  8$&       \nodata&$  78\pm 18$&$  
64\pm 20$
&$(  65\pm  8)$&$(  68\pm 10)$&$ 134\pm  7$&       \nodata&       \nodata&       
\nodata\\
MCT 2005\\
~~~$-$5112\dotfill&$ 103\pm  8$
&       \nodata&       \nodata&$  74\pm 11$&       \nodata&$  75\pm 18$&       
\nodata
&       \nodata&$ 113\pm  9$&$ 135\pm 12$&$( 168\pm 15)$&$( 150\pm 13)$&$( 
124\pm 18)$\\
MCT 2048\\
~~~$-$4504\tablenotemark{d}\dotfill&$ 122\pm 12$
&       \nodata&$  86\pm 21$&$  83\pm 15$&       \nodata&$  61\pm 59$&       
\nodata
&$  90\pm 28$&$ 118\pm 14$&$ 117\pm 17$&$ 149\pm 14$&$ 174\pm 16$&$( 172\pm 
15)$\\
NGC6905\\
~~~star\tablenotemark{e}\dotfill&$ 147\pm  8$
&$(  45\pm 39)$&       \nodata&$ 101\pm 39$&       \nodata&$  83\pm 43$&$  
62\pm 46$
&       \nodata&$ 139\pm 38$&$ 167\pm 38$&$ 205\pm 39$&       \nodata&$( 
154\pm 39)$\\
PG0919+272\dotfill&$  55\pm 27$
&$  34\pm 17$&$  31\pm 16$&$  43\pm 16$&$  45\pm 27$&$  83\pm 17$&$  59\pm 
17$
&$  64\pm 16$&$  81\pm 15$&$  65\pm 17$&$  87\pm 15$&$ 108\pm 16$&$( 106\pm 
15)$\\
PG0952+519\dotfill&$  52\pm  3$
&$  19\pm  8$&       \nodata&$  36\pm  7$&       \nodata&$  43\pm  8$&$  
53\pm  9$
&       \nodata&$  74\pm  7$&       \nodata&$( 125\pm  6)$&$( 154\pm  8)$&$( 
148\pm  6)$\\
PG1032+406\dotfill&$  -21\pm 27$
&$   3\pm 23$&$  11\pm 19$&$  15\pm 18$&$(  95\pm 24)$&$  20\pm 19$&$  -2\pm 
36$
&$  36\pm 18$&$  15\pm 17$&$  72\pm 17$&$  46\pm 20$&$  70\pm 18$&$(  94\pm 
17)$\\
PG1051+501\dotfill&$ 165\pm 15$
&       \nodata&       \nodata&$  81\pm 30$&       \nodata&$  64\pm 33$&       
\nodata
&       \nodata&$ 128\pm 27$&$ 142\pm 28$&       \nodata&       \nodata&$ 
185\pm 28$\\
PG1230+068\dotfill&$ 102\pm 15$
&$  40\pm  8$&$  45\pm 11$&$  72\pm  7$&       \nodata&$  82\pm 16$&$  86\pm  
9$
&$  79\pm 13$&$ 110\pm  7$&$  88\pm  8$&$ 105\pm 18$&$ 126\pm  7$&$( 139\pm  
7)$\\
PG1544+488\dotfill&$  84\pm  3$
&$  26\pm 11$&$  25\pm 13$&$  56\pm  8$&$  76\pm 34$&$  69\pm 10$&$  74\pm 
13$
&$  89\pm 11$&$  91\pm  7$&$  97\pm  9$&$( 124\pm 11)$&$( 132\pm  7)$&$( 
151\pm  8)$\\
PG1605+072\dotfill&$ 131\pm 15$
&       \nodata&       \nodata&$  72\pm 21$&       \nodata&$  99\pm 25$&       
\nodata
&       \nodata&$ 136\pm 20$&$ 163\pm 20$&$ 229\pm 20$&$( 166\pm 22)$&$( 
186\pm 20)$\\
PG1610+519\dotfill&$ 122\pm  9$
&       \nodata&       \nodata&$  81\pm 12$&       \nodata&       \nodata&       
\nodata
&       \nodata&$ 150\pm  9$&       \nodata&$ 256\pm  5$&       \nodata&$( 
228\pm  7)$\\
PG2158+082\dotfill&$ 117\pm  7$
&       \nodata&       \nodata&$ 100\pm 19$&       \nodata&$  97\pm 26$&       
\nodata
&       \nodata&$ 121\pm 19$&$ 175\pm 17$&       \nodata&$( 154\pm 22)$&       
\nodata\\
PG2317+046\dotfill&$  68\pm 20$
&$  49\pm 18$&$  29\pm 31$&$  61\pm 19$&       \nodata&$  50\pm 24$&$  28\pm 
21$
&$  48\pm 33$&$  71\pm 18$&$ 135\pm 18$&       \nodata&$ 132\pm 19$&$(  81\pm 
19)$\\
Ton 102\dotfill&$  51\pm 13$
&$  21\pm 15$&$  22\pm 16$&$  26\pm 16$&       \nodata&$  39\pm 16$&$  34\pm 
16$
&$  79\pm 16$&$  42\pm 17$&$  63\pm 16$&$  72\pm 15$&       \nodata&       
\nodata\\
Ton S227\dotfill&$  80\pm 25$
&$  27\pm 10$&$  26\pm 11$&$  43\pm 10$&       \nodata&$  21\pm 20$&$  27\pm 
11$
&$  30\pm 12$&$  58\pm  9$&       \nodata&$  85\pm 11$&$ 111\pm 12$&$ 121\pm  
9$\\
UV0904$-$02\tablenotemark{f}\dotfill&$  69\pm  4$
&$(  30\pm  9)$&       \nodata&$  43\pm  8$&       \nodata&$  59\pm 12$&$  
52\pm 10$
&       \nodata&$  67\pm  7$&$  65\pm 11$&$  75\pm  7$&$  84\pm  8$&$(  89\pm  
7)$\\
\enddata
\tablenotetext{a}{Equivalent widths are given in m\AA.  Values given in 
parentheses indicate
lines that were not used in the linear fits.}
\tablenotetext{b}{Numbers given below indicate (1) wavelengths in \AA, (2) 
line strengths in
terms of $\log (f\lambda)$ taken from Morton (2003), and (3) the 
equivalent width in m\AA\
for $N({\rm O~I})=10^{16}\,{\rm cm}^{-2}$ and $b=6\,{\rm km~s}^{-1}$.}
\tablenotetext{c}{{}Lines are arranged in order of increasing strength.}
\tablenotetext{d}{Name recognized by SIMBAD: 2MASS J20515997-4042465.}
\tablenotetext{e}{Central star of the planetary nebula NGC 6905.}
\tablenotetext{f}{Name recognized by SIMBAD: 2MASS J09070812-0306139.}
\end{deluxetable}

\clearpage
The sources of error mentioned above are straightforward to evaluate and 
would apply to just about any measure of an equivalent width.  However, with 
the subdwarf stars, we must also contend with the confusion produced by 
stellar lines.  We made no attempt to model such features, because in order 
to do so we would need to know the details of the stellar parameters for each 
star.  Instead, we regarded the influence of stellar features as random 
sources of error in our line measurements.  In order to estimate the 
amplitude of such errors, which can vary markedly from one star to the next 
and can change with wavelength, we measured for each target the variance of a 
large number of equivalent width measurements of imaginary, fake lines at 
wavelengths similar to the ones under study, but that were displaced away 
from known real interstellar lines, both atomic and molecular.  This variance 
was then used as a guide for estimating the errors that should arise from 
stellar features.

Figure~\ref{sample_spectra} shows samples of spectra covering the relevant 
wavelength regions for two stars.  These two cases illustrate strong 
differences in the degree of interference from stellar lines.  The first 
example, AGK+81~266, has many stellar lines that can either add an apparent 
absorption to an interstellar line or distort the continuum level that is 
measured on either side of the line.  For this target, these effects dominate 
over other sources of error and create $1\sigma$ uncertainties in $W_\lambda$ 
equal to 18.5 and 14.2$\,$m\AA\ for the Ar~I and O~I lines, respectively.  
This star also exhibits molecular hydrogen features of moderate strength, but 
here the lines are not strong enough to compromise the measurements of the 
atomic lines.  A far more favorable case for measuring interstellar features 
is presented by the star UV0904$-$02.  Here, uncertainties produced by random 
stellar features should create errors of only 3.6$\,$m\AA\ for the Ar~I line 
and 6.6$\,$m\AA\ for the O~I lines.

All of the errors discussed in this section were combined in quadrature to 
synthesize the overall errors in the equivalent widths.  Values for the 
equivalent widths of all lines and their associated uncertainties are listed 
for each of our targets in Table~\ref{EW_table}.  The columns in this table 
are arranged in a sequence from the weakest to the strongest lines.

\subsection{Derivations of [Ar~I/O~I] Values and Their 
Uncertainties}\label{derivation}

\subsubsection{Reference Abundances}\label{reference_abund}

Detailed discussions on various methods of measuring the protosolar and 
B-star abundances of Ar have been presented by Lodders (2008) and Lanz et
al. (2008).  We adopt a mean value for the recommended outcomes of the two,
$\log ({\rm Ar/H})+12=6.60\pm 0.10$.  (Since both determinations might be
subject to common errors, the error of the mean is not reduced below the
0.10$\,$dex errors specified by each of them.) This value is higher than the
solar photospheric value proposed by Asplund et al. (2009), and it remains so
even after one applies a correction for gravitational settling of $+0.07\,$dex 
 (Lodders 2003) to obtain a protosolar value of $\log ({\rm 
Ar/H})+12=6.47\pm 0.13$.  For O, we take the solar photospheric value given 
by Asplund et al. (2009) and again apply a 
$+0.07\,$dex settling correction to get $\log ({\rm O/H})+12=8.76\pm 0.05$.  
This value agrees remarkably well with the measurement $\log ({\rm 
O/H})+12=8.76\pm 0.03$ obtained for B-stars by Przybilla et al. 
 (2008).

We must now consider the prospect that some of the Ar and O atoms are 
incorporated into solid form within or on the surfaces of dust grains, and 
this effect might be large enough to distort our findings on the differences 
in ionization.  Unfortunately, we have no direct information about the 
depletion of gas-phase Ar, since ionization corrections (the object of the 
present study) can influence the outcome.  While in principle it would be 
beneficial if we could study $N({\rm Ar~I})$ along sight lines that penetrate 
dense media, where depletions are likely to dominate over ionization effects, 
this is not possible because the absorption lines are far too saturated (much 
more so than in the current study).

\onecolumn
\begin{figure}
\vspace*{-2cm}
\epsscale{.95}
\plotone{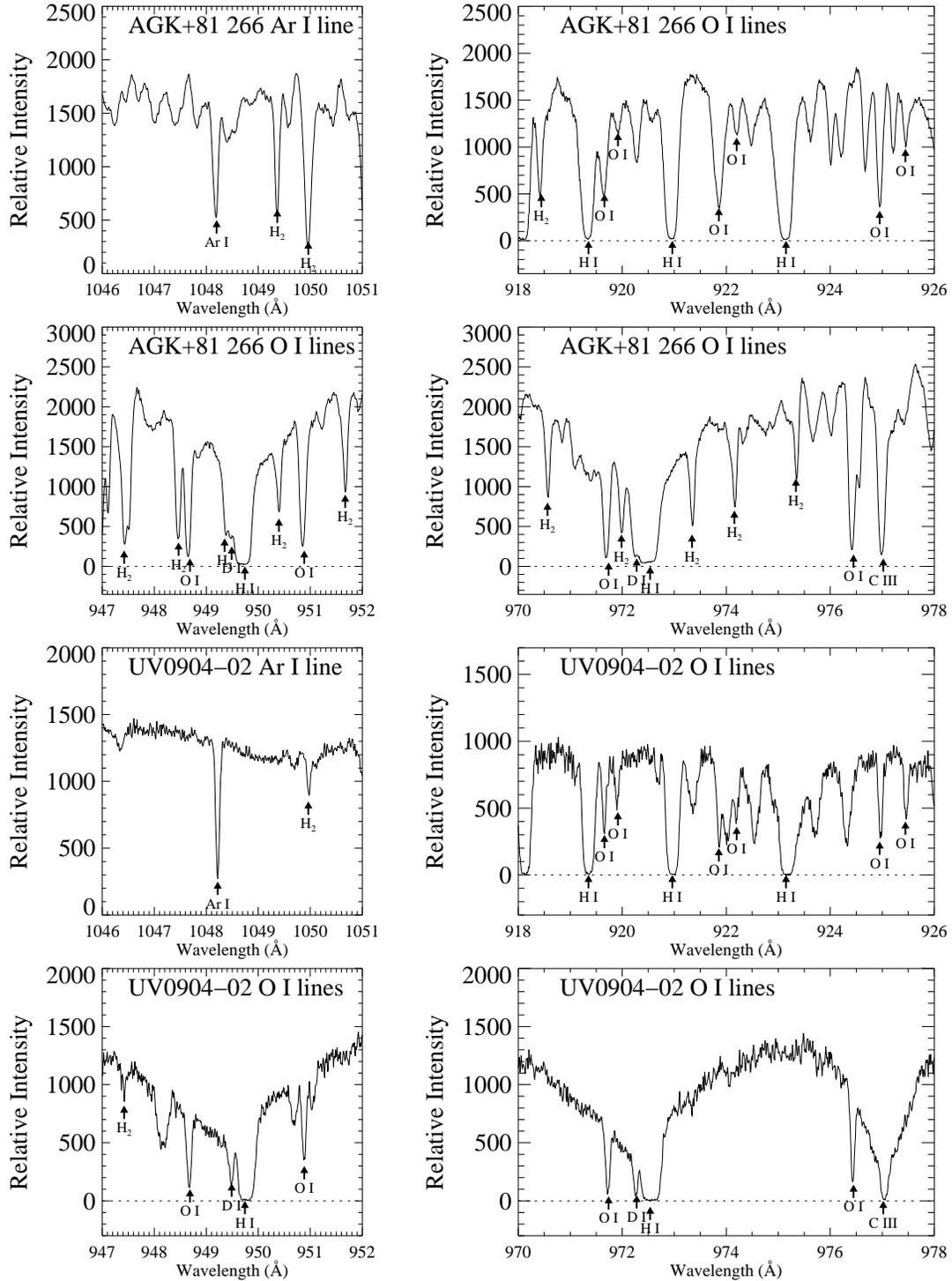}
\caption{Examples of {\it FUSE\/} spectra, where signals from all detector 
channels have been combined, for the two stars AGK+81~266 (top group of 
panels) and UV0904$-$02 (bottom group).  Various interstellar features are 
indicated.  The two panels showing the Ar~I feature best illustrate the large 
contrasts in the strengths of unidentified stellar features that can 
interfere with the interstellar ones.\label{sample_spectra}}
\end{figure}
\twocolumn

SJ98 presented a number of 
theoretical arguments that suggested Ar is not appreciably depleted in 
the low density ISM that we can observe.  However, it would be good to 
confirm this outlook by some independent, experimental means.  Fortunately, 
krypton is an element that can be observed in the ISM, and, like argon, is 
chemically inert.  It would be reasonable to expect that the capture of Kr 
onto interstellar dust grains, if it happens, would be similar to that of Ar.  
An advantage of studying Kr is that its interstellar features are weak 
 (Cartledge et al. 2008), which means that they can be used to measure 
column densities over sight lines that have high values of $n({\rm H}_{\rm 
tot})$ where element depletions should be generally very strong.  As with Ar, 
Kr has a photoionization cross section that is substantially larger than that 
of H (Sterling 2011).  Thus, any simple measure of the deficiencies 
of this element in the low density ISM could simply be a product of it being 
more easily photoionized than H.

A way to overcome the confusion from the offset produced by ionization is to 
compare differential capture rates of elements onto grains as the conditions 
that favor grain formation change.  For instance, Jenkins (2009) has
determined that for sight lines with $N({\rm H}^0)>10^{19.5}$, where
ionization corrections should be small, when $5\times 10^5$ O atoms are
removed from the gas phase, $1(\pm 1)$ atom of Kr vanishes.  Since O is more
abundant than Kr by a factor of $2.5\times 10^5$, any relative decrease in
the abundance of Kr in the gas phase would be about half that of O (but the
errors allow for this factor to range from zero to being equal to that of
oxygen).  If we accept the idea that Ar depletes in the same manner as Kr,
probably to within a factor of $\sqrt{m_{\rm Kr}/m_{\rm Ar}}$, and that in low
density media O shows very low depletions ($<0.1\,$dex) (Jenkins 2009), it is
reasonable to adopt a gas-phase abundance ratio that is virtually the same as
the protosolar ones, $\log ({\rm Ar/O})_\odot=-2.16\pm 0.11$.  If indeed
there is some mild depletion of O due to the formation of silicate dust
grains, we may understate the strength of the ionization of Ar.

\subsubsection{Interpretation of Line Strengths}\label{line_strengths}

\begin{figure}
\epsscale{1.25}
\plottwo{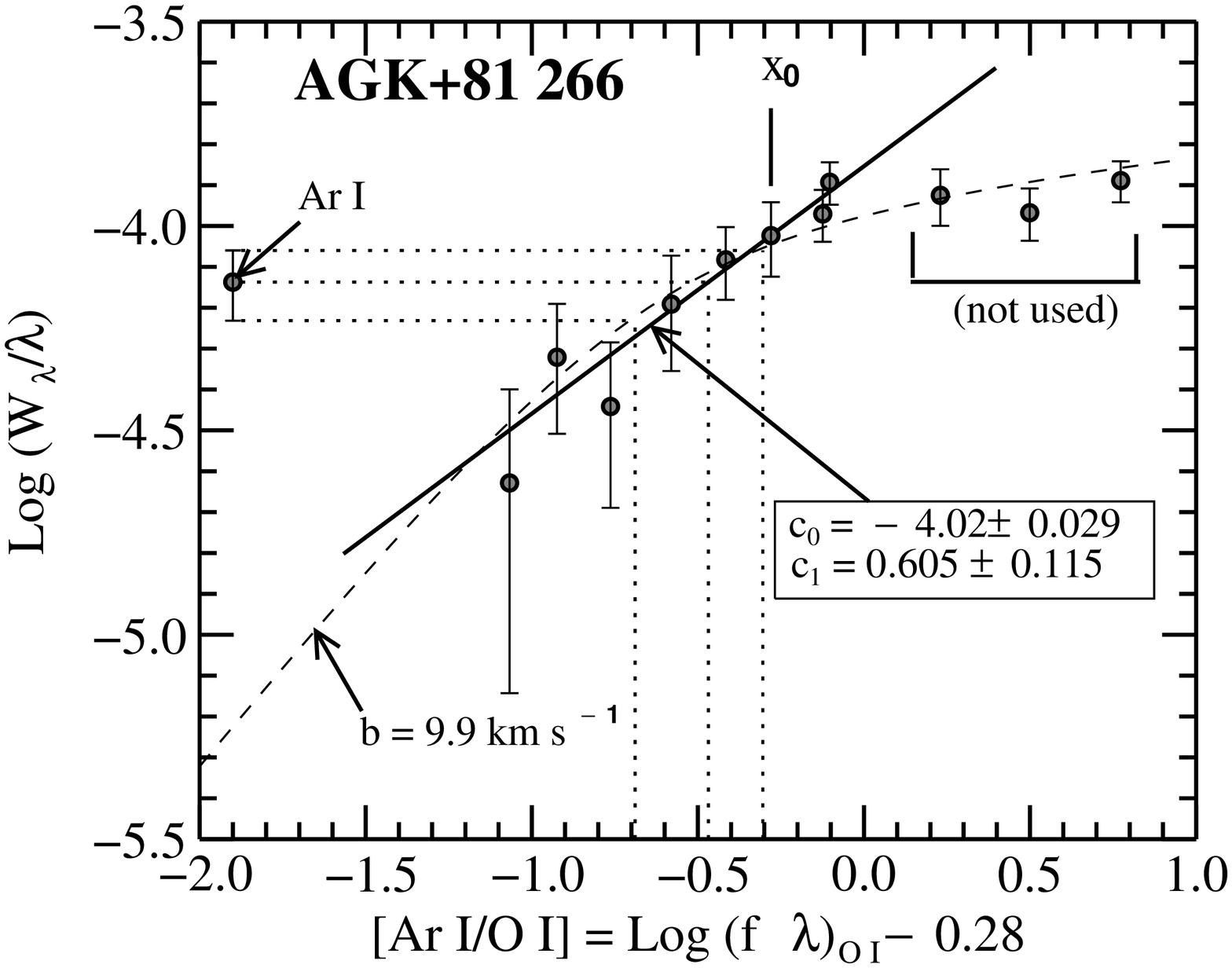}{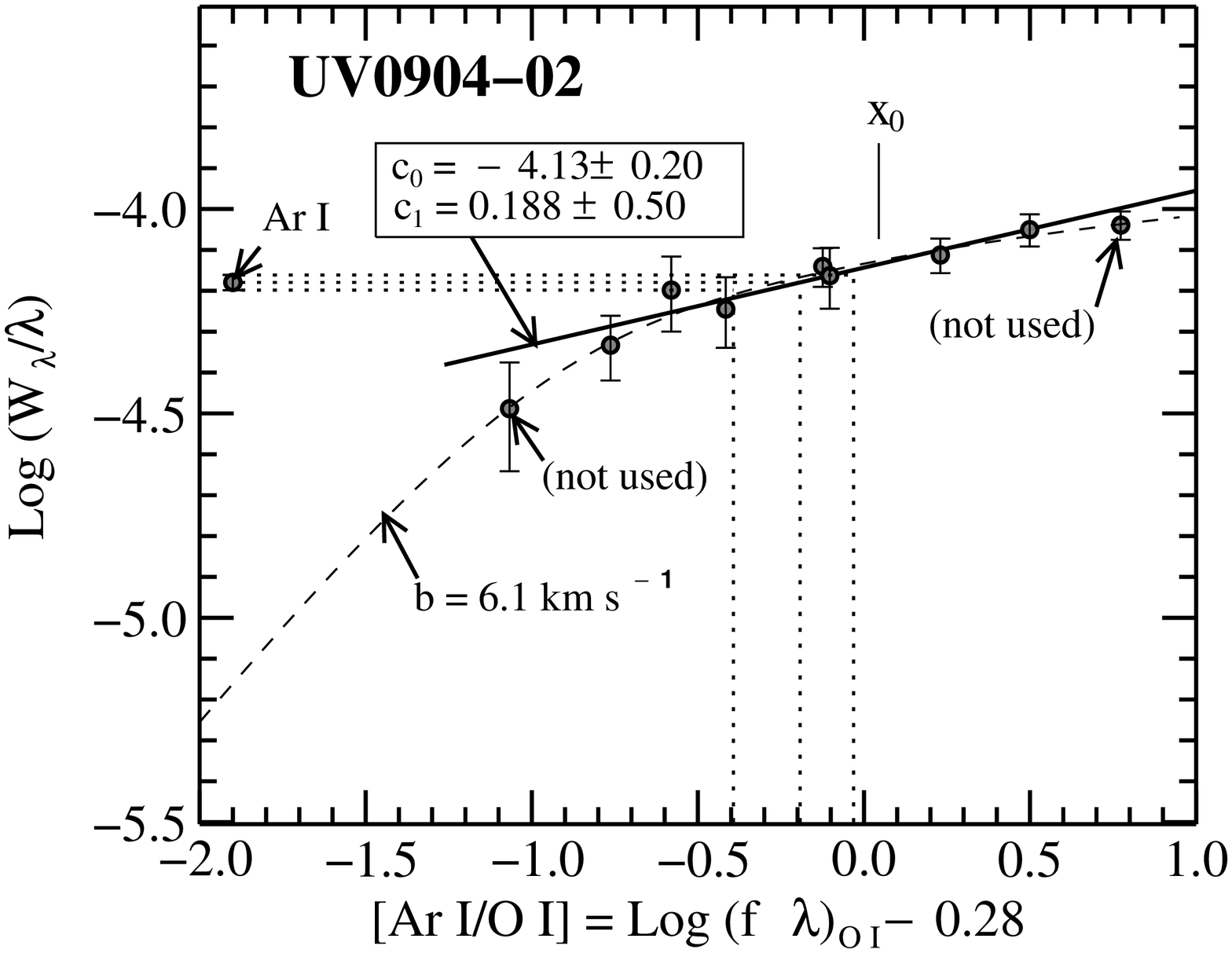}
\caption{Examples of how the values of [Ar~I/O~I] are derived for the two 
stars chosen for Fig.~\protect\ref{sample_spectra}.  The equality shown in 
each $x$-axis label applies to the horizontal projection of the Ar~I line 
strength onto the O~I curve of growth, as expressed in 
Eq.~\protect\ref{ar_o_eqn}.  We performed weighted least-squares linear fits 
for the logarithms of $W_\lambda/\lambda$ for lines of O~I that are most 
influential in establishing the trends ({\it solid lines\/}) for a comparison 
with the measurements of $\log W_\lambda/\lambda$ of the Ar~I line.  Values 
for $c_0$ and $c_1$ given in Eq.~\ref{linear_trend} that define these trends 
are shown in the boxes.  The best-fit curves of growth together with their 
$b$ values are also shown, but they are not used in the derivations.  O~I 
lines that had strengths that were well above or below those in the important 
portion of curve of growth were not included in the derivation of the best 
fits, and they are indicated here as ``(not used).''  These unused lines are 
identified with parentheses around the equivalent width values listed in 
Table~\protect\ref{EW_table}.\label{oi_fits}}
\end{figure}

We adopt the premise that the distribution of radial velocities of the 
neutral argon atoms is identical to that of neutral oxygen (but this is not 
exactly correct; we will revisit this issue later). Figure~\ref{oi_fits} 
shows examples of some standard curve-of-growth plots for the O~I lines 
appearing in the spectra of the same two targets that were featured in 
Fig.~\ref{sample_spectra}, AGK+81~266 and UV0904$-$02.  The former of the two 
illustrates an average amount of line saturation for relevant features, while 
the latter represents an extreme case of saturation caused by a low overall 
dispersion of radial velocities.  In principle, we could have derived values 
for $N$(O~I) from the best-fit curves of growth shown by the dashed curves in 
the figure panels and then assume that $N$(Ar~I) follows from the equivalent 
width of the one available line at 1048.220$\,$\AA\ assuming the same 
velocity dispersion parameter $b$ as that found for O~I. Instead, we used a 
much simpler approach that sidesteps the goal of deriving explicit values of 
$N$ and $b$ (whose errors are strongly correlated) and proceeds directly to 
an answer for just the ratio of the two column densities.  An advantage here 
is that we can make a straightforward analytical determination of the 
uncertainty of the outcome based on the errors of some linear fitting 
coefficients.

The comparison of Ar~I to O~I is based on the following principle.  If one 
could imagine the existence of a hypothetical O~I line with a transition 
strength $\log (f\lambda)_{\rm O~I}$ that is just right to produce a value of 
$W_\lambda/\lambda$ that exactly matches that of the Ar~I line, one could 
then derive the deficiency of Ar~I with respect to its expectation based on 
O~I, which we denote in logarithmic form as [Ar~I/O~I].  This quantity yields 
the logarithm of the ratio of the two neutral fractions relative to the solar 
abundance ratio and is given by the relation
\begin{eqnarray}\label{ar_o_eqn}
[{\rm Ar~I/O~I}]&=&\log (f\lambda)_{\rm O~I}-\log (f\lambda)_{\rm Ar~I}-\log 
({\rm Ar/O})_\odot\nonumber\\
&=&\log (f\lambda)_{\rm O~I}-0.28\pm 0.11,
\end{eqnarray}
where $\log (f\lambda)_{\rm Ar~I}=2.440\pm 0.004$ (Morton 2003).

\onecolumn
\twocolumn

To determine the strength of the hypothetical O~I line, we perform a weighted 
least-squares linear fit for $\log (W_\lambda/\lambda)$ vs. the quantity on 
the right-hand side of Eq.~\ref{ar_o_eqn} (the abscissa for each plot in 
Fig.~\ref{oi_fits}) for an appropriate selection of O~I lines.  [As with the 
Ar~I line the $f$-values of the O~I transitions are from Morton 
 (2003).] The lines that are chosen for this fit are ones 
that are situated not too far from the horizontal projection (shown by dotted 
lines in the figure) of $\log (W_\lambda/\lambda)$ for the single line of 
Ar~I onto the trend for the O~I lines.  With this restricted fit, we define a 
simple relationship in $\log (W_\lambda/\lambda)$ that is a good 
approximation to a relevant portion of the curve of growth.

In the two panels of Fig.~\ref{oi_fits}, these best-fit linear trends are 
shown by the straight solid lines.  They depict the relation between $y=\log 
W_\lambda/\lambda$ and $x=\log (f\lambda)-0.28$ according to the equation
\begin{equation}\label{linear_trend}
y=c_0+c_1(x-x_0),
\end{equation}
where
\begin{equation}\label{x_0}
x_0=\sum_i x_i\sigma (y)_i^{-2}\Big/ \sum_i \sigma (y)_i^{-2}
\end{equation}
represents a zero reference point that produces a vanishing covariance for 
the errors in the fitting coefficients $c_0$ and $c_1$.  A nominal value on 
the $x$ axis for the projection of $y_{\rm Ar~I}$ onto the linear relation is 
given by
\begin{equation}\label{aro_solution}
x_{\rm Ar~I}=x_0+{y_{\rm Ar~I}-c_0\over c_1}
\end{equation}
In the fraction part of this equation, the numerator and denominator have 
errors $(\sigma(y_{\rm Ar~I})^2+\sigma(c_0)^2)^{0.5}$ and $\sigma(c_1)$, 
respectively.  A conventional approach for deriving the error of the quotient 
is to add in quadrature the relative errors of the two terms, yielding the 
relative error of the quotient.  However, this scheme breaks down when the 
error in the denominator is not very much less than the denominator itself.  
A more robust way to derive the error of a quotient has been developed by 
Geary (1930); for a concise description of this method see 
Appendix~A of Jenkins (2009).  We use this method here; it 
is effective as long as there is little chance that the denominator minus its 
error could become very close to zero or be negative, i.e., ${\rm 
denom.}/\sigma({\rm denom.}) \gtrsim 3$.

The dotted lines in Figure~\ref{oi_fits} show schematically how the best fit 
values and the error ranges for [Ar~I/O~I] are derived.  Note that the 
horizontal and vertical segments for the error limits do not exactly 
intersect the best linear trend for the O~I lines because the error analysis 
allows for the uncertainty for the location of this line.  (But we point out 
that the intersections for the worst possible error in one direction for 
$y_{\rm Ar~I}$ do not occur at the locations for the worst possible 
deviations in the opposite direction for the trend line.)  Also, the final 
errors for $[{\rm Ar~I}/{\rm O~I}]=x_{\rm Ar~I}$ are not symmetrical about 
the best values.  On average, the upward error bounds are about 75\% as large 
as the negative ones.

The difference in atomic weights of Ar and O will cause the thermal 
contributions to the Doppler broadenings of these two elements to differ from 
each other.  The impact of this effect on our results for the column density 
ratios is small however.  For instance, if we consider that we are viewing 
absorption lines arising from the warm, neutral medium (WNM) and there were 
no bulk motions of the gas, the line broadening parameters $b_{\rm therm.}$ 
caused by thermal Doppler broadening for $T=7000\,$K would be 2.7 and 
$1.7\,{\rm km~s}^{-1}$ for O~I and Ar~I, respectively.  For a typical 
observation, such as the one for AGK+81~266 illustrated in the left-hand 
panel of Fig.~\ref{oi_fits}, we find that the observed curve of growth for 
the O~I lines, yielding an apparent $b_{\rm obs.}=9.9\,{\rm km~s}^{-1}$, 
indicates that kinematic effects arising from turbulent motions (or multiple 
velocity components) should be the most important contribution to the 
broadening since $b_{\rm turb.}=\sqrt{b_{\rm obs.}^2-b({\rm O~I})_{\rm 
therm.}^2}=9.53\,{\rm km~s}^{-1}$ is only slightly smaller than $b_{\rm 
obs.}$.  The curve of growth that characterizes the saturation of the Ar~I 
line would conform to a slightly lower velocity dispersion parameter compared 
to that observed for O~I, $b_{\rm obs.}=\sqrt{b_{\rm turb.}^2+b({\rm 
Ar~I})_{\rm therm.}^2}=9.68\,{\rm km~s}^{-1}$,  because the higher atomic 
weight of Ar causes the thermal contribution to be smaller.  For the observed 
equivalent width of the Ar~I line, the error in the ratio of column densities 
caused by our assumption that the values of $b_{\rm obs.}$ of the two 
elements are identical will create an underestimate of $[{\rm 
Ar~I/O~I}]=-0.012\,$dex.  A similar calculation for the more extreme line 
saturation exhibited by UV0904$-$02 shown in the right-hand panel of 
Fig.~\ref{oi_fits} indicates that the perceived outcome for [Ar~I/O~I] could 
be too low by $-0.12\,$dex.  Cases showing this much saturation are rare for 
our collection of sight lines.

If the kinematic line broadening is not an approximately Gaussian form that 
one would expect from pure turbulent broadening, but instead results from 
distinct and well-separated narrow components, the errors in [Ar~I/O~I] could 
be larger than those evaluated above.  However, the recordings of Ar~I lines 
for 9 different stars made by {\it IMAPS\/} at a resolution of $4\,{\rm 
km~s}^{-1}$ that were shown by SJ98 reveal profiles that, while not exactly 
Gaussian, are nevertheless generally smooth and devoid of any narrow spikes 
that are isolated from each other.  Thus, the numerical estimates presented 
in the above paragraph should be reasonably accurate.

One might question whether or not errors in the adopted $f$-values could 
cause global systematic errors in the evaluations of [Ar~I/O~I].  While 
Morton (2003) listed a very small uncertainty for the 
$f$-value of the Ar~I line at 1048.220$\,$\AA, he did not specify errors for 
the O~I lines.  Nevertheless, empirical evidence from high quality {\it 
FUSE\/} observations of WD stars by H\'ebrard et al. (2002), 
Sonneborn et al. (2002) and Oliveira et al. 
(2003) all showed curves of growth that are 
remarkably well behaved for the same lines that are used in the present 
study.  While one could still pose the objection that all of the O~I lines 
could collectively have a systematic error of a certain magnitude, and yet 
could still yield acceptable curves of growth, this seems unlikely: the 
values of [O~I/H~I] derived by Sonneborn et al. (2002) and Oliveira et al.
(2003) are generally consistent with those found elsewhere in the
ISM based on measurements of the
intersystem O~I line at 1355.6$\,$\AA\ (Jenkins 2009).
[H\'ebrard et al. (2002) did not attempt to measure
$N$(H~I).]

\subsubsection{Outcomes}\label{outcomes}

Table~\ref{aro_results} shows the outcomes of our analysis for all of the 
targets in the survey, along with the applicable Galactic coordinates and 
apparent magnitudes of the stars.  The last two columns present some cautions 
in numerical form.  First, Column (9) lists for the fraction part of 
Eq.~\ref{aro_solution} the denominator divided by its error.  If this number 
is less than about 3, the upper limit for [Ar~I/O~I] should not be trusted.  
Second, Column (10) lists the probability of obtaining a worse fit of the O~I 
lines to the linear trend (i.e., a higher value for $\chi^2$), given the 
errors that we derived. A plot of the frequency of all of these numbers shows 
a distribution that is consistent with a uniform distribution between 0 and 
1, which indicates that our error estimates for $W_\lambda$ of the O~I lines 
are neither too conservative nor too generous.  Thus, any individual case 
where this probability is low should not be considered as a real anomaly.

\begin{figure}[h!]
\epsscale{1.0}
\plotone{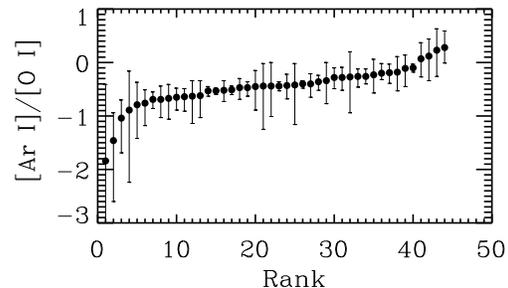}
\caption{Outcomes for [Ar~/O~I] sorted according to their best 
values.\label{sorted_results}}
\end{figure}
\begin{figure}[h!]
\vspace{-.3cm}
\plotone{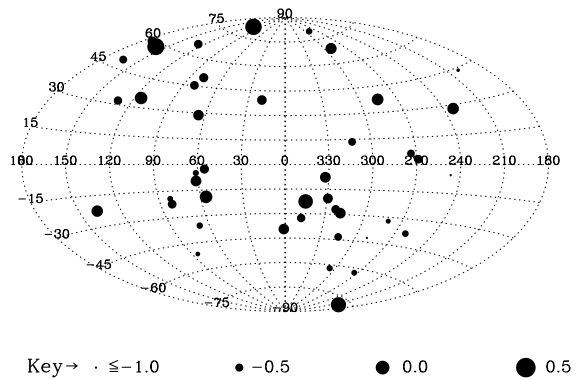}
\caption{Locations of the stars and their values of [Ar~I/O~I] shown on an 
Aitoff projection of Galactic coordinates.  The sizes of the black circles 
indicate the values of [Ar~I/O~I] according to the key shown at the 
bottom.\label{sky_display}}
\end{figure}

Figure~\ref{sorted_results} shows a graphic representation of all of the 
results and their uncertainties.  They were ranked and then arranged in order 
of the best values of [Ar~I/O~I] to make it easier to see the dispersion of 
results and also to show that the more extreme deviations often represent 
cases where the errors are somewhat larger than normal.
\onecolumn
{\setlength{\textheight}{9in}
\begin{deluxetable}{
l   
c   
c   
c   
c   
c   
c   
c   
c   
c   
}
\tabletypesize{\footnotesize}
\tablecolumns{10}
\tablewidth{0pt}
\tablecaption{Results for [Ar I/O I]\tablenotemark{a}\label{aro_results}}
\vspace{-.5cm}
\tablehead{
\colhead{} & \multicolumn{2}{c}{Galactic Coord.}\tablenotemark{b} &
\colhead{} & \colhead{} & \multicolumn{3}{c}{[Ar I/O I]}\\
\cline{2-3} \cline{6-8}
\colhead{Star} & \colhead{$\ell$} & \colhead{$b$} &
\colhead{} & \colhead{} & \colhead{Lower} & \colhead{Best} &  \colhead{Upper} 
&
\colhead{Denom. Val.}&\colhead{Prob. of}\\
\colhead{Name} & \colhead{(deg.)} & \colhead{(deg.)} & 
\colhead{$m_B$\tablenotemark{b}} &
\colhead{$m_V$\tablenotemark{b}} & \colhead{Limit} & \colhead{Value} &
\colhead{Limit} & \colhead{/Error\tablenotemark{c}} &
\colhead{Worse Fit}\\
\colhead{(1)}& \colhead{(2)}& \colhead{(3)}& \colhead{(4)}&
 \colhead{(5)}& \colhead{(6)}& \colhead{(7)}& \colhead{(8)}& \colhead{(9)}&
\colhead{(10)}
}
\startdata
2MASSJ15265306+7941307&$115.06$&$34.93$&$11.30$&$11.60$&$-0.45$&$-0.11$&$0.14
$&$4.74$&$0.26$\\
AA 
Dor\dotfill&$280.48$&$-32.18$&$10.84$&$11.14$&$-1.18$&$-0.76$&$-0.51$&$2.60$&
$0.70$\\
AGK+81 
266\dotfill&$130.67$&$31.95$&$11.60$&$11.94$&$-0.69$&$-0.47$&$-0.30$&$5.25$&$
0.89$\\
BD+18 
2647\dotfill&$289.48$&$80.14$&$11.48$&$11.82$&$-1.14$&$-0.63$&$-0.34$&$3.76$&
$0.66$\\
BD+25 
4655\dotfill&$81.67$&$-22.36$&$9.39$&$9.68$&$-1.25$&$-0.44$&$-0.02$&$4.68$&$0
.80$\\
BD+28 
4211\dotfill&$81.87$&$-19.29$&$10.17$&$10.51$&$-1.03$&$-0.69$&$-0.44$&$7.42$&
$0.07$\\
BD+37 
442\dotfill&$137.07$&$-22.45$&$9.69$&$9.92$&$-0.32$&$-0.20$&$-0.02$&$2.54$&$0
.04$\\
BD+39 
3226\dotfill&$65.00$&$28.77$&$9.89$&$10.18$&$-0.49$&$-0.28$&$-0.10$&$3.42$&$0
.69$\\
CPD$-$31 
1701\dotfill&$246.46$&$-5.51$&$10.22$&$10.56$&$-1.69$&$-1.04$&$-0.70$&$4.68$&
$0.42$\\
CPD$-$71 
172\dotfill&$290.20$&$-42.61$&$10.90$&$10.68$&$-2.60$&$-1.46$&$-0.94$&$2.28$&
$0.77$\\
EC11481-2303\dotfill&$285.29$&$37.44$&$11.49$&$11.76$&$-0.53$&$-0.18$&$0.11$&
$4.81$&$0.43$\\
Feige 
34\dotfill&$173.32$&$58.96$&$10.84$&$11.18$&$-1.16$&$-0.42$&$0.02$&$3.43$&$0.
82$\\
HD113001\dotfill&$110.97$&$81.16$&$9.65$&$9.65$&$-0.26$&$0.23$&$0.63$&$4.12$&
$0.42$\\
JL 
119\dotfill&$314.61$&$-43.36$&$13.22$&$13.49$&$-0.63$&$-0.53$&$-0.45$&$5.68$&
$0.16$\\
JL 
25\dotfill&$318.63$&$-29.17$&$13.09$&$13.28$&$-0.94$&$-0.27$&$0.20$&$3.54$&$0
.29$\\
JL 
9\dotfill&$322.60$&$-27.04$&$12.96$&$13.24$&$-0.68$&$-0.43$&$-0.22$&$5.12$&$0
.24$\\
LB 
1566\dotfill&$306.36$&$-62.02$&$12.81$&$13.13$&$-0.89$&$-0.65$&$-0.48$&$3.06$
&$0.82$\\
LB 
1766\dotfill&$261.65$&$-37.93$&\nodata&$12.34$&$-1.03$&$-0.62$&$-0.34$&$4.71$
&$0.07$\\
LB 
3241\dotfill&$273.70$&$-62.48$&$12.45$&$12.73$&$-0.86$&$-0.69$&$-0.55$&$7.98$
&$0.01$\\
LS 
1275\dotfill&$268.96$&$2.95$&$10.94$&$11.40$&$-1.01$&$-0.44$&$0.00$&$4.69$&$0
.77$\\
LSE 
234\dotfill&$329.44$&$-20.52$&\nodata&$12.63$&$-0.77$&$-0.34$&$0.00$&$3.70$&$
0.38$\\
LSE 
259\dotfill&$332.36$&$-7.71$&\nodata&$12.54$&$-0.40$&$-0.26$&$-0.12$&$2.52$&$
0.08$\\
LSE 
263\dotfill&$345.24$&$-22.51$&$11.40$&$11.30$&$-0.27$&$0.07$&$0.37$&$6.32$&$0
.24$\\
LSE 
44\dotfill&$313.37$&$13.49$&$12.21$&$12.45$&$-0.73$&$-0.52$&$-0.34$&$5.53$&$0
.11$\\
LSII +18 
9\dotfill&$55.20$&$-2.65$&$11.81$&$12.13$&$-0.65$&$-0.40$&$-0.21$&$2.43$&$0.6
6$\\
LSII +22 
21\dotfill&$61.17$&$-4.95$&$12.23$&$12.58$&$-1.06$&$-0.67$&$-0.41$&$4.04$&$0.
91$\\
LSIV +10 
9\dotfill&$56.17$&$-19.01$&$11.71$&$11.98$&$-0.18$&$-0.10$&$-0.03$&$11.01$&$0
.21$\\
LSS 
1362\dotfill&$273.67$&$6.19$&$12.27$&$12.50$&$-0.60$&$-0.53$&$-0.48$&$7.05$&$
0.65$\\
MCT 
2005$-$5112\dotfill&$347.71$&$-32.68$&$15.20$&\nodata&$-0.63$&$-0.47$&$-0.36$
&$3.82$&$0.46$\\
MCT 
2048$-$4504\dotfill&$1.10$&$-39.51$&$14.85$&$15.20$&$-0.52$&$-0.28$&$-0.09$&$
4.97$&$0.97$\\
NGC6905  
star\dotfill&$61.49$&$-9.57$&$16.30$&$14.50$&$-0.46$&$-0.26$&$-0.12$&$2.56$&$
0.75$\\
PG0919+272\dotfill&$200.46$&$43.89$&$12.27$&$12.69$&$-2.24$&$-0.89$&$-0.16$&$
3.83$&$0.64$\\
PG0952+519\dotfill&$164.07$&$49.00$&$12.47$&$12.80$&$-0.60$&$-0.51$&$-0.44$&$
5.11$&$0.90$\\
PG1032+406\dotfill&$178.88$&$59.01$&$10.80$&\nodata&$-3.79$&$-1.84$&$-0.41$&$
2.71$&$0.30$\\
PG1051+501\dotfill&$159.61$&$58.12$&$14.59$&\nodata&$-0.01$&$0.28$&$0.59$&$2.
77$&$0.61$\\
PG1230+068\dotfill&$290.42$&$68.85$&$12.25$&\nodata&$-0.57$&$-0.23$&$0.06$&$7
.58$&$0.03$\\
PG1544+488\dotfill&$77.54$&$50.13$&$12.80$&\nodata&$-0.48$&$-0.40$&$-0.34$&$5
.41$&$0.56$\\
PG1605+072\dotfill&$18.99$&$39.33$&$13.01$&$12.84$&$-0.52$&$-0.36$&$-0.24$&$5
.31$&$0.88$\\
PG1610+519\dotfill&$80.50$&$45.31$&$13.73$&\nodata&$-0.53$&$-0.44$&$-0.36$&$1
0.93$&$0.22$\\
PG2158+082\dotfill&$67.58$&$-35.48$&$12.66$&\nodata&$-0.91$&$-0.64$&$-0.49$&$
2.78$&$0.16$\\
PG2317+046\dotfill&$84.84$&$-51.10$&\nodata&$12.87$&$-1.42$&$-0.79$&$-0.37$&$
4.20$&$0.04$\\
Ton 
102\dotfill&$127.05$&$65.78$&$13.29$&$13.54$&$-0.89$&$-0.45$&$-0.15$&$2.96$&$
0.31$\\
Ton 
S227\dotfill&$201.39$&$-77.81$&$11.60$&$11.90$&$-0.34$&$0.12$&$0.44$&$9.06$&$
0.36$\\
UV0904$-$02\dotfill&$232.98$&$28.12$&$11.64$&$11.96$&$-0.39$&$-0.19$&$-0.03$&
$3.78$&$0.93$\\
\enddata
\tablenotetext{a}{$[{\rm Ar~I/O~I}]\equiv {N({\rm Ar~I})/N({\rm O~I})\over 
({\rm Ar/O})_\odot}$.  The range of possible values shown here do not include 
an overall systematic error that arises from the uncertainties in the 
constants that are included in Eq.~\protect\ref{ar_o_eqn}, which amounts to a 
combined error of 0.11~dex.}
\tablenotetext{b}{Coordinates and apparent magnitudes were supplied by the 
SIMBAD database.}
\tablenotetext{c}{The relevance of this quantity is discussed in the text 
that follows Eq.~\protect\ref{aro_solution}.  If the listed value is less 
than about 3, the positive value of the error quotient may be misleading.}
\end{deluxetable}
}
\twocolumn

Figure~\ref{sky_display} shows the locations of the targets in the sky and 
their respective values of [Ar~I/O~I].  While no obvious regional trends seem 
to be evident, we can perform a test to determine whether or not the 
variability of the outcomes exceeds what we would have expected from our 
errors (assuming that they are correct).  We do this by computing a value for 
$\chi^2$, where the choice for the error of each case depends on whether a 
test value is above or below the measurement outcome.  This procedure 
properly takes into account the asymmetries of the errors.  Adopting this 
method, we find that a minimum $\chi^2$ of 66.9 for the 44 measurements 
occurs at a value $[{\rm Ar~I/O~I}]=-0.438$.  This minimum for the $\chi^2$ 
with 43 degrees of freedom is greater than what we would have expected from 
our errors alone (the probability of obtaining by chance a higher value for 
$\chi^2$ is only 1\%).  If we were to propose that the real variability 
across the sky is $\sigma=0.11~{\rm dex}$, and we add this value in 
quadrature to all of the experimental errors, the minimum $\chi^2$ drops to 
42.4, which makes the probability of a worse fit equal to 50\%.  The location 
of this new minimum is at $[{\rm Ar~I/O~I}]=-0.427$.

The discussion above has considered only the random fluctuations arising from 
either the measurements or the true variability in [Ar~I/O~I] in different 
sight lines.  We must not lose sight of the fact that there can be an overall 
systematic error of 0.11~dex for the entire collection.  This global error 
arises from the uncertainties of the value for $\log ({\rm Ar/O})_\odot$ that 
went into Eq.~\ref{ar_o_eqn}.  It is much larger than the error in the 
weighted mean value for all of the measurements.

\section{ARE THE STARS BEYOND THE BOUNDARY OF THE LOCAL 
BUBBLE?}\label{beyond}

As discussed in Section~\ref{strategy}, our objective is to sample 
interstellar material that is beyond the edge of the Local Bubble, enough so 
that our measurements are not heavily influenced by the very low density gas 
within this cavity.  In principle, we could compare the three-dimensional locations of the 
target stars with maps that outline the boundary of the Local Bubble 
 (Vergely et al. 2010; Welsh et al. 2010; Reis et al. 2011) to indicate
whether or not we are primarily sampling gas in the surrounding denser
medium.  However, the distances to our targets are uncertain, which makes
this approach unworkable.  Instead, we adopt a definition of the boundary
proposed by Sfeir et al. (1999) (and one that was also used by Lehner et al.
(2003)), who linked its location to the sudden onset of Na~I D-line
absorption that crossed a threshold $W_\lambda({\rm D2})=20\,$m\AA.  This
threshold is equivalent to $N({\rm H~I})\approx 2\times 10^{19}{\rm cm}^{-2}$
 (Ferlet et al. 1985), which in turn corresponds to $N({\rm O~I})$ that is
slightly greater than $10^{16}{\rm cm}^{-2}$.  Lehner et al.  (2003) found
that a typical velocity dispersion parameter $b({\rm O~I})=6\,{\rm km~s}^{-1}$
occurred inside the Local Bubble.  Using these two parameters for O~I, $(N,
~b)=(10^{16}{\rm cm}^{-2},~6\,{\rm km~s}^{-1})$, we can compute the equivalent
widths for all of the O~I lines when the boundary is crossed.  The third row
of numbers in the column headings in Table~\ref{EW_table} shows the values of
$W_\lambda$ (in m\AA) for all but the strongest three lines.  By comparing
these values with the entries that show our measurements, particularly the
weaker but securely measured ones, we can ascertain that our targets are
beyond the edge of the Local Bubble.

\section{INTERPRETATION: FUNDAMENTAL PROCESSES AND 
EQUATIONS}\label{fundamentals}

In this section, we address the basic physical processes that relate our 
findings on [Ar~I/O~I] to the ionization balance of the gas and the resulting 
degree of partial ionization.  Our discussion about the means of ionizing the 
atoms will focus mainly on the primary ionization by photons, along with the 
effects of collisional ionizations caused by secondary electrons that 
originate from these primary ionizations.  These two processes are the most 
important sources of ionization, and they represent one side of the balance 
between recombinations with free electrons and charge exchange reactions 
between various constituents of the medium.  For completeness, we will also 
cover other means of ionizing atoms and creating free electrons, such as 
cosmic ray ionizations, the ionizations of inner shell electrons of heavy 
elements by X-rays, the creation of ionizing photons when helium ions 
recombine, and the nearly complete ionization of many elements that have 
first ionization potentials below that of hydrogen.

\subsection{Direct and Secondary Ionizations of H and Ar}\label{ionization}

\subsubsection{Photoionization}\label{photoionization}

The primary photoionization cross sections for neutral Ar are larger than 
those for H by about one order of magnitude at low energies, and the ratio 
increases substantially at higher energies.  Various secondary ionizing 
processes initiated by the primary photoionizations of H and He likewise have 
a stronger effect on Ar than on H.  This contrast in ionization rates is the 
fundamental tool that we use in the interpretation of the Ar data to quantify 
the photoionization of H and the subsequent creation of free electrons.  For 
the collective effect of all of these ionization channels, we can construct a 
simple formalism based on arguments created by SJ98.  They defined a quantity 
based on ionization rates $\Gamma$ and recombination coefficients $\alpha$ 
for the two elements,
\begin{equation}\label{p_ar}
P_{\rm Ar}= {\Gamma({\rm Ar}^0)\alpha({\rm H}^0,T)\over \Gamma({\rm 
H}^0)\alpha({\rm Ar}^0,T)}~.
\end{equation}
SJ98 considered only primary photoionizations of these two elements.  Going 
beyond their development, we construct a more comprehensive picture by 
considering some refinements in the calculations of $\Gamma$ for both ${\rm 
H}^0$ and ${\rm Ar}^0$.

First, we start with the primary photoionization rates $\Gamma_p=\int 
F(E)\sigma(E)dE$, where $F(E)$ is the ambient photon flux as a function of 
energy $E$ and $\sigma(E)$ is the photoionization cross section for either 
H$^0$
 (Spitzer 1978, pp. 105-106) or Ar$^0$ (Marr \& West 1976).  
Next, we include secondary ionizations with rates $\Gamma_s$ that are created 
by the collisions from energetic electrons that are liberated by the primary 
photoionizations of H and He.  Added to this are the effects from photons 
with energies above about 300$\,$eV, which can interact with the abundant 
heavy elements in the ISM to produce additional energetic electrons that will 
ionize H and Ar with a rate that we identify as $\Gamma_{s^\prime}$.  These 
electrons arise from the primary ionizations of the inner electronic shells, 
and they are supplemented by one or more additional electrons from the Auger 
process.  Finally, we must acknowledge that recombinations of singly- and 
doubly-ionized He ions with electrons create additional photons when 
de-excitation occurs in the lower stage of ionization, either He$^0$ or 
He$^+$.  The importance of not overlooking secondary electrons and 
recombination photons from the He ionizations is underscored by the fact that 
while the abundance of He is only 1/10 that of hydrogen, its primary 
ionization cross section is 6 to 100 times that of H over the energy range 25 
to 4000$\,$eV.  Most of the recombination photons are capable of ionizing 
both H and Ar, and they supplement the other sources of ionization with rates 
$\Gamma_{He^0}$ and $\Gamma_{He^+}$.  In short, we consider that the total 
ionization rates for the two elements in Eq.~\ref{p_ar} each consist of five 
contributions,
\begin{equation}\label{Gammas}
\Gamma=\Gamma_p+\Gamma_s+\Gamma_{s^\prime}+\Gamma_{He^+}+\Gamma_{He^0}
\end{equation}
The details of how we compute $\Gamma_s$, $\Gamma_{s^\prime}$, 
$\Gamma_{He^+}$ and $\Gamma_{He^0}$ are discussed in Appendices~\ref{gamma_s} 
and \ref{gamma_he}.

The quantity $P_{\rm Ar}$ defined in Eq.~\ref{p_ar} provides a means for 
evaluating how large the neutral fraction of Ar should be relative to that of 
H according to the formula
\begin{equation}\label{ar_h}
[{\rm Ar~I/H~I}]=\log\left[ {1+n({\rm H}^+)/n({\rm H}^0)\over 1+P_{\rm Ar} 
n({\rm H}^+)/n({\rm H}^0)}\right] .
\end{equation}
For a more accurate formulation of [Ar~I/H~I] that will be developed later in 
Section~\ref{equilibrium}, we will introduce a more refined parameter 
$P^\prime_{\rm Ar}$ that will be based on a calculation that is more 
elaborate than the one shown in Eq.~\ref{p_ar}.  This new parameter will be 
substituted for $P_{\rm Ar}$ in Eq.~\ref{ar_h}.  Under most conditions, the 
differences between $P^\prime_{\rm Ar}$ and $P_{\rm Ar}$ are small.

\begin{figure}
\epsscale{.8}
\plotone{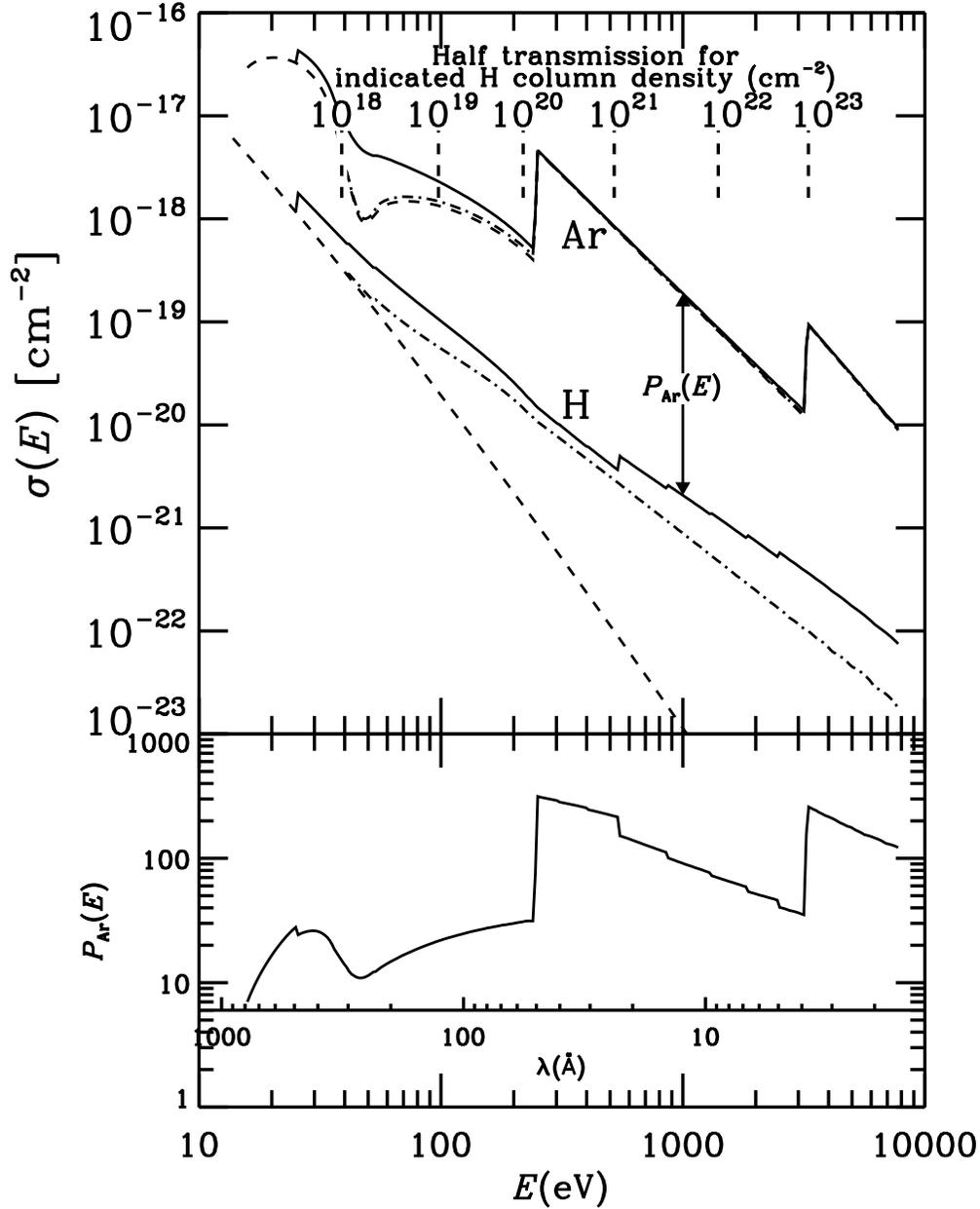} 
{\renewcommand{\baselinestretch}{.95}
\caption{{\it Upper panel:\/} Various combinations of effective 
photoionization cross sections for neutral hydrogen and argon as a function 
of energy $E$.  The dashed lines depict the absorption cross sections for the 
primary ionization rates $\Gamma_p$ that would apply to monoenergetic 
radiation at the energy $E$ shown on the $x$-axis.  The effective cross 
sections that arise by including the additional effects from ionizations by 
secondary electrons produced by H and He (related to $\Gamma_s$) are shown by 
the dot-dash lines.  These secondary ionization efficiencies were calculated 
according to the principles outlined in Section~\protect\ref{electrons_H_He} 
of the Appendix for an electron fraction $x_e=0.05$.; see 
Eqs.~\protect\ref{Hphi} and \protect\ref{Hephi}.  These two ionization 
processes can be further enhanced by photons that come from recombinations of 
He ions with free electrons (Appendix~\protect\ref{gamma_he}) and electrons 
from inner shell ionizations of heavy elements 
(Appendix~\protect\ref{heavy_elem}), leading to total effective cross 
sections depicted by the solid lines.  Across the top of the plot are markers 
showing the energies at which half-intensity penetration depths for the 
photons occur for various hydrogen column densities, assuming the gases have 
a solar abundance ratio for various elements.  The quantity $P_{\rm Ar}(E)$ 
defined in Eq.~\ref{p_ar} for a specific energy $E$ is approximately equal to 
the ratio of the total effective photoionization cross sections that give 
rise to $\Gamma({\rm H}^0,~{\rm Ar}^0)$ defined in Eq.~\protect\ref{Gammas}, 
since the recombination rates for the two elements are about the same for all 
temperatures.  {\it Bottom panel:\/} A plot of $P_{\rm Ar}(E)$ as a function 
of energy.\label{p_ar_plot}}}
\end{figure}

Figure~\ref{p_ar_plot} shows for both H and Ar the monoenergetic cross 
sections for primary ionization (dashed lines), the effective enhancements 
arising from the secondary ionizations $\Gamma_s$ (dash-dot lines) for 
$x_e=0.05$, and the additional effects from $\Gamma_{s^\prime}$, 
$\Gamma_{He^+}$ and $\Gamma_{He^0}$, all of which give total ionization rates 
shown by the solid lines. For H, secondary ionizations outweigh the primary 
ones for photon energies above about 100~eV.  By contrast, we find that for 
Ar the enhancement at these energies is small compared to its much higher 
primary cross section.  A more narrowly defined version of $P_{\rm Ar}$, 
which we denote as $P_{\rm Ar}(E)$, applies to an irradiation of the gas by 
photons of a given energy $E$, rather than from a combination of fluxes over 
a broad energy range.  The definition of $P_{\rm Ar}(E)$ is illustrated in 
the upper panel of the figure, and its dependence on $E$ is shown in the 
bottom panel.

\onecolumn
\twocolumn

\subsubsection{Other Ionization Mechanisms}\label{other_ionization}

As discussed earlier in Section~\ref{intro}, neutral atoms in the ISM are 
subjected to ionization by cosmic ray particles.  These ionizations, with a 
rate $\zeta_{\rm CR}$, add to the effects of primary and secondary 
ionizations from photons discussed above.  To obtain proper values of 
$\zeta_{\rm CR}$ that apply to the diffuse medium, we must apply corrections 
to the values of $\zeta_{\rm CR}$ that were measured for the more dense media 
that have appreciable concentrations of molecules.  The details of this 
computation are discussed in Appendix~\ref{cr_ioniz}.

In the WNM, the main source of free electrons is from the 
ionization of H and He.  However, a small number of additional electrons 
arise from other atoms that can be ionized by starlight photons less 
energetic than the ionization potential of H.  For the combined effect from 
these elements, we adopt an estimate $n({\rm M}^+)$ equal to $1.5\times 
10^{-4}$ times the density of hydrogen based on the assumption that the gas 
we are viewing is in a regime where the element depletions are relatively 
modest.

\subsection{Recombination}\label{recombination}

The radiative recombination coefficients for free electrons and ions to 
create neutral hydrogen and singly-ionized helium, $\alpha({\rm H}^0,T)$ and 
$\alpha({\rm He}^+,T)$, are taken from
Spitzer (1978, pp. 105-107).  For $\alpha({\rm H}^0)$, we 
excluded recombinations to the lowest electronic level, since they generate 
Lyman limit photons that can re-ionize hydrogen atoms over a short distance 
scale.  Recombination coefficients for He$^0$ were taken from Aldrovandi \& 
P\'equignot (1973) and those for Ar$^0$ from Shull \& Van Steenberg 
 (1982).  The results for Ar$^0$ given by 
Shull \& Van Steenberg (1982) agree well 
with the radiative recombination coefficients listed by Aldrovandi \& 
P\'equignot (1974).  The minimum 
temperature where dielectronic recombination for Ar$^0$ becomes important is 
$2.5\times 10^4\,$K
 (Aldrovandi \& P\'equignot 1974), which is above temperatures that we 
will consider, hence we can ignore this process.

In addition to recombining with a free electron, an ion can also be 
neutralized by colliding with a dust grain and removing an electron from it 
 (Snow 1975; Draine \& Sutin 1987; Lepp et al. 1988; Weingartner \& Draine
2001a).  The operation of this effect on protons is important for regulating
the fraction of free electrons in the cold neutral medium (CNM), but it is of
lesser significance for the WNM [cf. Figs. 16.1 and 16.2 of
Draine (2011)], which probably dominates the sight lines in our study.  The
rate constant $\alpha_g$ for this process is normalized to the local
hydrogen density $n({\rm H_{tot}})\equiv n({\rm H}^0)+n({\rm H}^+)$, and it
depends on several physical parameters that influence the charge on the
grains, such as the electron density $n(e)$, the rate of photoelectric
emission that is driven by the intensity $G$ of the local radiation field
between 6 and 13.6~eV, and the temperature $T$.  For our equilibrium
equations in Section~\ref{equilibrium}, we have adopted parametric fits for
$\alpha_g({\rm H}^0,n(e),G,T)$ and $\alpha_g({\rm He}^0,n(e),G,T)$ from
Weingartner \& Draine  (2001a).  They do not supply fit coefficients 
for $\alpha_g({\rm Ar}^0,n(e),G,T)$, but since the ionization potential of 
Ar$^0$ is close to that of H$^0$, it is reasonable to adopt the hydrogen rate 
coefficient and divide it by the square-root of the atomic weight (40) of Ar.  
Throughout our analysis, we set $G=1.13$, which is the value recommended for 
the general ISM by Weingartner \& Draine (2001a).

\subsection{Charge Exchange}\label{charge_exchange}

In addition to the recombination processes mentioned in the previous section, 
charge exchange reactions with neutral hydrogen can also lower the ionization 
state of an atom.  With reference to such reactions for element X, ${\rm 
X}^++{\rm H}^0\rightarrow {\rm X}^0+{\rm H}^+$ and ${\rm X}^{++}+{\rm 
H}^0\rightarrow {\rm X}^++{\rm H}^+$, we adopt the notation $C^\prime({\rm 
X}^+,T)$ and $C^\prime({\rm X}^{++},T)$ for the respective rate constants.  
For our calculations of $C^\prime({\rm He}^+,T)$, $C^\prime({\rm 
He}^{++},T)$, $C^\prime({\rm Ar}^+,T)$, and $C^\prime({\rm Ar}^{++},T)$, we 
adopted Kingdon \& Ferland's (1996) fits to the 
calculations from various sources (see their Table~1 for the coefficients and 
references).

Since the ionization potential of neutral oxygen is close to that of 
hydrogen, the rate constant for charge exchange of these two species is large 
and reverse endothermic reaction is not negligible, except at very low 
temperatures.  The rate constant $C({\rm O}^+,T)$ for the reaction ${\rm 
O}^0+{\rm H}^+\rightarrow {\rm O}^++{\rm H}^0$ can be obtained from $C^\prime 
({\rm O}^+,T)$ by the principle of detailed balancing.  However, in doing so, 
one must treat the three fine-structure levels of the ground state of O$^0$ 
separately, since their energy separations are comparable to the differences 
in the ionization potentials of H and O.  The large rate constants in both 
directions (Stancil et al. 1999) assures that the ionization fraction 
of O is locked very close to a value 8/9 times that of H for $T\gtrsim 
10^3\,$K.  This is a key principle that allows us to use O as a substitute 
for H in the present study or Ar vs. H fractional ionizations.

For completeness, we should also consider charge exchange reactions between 
Ar and He, even though the abundance of He is much lower than that of H.  The 
charge transfer recombination reaction ${\rm Ar}^{++}+{\rm He}^0\rightarrow 
{\rm Ar}^++{\rm He}^+$ has a rate constant $D^\prime({\rm 
Ar}^{++},T)=1.3\times 10^{-10}{\rm cm}^3{\rm s}^{-1}$ according to Butler \& 
Dalgarno (1980), which they claim to be 
constant for $T\geq 10^3\,$K.  The charge transfer ionization reaction with 
He$^+$, i.e., ${\rm Ar}^0+{\rm He}^+ \rightarrow {\rm Ar}^++{\rm He}^0$, has 
a rate constant $D({\rm Ar}^+,T)<10^{-13}{\rm cm}^3{\rm s}^{-1}$ according to 
Albritton (1978), so we will ignore this process in the 
equation for the ionization balance for Ar (Eq.~\ref{equi1} below).

\subsection{Equilibrium Equations}\label{equilibrium}

In a medium where both hydrogen and helium are partly ionized, the densities 
of an element X in its 3 lowest levels of ionization X$^0$, X$^+$ and 
X$^{++}$ are governed by the equilibrium equations\footnote{The
development here follows that given by Eqs. 12--19 of SJ98, except that we 
have added the He charge exchange recombination reaction as an additional 
channel for reducing the ionization of element X in its doubly charged form.  
We have also added cosmic ray ionizations and have implicitly included the 
various kinds of secondary photoionizations in the definition of $\Gamma$, as 
indicated in Eq.~\protect\ref{Gammas}.  We have corrected 
Eqs.~\protect\ref{equi1} and \protect\ref{n0} here to include a missing 
$n({\rm H^+})$ term, which was a typographical oversight in the equations of 
SJ98.}
\begin{eqnarray}\label{equi1}
&&\Big[ \Gamma({\rm X}^0)+\zeta_{\rm CR}({\rm X}^0)+C({\rm X}^+,T)n({\rm 
H^+})\Big] n({\rm X}^0)\nonumber\\
&=&\Big [\alpha({\rm X}^0,T)n(e)+C^\prime({\rm X}^+,T)
n({\rm H}^0)\nonumber\\
&+&\alpha_g({\rm X}^0,n(e),G,T)n({\rm H_{tot}})\Big] n({\rm X}^+)
\end{eqnarray}
and
\begin{eqnarray}\label{equi2}
&&\Gamma({\rm X}^+)n({\rm X}^+)=\Big[ \alpha({\rm X}^+,T)n(e)+C^\prime({\rm
X}^{++},T)n({\rm H}^0)\nonumber\\
&+&D^\prime({\rm X}^{++},T)n({\rm He}^0)\Big] n({\rm
X}^{++})
\end{eqnarray}
where $\Gamma({\rm X}^y)$ is the photoionization rate of element X in
its ionization state $y$ (neutral, +, or ++) and $\alpha({\rm X}^y,T)$ is the 
recombination rate of the $y+1$ state with free electrons as a function of 
temperature $T$.  The simultaneous solution to these two equations yields the 
fractional abundances in the three ionization levels
\begin{eqnarray*}\label{n0}
f_0({\rm X},T) \equiv {n({\rm X}^0)\over n({\rm X}^0)+n({\rm
X}^+)+n({\rm X}^{++})}=\Bigg( 1\, +
\end{eqnarray*}
\begin{mathletters}
\begin{equation}
{\Big[\Gamma({\rm X}^0)+\zeta_{\rm CR}({\rm X}^0)+C({\rm 
X}^+,T)n({\rm H^+})\Big]
\Big[\Gamma({\rm X}^+)+Y\Big]\over \Big[\alpha({\rm X}^0,T)n(e) +
C^\prime({\rm X}^+,T)n({\rm H}^0)+\alpha_g({\rm X}^0,n(e),G,T)n({\rm 
H_{tot}})\Big]Y}\Bigg)^{-1}
\end{equation}
with
\begin{equation}
Y=\alpha({\rm X}^+,T)n(e)+C^\prime({\rm X}^{++},T)n({\rm
H}^0)+D^\prime({\rm X}^{++},T)n({\rm He}^0)~,
\end{equation}
\end{mathletters}
\begin{equation}\label{n++}
f_{++}({\rm X},T) \equiv {n({\rm X}^{++})\over n({\rm X}^0)+n({\rm
X}^+)+n({\rm X}^{++})}={1-f_0({\rm X},T)\over 1+Y/\Gamma({\rm X}^+)}~,
\end{equation}
and
\begin{equation}\label{n+}
f_+({\rm X},T)\equiv {n({\rm X}^+)\over n({\rm X}^0)+n({\rm X}^+)+n({\rm
X}^{++})}=1-f_0({\rm X},T)-f_{++}({\rm X},T)~.
\end{equation}

\subsection{Electron Density and the Ionization Fractions of H and 
He}\label{electron_density}

Before the ionization fractions of Ar can be derived, we must determine not 
only the ionization balance of hydrogen, but also that for helium.  We do 
this by solving
Eqs.~\ref{n0}--\ref{n+} (substituting He for X and eliminating the $D^\prime$ 
term) along with the equation for the hydrogen ionization balance,
\begin{mathletters}
\begin{equation}\label{H_ioniz_balance}
{n({\rm H}^+)\over n({\rm H}^0)}={\Gamma({\rm H}^0)+\zeta_{\rm CR}({\rm 
H}^0)+Z\over \alpha({\rm
H}^0,T)n(e)+\alpha_g({\rm X}^0,n(e),G,T)n({\rm H_{tot}})-Z}~,
\end{equation}
with
\begin{equation}\label{Z}
Z=0.1n({\rm H}^0)\big[C^\prime({\rm He}^+,T)f_+({\rm He},T)+C^\prime({\rm
He}^{++},T)f_{++}({\rm He},T)\big]~,
\end{equation}
\end{mathletters}
and the constraints
\begin{equation}\label{He_total}
n({\rm He}^0)+n({\rm He}^+)+n({\rm He}^{++})=0.1n({\rm H}^0)\left[1+{n({\rm
H}^+)\over n({\rm H}^0)}\right]
\end{equation}
and
\begin{equation}\label{n(e)}
n(e)=n({\rm H}^+)+n({\rm He}^+)+2n({\rm He}^{++})+n({\rm M}^+)~.
\end{equation}

Since the hydrogen ionization balance depends on $n(e)$ which in turn is 
influenced by the ionization fractions of He (which are also influenced by 
$n(e)$), we must solve
Eqs.~\ref{H_ioniz_balance}-\ref{n(e)} iteratively to obtain a solution.  We 
found that these equations converged very well if we kept $n({\rm H_{tot}})$ 
and the ionization rate of H pegged to a certain value.\footnote{Fixing 
$n(e)$ instead of $n({\rm H_{tot}})$ produces unstable, oscillating solutions 
under certain circumstances.}  Starting with a zero helium ionization rate, 
the iterations progressed slowly to successively higher rates until the 
final, correct value was reached and the ionization fractions had stabilized.

After obtaining the final results for the coupled hydrogen and helium 
ionization balances, we can solve for $f_0({\rm Ar})$ using
Eq.~\ref{n0}.  This result can then be used to derive a more accurate value 
for $P_{\rm Ar}$,
\begin{equation}\label{Pprime_ar}
P^\prime_{\rm Ar}={n({\rm H}^0)\over n({\rm H}^+)}\left( {1\over f_0({\rm 
Ar})}-1\right)~,
\end{equation}
which can be substituted for $P_{\rm Ar}$ in Eq.~\ref{ar_h} to obtain a 
solution for [Ar~I/H~I] that makes use of all of the physical processes that 
were discussed in Sections~\ref{ionization}--\ref{charge_exchange}.

\section{OUTCOME FROM KNOWN SOURCES OF 
PHOTOIONIZATION}\label{known_photoionization}

\subsection{External Radiation}\label{external_radiation}

Over many decades, the diffuse, soft X-ray background has been measured by a 
large number of different experiments [for a review, see McCammon \& Sanders 
 (1990)].  Most of the emission below 1$\,$keV 
arises from hot ($T>10^6\,$K) gas in the Galactic disk and halo, with 
radiation from extragalactic sources dominating at higher energies 
 (Chen et al. 1997; Miyaji et al. 1998; Moretti et al. 2009).  Much of the
literature on the diffuse radiation shows a distinction between contributions
from a local component with little foreground absorption and more distant
emissions with varying levels of absorption.  The local background was once
identified as having originated from hot gas in the Local Bubble (Sanders et
al. 1977; Hayakawa et al. 1978; Fried et al. 1980), but in recent years it
has been recognized to be strongly contaminated, or completely dominated, by
X-rays arising from charge exchange produced by the interaction of the solar
wind with incoming interstellar atoms (Cravens 2000; Lallement 2004; Pepino
et al. 2004; Koutroumpa et al. 2006, 2007, 2009; Peek et al. 2011; Crowder et
al. 2012).  For this reason, we ignore the weakly absorbed, nearly isotropic
portion of the X-ray background and focus our attention to the component that
exhibits a pattern in the sky that clearly shows absorption by gas in the
Galaxy.

Kuntz \& Snowden (2000) have performed a detailed 
investigation of the nonlocal component, which they call the transabsorption 
emission (TAE).  They describe the strength and spectral character of the TAE 
in terms of emissions from optically thin plasmas at two different 
temperatures.  They define a soft component that has a mean intensity over 
the sky $I=2.6\times 10^{-8}{\rm erg~cm}^{-2}{\rm s}^{-1}{\rm sr}^{-1}$ over 
the interval $0.1 < E < 2\,$keV and a spectrum consistent with the emission 
from a plasma at a temperature $T=10^{6.06}\,$K, and this flux is accompanied 
by a hard component with $I=8.5\times 10^{-9}{\rm erg~cm}^{-2}{\rm 
s}^{-1}{\rm sr}^{-1}$ over the same energy interval with $T=10^{6.46}\,$K.  
To translate the sum of these two components into a distribution of the 
photon flux as a function of energy, $F(E)$, we calculate synthetic flux 
representations using the CHIANTI database and software (Version~6.0) 
 (Dere et al. 1997, 2009), after normalizing the 
emission measures  to give the intensities stated above (we find that ${\rm 
EM}=10^{16.37}$ and $10^{15.81}{\rm cm}^{-5}$ for the soft and hard 
components, respectively).  We supplement the TAE result with an underlying 
power-law extragalactic emission of the form $10.5\,{\rm phot~cm}^{-2}{\rm 
s}^{-1}{\rm sr}^{-1}{\rm keV}^{-1}E({\rm keV})^{-1.46}$ (Chen et al. 1997).

The ISM is opaque to X-rays at the lowest energies.  The energies at which 
half of the X-rays are absorbed for various column densities are shown in the 
top portion of Figure~\ref{p_ar_plot}, which were derived from the 
calculations of Wilms et al. (2000).  At energies of 
around 100$\,$eV where the ISM is neither completely opaque or transparent 
for $N({\rm H}^0)\approx {\rm few}\,\times 10^{19}{\rm cm}^{-2}$, 
uncertainties in the layout of emitting and absorbing regions make it 
difficult to calculate with much precision how far the X-rays can penetrate 
the typical gas volumes that were sampled in our survey of Ar~I and O~I.  
Thus, rather than implement an elaborate attenuation function that would be 
difficult to explain (and perhaps not especially correct at our level of 
understanding), we apply a simplification that all of the X-rays are 
transmitted above some threshold energy and none below it.  The threshold 
that we adopted was 90$\,$eV, on the assumption that in some directions the 
gas can view the unattenuated X-ray sky through $N({\rm H}^0)$ slightly less 
than $10^{19}{\rm cm}^{-2}$.

\begin{figure}
\epsscale{0.7}
\plotone{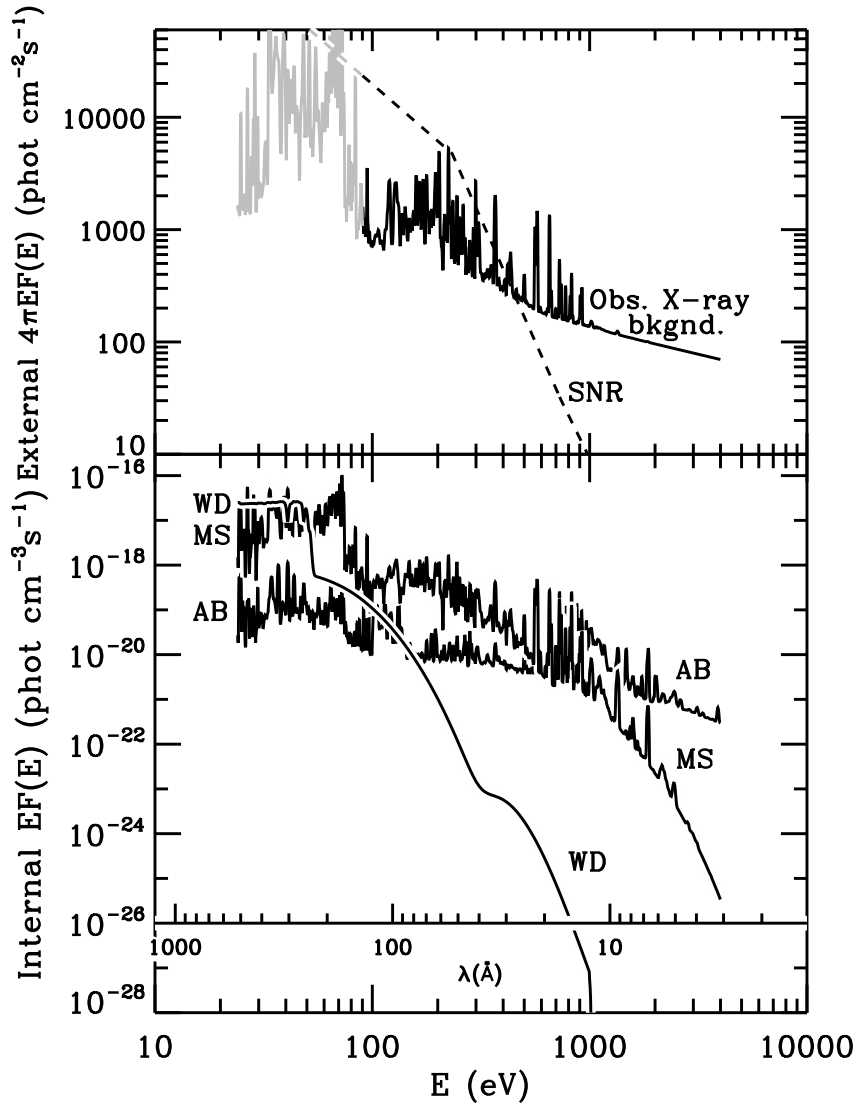}
\caption{{\it Upper panel:\/} External fluxes arising from the known X-ray 
background radiation (solid line) and a hypothetical, time-averaged flux 
(over a recombination time of $\approx 1\,$Myr) from 3 supernova remnants at 
a distance of 100$\,$pc (dashed line).  The gray portion of the solid curve 
is in an energy range where the opacity to X-rays is high, and thus this 
radiation is not likely to penetrate into much of the gas that we observe. 
{\it Lower panel:\/} Internal average rates per unit volume for the injection 
of ionizing photons by the coronae of main-sequence stars (MS), active 
binaries (AB) and the photospheres of WD stars 
(WD).  (A machine readable table of all 5 flux distributions vs. $E$
shown in this figure is available in the online version of the
{\it Astrophysical Journal.})\label{flux_plot}}
\end{figure}

The upper panel of Figure~\ref{flux_plot} shows our synthesis of the sum of 
the extragalactic power-law emission and TAE synthesis described above.  For 
the purpose of calculating $\Gamma$ for various elements, we make use of only 
the flux depicted by the dark trace in the figure, i.e., that which starts at 
the cutoff energy (90$\,$eV) and ends at an energy beyond which no 
appreciable additional ionization occurs.

\subsection{Internal Radiation Sources}\label{internal_radiation}
Embedded within the ISM are sources of EUV and X-ray radiation that can make 
additional contributions to $\Gamma$.  We can make estimates for their 
average space densities and the character of their emissions, but one 
uncertainty that remains is how well the ensuing photoionizations are 
dispersed throughout the ambient gas.  At one extreme representing minimum 
dispersal, we envision the classical Str\"omgren Spheres that surround 
sources that are not moving rapidly and that emit most of their photons with 
barely enough energy to ionize hydrogen.  These photons have a short mean 
free path in a neutral medium.  Under these circumstances, the zone of 
influence of the source is sharply bounded, and the resulting ionization is 
nearly total inside the region and zero outside it.  At the other extreme, 
one can imagine that the photons, ones that have relatively high energy, can 
travel over a significant fraction of the inter-source distances before they 
are absorbed.  In addition, the sources themselves could move rapidly enough 
that they never have a chance to establish a stable condition of ionization 
equilibrium.  (This issue will be investigated quantitatively in 
Section~\ref{recomb_time}.)  These conditions could lead to the ionization 
being more evenly distributed and not necessarily complete.  If the sources 
have large enough velocities, we can even imagine a picture where there is a 
random network of ``fossil Str\"omgren trails'' (Dupree \& Raymond 1983) that
ultimately might blend together.  More extreme manifestations of such trails
in denser media may have already been discovered by McCullough \& Benjamin
(2001) and Yagi et al.  (2012), who observed faint, but straight and narrow 
lines of H$\alpha$ emission in the sky (but were unable to identify the 
sources that created them).

It is difficult to establish where in the continuum between the two extremes 
for the dispersal of ionization the true effects of embedded EUV and X-ray 
sources are to be found.  For our treatment of the influence of these 
production sites for ionizing radiation, we will adopt the simplified premise 
that all of their photons are available to create a uniform but weak level of 
ionization everywhere.  This picture is not entirely correct, since one 
should expect that very near the sources some fraction of the ionizing 
photons are ``wasted'' by creating localized regions with much higher than 
usual levels of ionization.  Such regions dissipate most of their ionization 
rapidly because their recombination times are short.  For this reason, our 
making use of a calculated average production rate of ionizing photons per 
unit volume will lead to an equilibrium equation that overestimates the 
space- and time-averaged level of ionization.  (Later, it will be shown that 
this overestimate is of no real consequence because we obtain an answer that 
is still below that needed to explain the overall average ionization level.)

In the sections that follow, we consider three classes of sources that are 
randomly distributed throughout the neutral ISM and can in principle help to 
ionize it: main-sequence stars, active X-ray binaries, and WD stars.  
Luminous, early-type stars contribute large amounts of ionizing radiation, 
but most of their radiation completely ionizes the surrounding media and 
makes the gas virtually invisible in the Ar~I and O~I lines.  These stars 
also tend to be clustered inside the dense clouds of gas that led to their 
formation.

\onecolumn
\twocolumn

The question may arise as to whether or not, by considering the embedded 
sources as a separate contribution that adds to the external radiation 
background discussed earlier, we are possibly ``double counting'' some of the 
photons that could ionize the ISM.  Kuntz \& Snowden (2001) have computed the
probable relative contribution of Galactic point sources that were not
explicitly taken out of their measurements of the diffuse X-ray background,
and they concluded that this contamination was, at most, only about $2-10\%$
of the radiation that was thought not to arise from the Local Bubble.  Within
their highest energy bands ($0.73-2.01\,$keV), they state that the
contamination could be as high as 51\%.

\subsubsection{Main-sequence Stars}\label{ms}

For various kinds of point sources, the ratio of the emission of X-rays to 
photons in the $V$ band is usually defined by the relation
\begin{equation}\label{f_X/f_V}
\log(f_X/f_V)=\log f_X+0.4m_V+5.37~,
\end{equation}
where $f_X$ is the apparent X-ray flux of the source over a specified energy 
interval, expressed in ${\rm erg~cm}^{-2}{\rm s}^{-1}$, and $m_V$ is its 
apparent visual magnitude.\footnote{Numerical 
values for this formula are not universal, since the adopted X-ray energy 
interval depends on which instrument was used.  For instance, surveys that 
used the {\it Einstein Observatory\/}, e.g. those reported by Maccacaro et 
al. (1988) or Stocke et al. (1991), used the passband $0.3<E<3.5\,$keV for
$f_X$, whereas those based on the {\it ROSAT\/} All-Sky Survey (RASS)
 (Krautter et al. 1999; Ag\"ueros et al. 2009) measured $f_X$ over 
the interval $0.1<E<2.4\,$keV.}  The total output of X-rays from a source with an 
absolute magnitude $M_V=0$ should be $5.11\times 10^{34}(f_X/f_V)\,{\rm 
erg~s}^{-1}$.  Within each spectral class, there is a large dispersion in the 
measured values of $\log(f_X/f_V)$, typically of order 1$\,$dex, which is 
probably attributable to differences in stellar rotation velocities 
 (Audard et al. 2000; Feigelson et al. 2004), variations in 
foreground absorption by the ISM, and time variability of the X-ray emission.  
While the most significant X-ray flares from stars can create spectacular 
increases in flux, their time-averaged effect has been estimated by Audard et 
al. (2000) to amount to only about 10\% of the steady 
emission.  On the basis of white-light monitoring of dwarf stars, Walkowicz 
et al. (2011) found that the duty cycle of 
flaring events is only of order a few percent.

For any given stellar spectral class with a characteristic absolute magnitude 
$M_V$ that spans a range $\Delta M_V$ along the main sequence and has 
luminosity function $\phi(M_V)\,{\rm stars~mag}^{-1}{\rm pc}^{-3}$, the 
energy density of X-rays per unit volume is given by
\begin{eqnarray}\label{F(X)}
&&F(E)_{\rm MS}=1.74\times 10^{-21}\phi(M_V)\Delta M_V\nonumber\\
&\times& 10^{\log (f_X/f_V)-0.4M_V}{\rm erg~cm}^{-3}{\rm s}^{-1}~.
\end{eqnarray}
If we use the mean values of $\log (f_X/f_V)$ for different spectral classes 
listed by Ag\"ueros et al. (2009) for the X-ray band $0.1<E<2.4\,$keV, 
obtain values of $M_V$ for these classes from Schmidt-Kaler 
(1982), define a luminosity function for stars in the 
disk of our Galaxy from the formula given by Bahcall \& Soneira 
(1980), and then sum the results over all 
spectral types A-M, we obtain an average energy density equal to $1.90\times 
10^{-28}{\rm erg~cm}^{-3}{\rm s}^{-1}$.  While we acknowledge that the 
spectral character of coronal emissions can vary for different stars along 
the main sequence, in the interest of simplicity we adopted for all cases a 
spectrum based on the differential emission measure (DEM) for the coronal 
emission from the quiet Sun, as defined in the CHIANTI database.  To 
obtain a final photon emission rate per unit volume and energy in the ISM, we 
made use of this spectrum and normalized its energy output over the 
$0.1-2.4\,$keV band to the energy density factor given above.  The result is 
shown by the spectrum labeled ``MS'' in the lower panel of 
Figure~\ref{flux_plot}.

\subsubsection{Active Binaries and Cataclysmic Variables that Emit 
X-rays}\label{ab}

At high Galactic latitudes, most of the X-ray radiation above several keV 
originates from extragalactic sources.  Near the direction toward the 
Galactic center, however, Revnivtsev et al. (2009) 
found that at 4$\,$keV about half of the X-ray emission arises from point 
sources in the Galaxy that can be resolved by the {\it Chandra X-ray 
Observatory\/}, with the remainder coming from either a diffuse Galactic (hot 
gas) emission or an extragalactic contribution.  To estimate the average 
emission per unit volume from these sources, we integrate the $2-10\,$keV 
X-ray emission over the entire luminosity function $\phi(\log L_{2-10\,{\rm 
keV}})_{\rm AB}$ specified by Sazonov et al. (2006) 
to obtain the total energy output
\begin{eqnarray}\label{L_AB}
&&L(2-10\,{\rm keV})_{\rm AB}\nonumber\\
&=&\int_{22.5}^{34.0} L_{2-10\,{\rm keV}}\phi(\log 
L_{2-10\,{\rm keV}})_{\rm AB}\nonumber\\
&\times& d\,(\log L_{2-10\,{\rm keV}})= 2.13\times 
10^{38}{\rm erg~s}^{-1}
\end{eqnarray}
for all of the sources in our Galaxy.\footnote{Note that the upper bound for 
the integration is at $10^{34}{\rm erg~s}^{-1}$.  While the energy output 
over the whole Galaxy for neutron star and black hole X-ray binaries with 
individual outputs $L_{2-10\,{\rm keV}}>10^{34}{\rm erg~s}^{-1}$ is 
substantial, these objects are so few in number that they can no longer be 
considered as embedded sources.  The brightest such object in the sky, 
Sco~X$-$1, is at a distance of 2.8$\,$kpc and creates a flux at our location of 
only $2.8\times 10^{-7}{\rm erg~cm}^{-2}{\rm s}^{-1}$ over the $2-10\,$keV 
band (Grimm et al. 2002), which is substantially lower than the 
extragalactic background flux integrated over the whole sky.} We may convert 
this value to an average volume emissivity by multiplying it by the stellar 
mass density at our location ($0.04\,M_\odot~{\rm pc}^{-3}$) divided by 
the total stellar mass of the Galaxy ($7\times 10^{10}M_\odot$), where 
both of these numbers were those adopted by Sazonov et al. (2006), and
ultimately obtain a value $4.15\times 10^{-30}{\rm erg~cm}^{-3}{\rm s}^{-1}$.
 To define the spectral shape for the radiation emitted by these sources, we
used the CHIANTI software to compute the emission from a plasma with
an average of the DEM functions expressed by Sanz-Forcada et al. (2002,
2003)\footnote{These authors describe their DEM functions in terms of
$N(e)N({\rm H}^+)dV/d(\log T$), whereas the convention in CHIANTI
assumes that the DEM is defined in terms of $N(e)N({\rm H}^+)dV/dT$.}
that were constructed from their {\it Extreme-Ultraviolet Explorer\/}
observations of various active
binary sources.  This spectrum was then normalized such that the flux in the
$2-10\,$keV band matched the volume emissivity described above.  The emission
from active binaries that we derived $F(E)_{\rm AB}$ is shown by the curve
labeled ``AB'' in the lower panel of Figure~\ref{flux_plot}.

\subsubsection{White Dwarf Stars}\label{wd}

WD stars that are hot enough to emit significant fluxes in the EUV 
spectral range are much less numerous than the main-sequence stars considered 
in Section~\ref{ms}.  However their photospheres generate outputs in the EUV 
region that exceed by far the coronal emissions from individual main-sequence 
stars.  This fact is demonstrated by the actual observations of the local EUV 
sources compiled by Vallerga (1998), where he found that a 
vast majority of the detected objects were nearby hot WD 
stars.\footnote{The spectrum shown by Vallerga (1998) 
indicates that the $m_V=1.5$ B2~II star Adhara ($\epsilon$~CMa) dominates the 
local flux at energies below the He~I ionization edge at 24.6$\,$eV.  This is 
an atypical situation, since $N({\rm H~I})\lesssim 10^{18}{\rm cm}^{-2}$ 
toward this star (Gry \& Jenkins 2001).}  Krzesinski et al. 
 (2009) have measured the luminosity function for 
DA WDs $\phi(M_{\rm bol})_{\rm WD}$, expressed in terms of (${\rm 
stars}~M_{\rm bol}^{-1}{\rm pc}^{-3}$), in our part of the Galaxy using 
results from the {\it Sloan Digital
Sky Survey Data Release 4\/} (SDSS DR4) database.  If we combine this information with the 
theoretical computations of stellar atmosphere fluxes $F_\lambda$ (equal to 4 
times the Eddington flux $H_\lambda$), expressed in the units ${\rm 
erg~cm}^{-2}{\rm s}^{-1}{\rm cm}^{-1}$ (Rauch 2003; Rauch et al. 2010),
convert it to a physical flux at the stellar surface ${\mathcal F}(E)=
7.74\times 10^7\pi F_\lambda E({\rm eV})^{-3}{\rm phot~cm}^{-2}{\rm s}^{-
1}{\rm eV}^{-1}$, and assume each star has a radius $r_*=0.014r_\odot$ 
(Liebert et al. 1988), we can obtain a total emissivity per unit volume,
\begin{eqnarray}
&&F(E)_{\rm WD}=4\pi r_*^2 (1.65\times 10^{-34})\nonumber\\
&\times& \int{\mathcal F}(E)\phi(M_{\rm 
bol})_{\rm WD}d\, M_{\rm bol}~{\rm phot~cm}^{-3}{\rm s}^{-1}{\rm eV}^{-1}~.
\end{eqnarray}
A conversion from $M_{\rm bol}$ to $T_{\rm eff}$ is shown in the plot of the 
WD luminosity function presented by Krzesinski et al. (2009).  The model
fluxes for stars with $T_{\rm eff}<40,000\,$K were assumed to arise from
stars with pure hydrogen atmospheres, but stars with temperatures above this
limit are known to have significant abundances of metals in their atmospheres
because radiative levitation can overcome diffusive settling (Dupuis et al.
1995; Marsh et al. 1997; Schuh et al. 2002).  Thus, for $T_{\rm eff}=50,000\,
$K and above, we used the fluxes for model atmospheres with [$X$]~=~[$Y$]~=~0 and
$[Z]=-1$, which have significant sources of opacity that reduce the
radiation at energies $E>54\,$eV.

As a check on the calculations described above, we can compute the X-ray 
energy outputs as a function of $T_{\rm eff}$ over the $0.1-0.28\,$keV range, 
synthesize a luminosity function as a function of $L_X$ in this band, and 
then compare the results to X-ray luminosity function derived by Fleming et 
al. (1996) from a RASS survey of WDs.  Our 
source densities at the high end of the luminosity distribution compare 
favorably with the distribution shown by Fleming et al., but we predict a 
somewhat greater number of sources with $L_X\lesssim 10^{31}{\rm erg~s}^{-1}$ 
due to a large space density of stars with $30,000\leq T_{\rm eff}\leq 
40,000\,$K.

\subsection{Interpretation of Internal 
Ionizations}\label{interpretation_internal}

The treatments of the ionizations that arise from external and internal 
sources differ in a fundamental way.  External radiation above some energy 
threshold is regarded as not being consumed by ionizing the gas, i.e., it is 
assumed to be unattenuated and thus each gas constituent is ionized at an 
appropriate rate $\Gamma$, as defined in Eq.~\ref{Gammas}, that is simply 
proportional to $\int \sigma(E)F(E)dE$.  Here, $\sigma(E)$ is the effective 
cross section for the combination of the different ionization channels, as 
depicted for Ar and H by the solid lines in Fig.~\ref{p_ar_plot}.

As indicated in the beginning of Section~\ref{internal_radiation}, for the 
internally generated radiation we switch to a very different concept and 
adopt the simplified premise that all of the photons emitted by embedded 
sources are used up by ionizing the various gas constituents that surround 
them.  Thus the primary ionization rate for each kind of atom (or ion) $X$ is 
given by
\begin{equation}\label{Gamma_pX}
\Gamma_p(X)=\int F(E)y(X,E)n(X)^{-1}dE
\end{equation}
where $F(E)$ in this equation is the sum of all of the photon generation 
rates per unit volume specified in Sections~\ref{ms} through \ref{wd}.  Each 
of the different species represented by $X$ must compete with others for the 
photons that are consumed.  Thus the equation includes a term $y(X,E)$, which 
is a sharing function for the ionization rate that is represented by the 
relative probability that any photon with an energy $E$ will interact with a 
given species $X$,
\begin{equation}\label{y(X,E)}
y(X,E)={n(X)\sigma(X,E)\over\sum\limits_{X^\prime}n(X^\prime)\sigma(X^\prime,
E)}~,
\end{equation}
where $n(X)$ is the number density of $X$, $\sigma(X,E)$ is the 
photoionization cross section of $X$ at an energy $E$ (likewise for 
$X^\prime$), and the sum in the denominator covers all of the major species 
competing for photons, i.e., $X^\prime={\rm H}^0$, ${\rm He}^0$, ${\rm 
He}^+$, and heavy elements whose inner shells respond to the more energetic 
X-rays.  For either the external or internal radiations, the secondary 
ionization rates $\Gamma_s$ and $\Gamma_{s^\prime}$ follow in proportion to 
the primary ones according to the descriptions given in Appendix~\ref{gamma_s}.  
The ionizations from helium recombinations $\Gamma_{\rm He}^+$ and 
$\Gamma_{\rm He}^0$ are treated as internal sources of ionization, and their 
rates are driven by the local densities of He$^{++}$, He$^+$, and electrons, 
as described in Appendix~\ref{gamma_he}.

\subsection{Predicted Level of Ionization}\label{predicted_level}
 
Given the computed rates of ionization in the previous sections, an 
evaluation of the electron fraction $x_e=n(e)/n({\rm H}_{\rm tot})$ will 
depend on both the temperature $T$ and density $n({\rm H}_{\rm tot})$ for the 
gas.  This fraction is higher than $n({\rm H}^+)/n({\rm H}^0)$ because some 
electrons arise from the ionization of He.  The coupling of the H and He 
ionization fractions is governed by Eqs.~\ref{H_ioniz_balance} through 
\ref{n(e)}.  Our observational constraint, which must ultimately agree with 
the ionization calculations, arises from the values for [Ar~I/O~I], which 
respond to $n({\rm H}^+)/n({\rm H}^0)$  in accord with Eq.~\ref{ar_h}.  While 
the use of $P_{\rm Ar}$ in this equation will give an approximate value for 
[Ar~I/O~I], a more accurate result emerges by replacing $P_{\rm Ar}$ by 
$P^\prime_{\rm Ar}$, the derivation of which was described in 
Sections~\ref{equilibrium} and \ref{electron_density}.  Our goal will be to 
explore parameters that will match a computed value for [Ar~I/O~I] to the 
representative observed value $[{\rm Ar~I/O~I}]=-0.427\pm0.11$.

We have no direct knowledge about the local values of $n({\rm H}_{\rm tot})$ 
that apply to the gas in front of the stars in this survey.  We must 
therefore rely on general estimates that have appeared in the literature.  We 
can draw upon two resources.  First, surveys of 21-cm emission indicate the 
amounts of H~I in the disk at a Galactocentric distance of the Sun, but 
difficulties in interpreting the outcomes arise from self-absorption effects 
and ambiguities in distinguishing between cold, dense clouds 
(CNM) and their surrounding WNM.  Ferri\`ere (2001) lists values 
for $n({\rm H}^0)_{\rm WNM}$ that are in the range $0.2-0.5\,{\rm cm}^{-3}$.  
Dickey \& Lockman (1990) state a value of 
$0.57\,{\rm cm}^{-3}$, but it is not clear whether this excludes 
contributions from the CNM.  Kalberla \& Kerp (2009) estimate that the
midplane density of the WNM is $0.1\,{\rm cm}^{-3}$, but this low value is
difficult to reconcile with measurements of the mean thermal pressures $nT=
3800\,{\rm cm}^{-3}$K by Jenkins \& Tripp  (2011) and a general recognition
that $T_{\rm WNM}<10^4\,$K.

\begin{deluxetable}{ 
l   
c   
c   
c   
c   
c   
c   
}
\tablecolumns{7}
\tablewidth{0pt}
\tablecaption{Ionization Rates from Known 
Sources\tablenotemark{a}\label{known_source_rates}}
\tablehead{
\colhead{Ionization} & \colhead{Rate} & \multicolumn{5}{c}{Relative 
Contribution\tablenotemark{b}}\\
\cline{3-7}
\colhead{Source\tablenotemark{c}} & \colhead{($10^{-17}\,{\rm s}^{-1}$)} & 
\colhead{$\Gamma_p$} & \colhead{$\Gamma_{s,{\rm H}^0}$} & 
\colhead{$\Gamma_{s,{\rm He}^0}$} & \colhead{$\Gamma_{s,{\rm He}^+}$} & 
\colhead{$\Gamma_{s^\prime}$}\\
\colhead{(1)}& \colhead{(2)}& \colhead{(3)}& \colhead{(4)}&
\colhead{(5)}& \colhead{(6)}& \colhead{(7)} 
}
\startdata
$\Gamma({\rm H}^0)_{\rm ext.}$& 5.26& 0.20& 0.22& 0.52& 0.013& 0.048\\
$\Gamma({\rm H}^0)_{\rm MS}$& 6.72& 0.26& 0.044& 0.57& 0.11& 0.012\\
$\Gamma({\rm H}^0)_{\rm AB}$& 0.649& 0.035& 0.015& 0.52& 0.22& 0.21\\
$\Gamma({\rm H}^0)_{\rm WD}$& 2.19& 0.92& 0.030& 0.044& 6.3e-3& 9.4e-6\\
$\Gamma_{He^+}({\rm H}^0)$& 0.166&\nodata&\nodata&\nodata&\nodata&\nodata\\
$\Gamma_{He^0}({\rm H}^0)$& 4.94&\nodata&\nodata&\nodata&\nodata&\nodata\\
$\zeta_{\rm CR}({\rm H}^0)$& 12.3&\nodata&\nodata&\nodata&\nodata&\nodata\\
Total $\Gamma({\rm H}^0)$& {\bf 
32.3}&\nodata&\nodata&\nodata&\nodata&\nodata\\[5pt]
$\Gamma({\rm He}^0)_{\rm ext.}$& 31.4& 0.87& 0.035& 0.084& 2.2e-3& 7.5e-3\\
$\Gamma({\rm He}^0)_{\rm MS}$& 25.7& 0.80& 0.012& 0.15& 0.028& 3.0e-3\\
$\Gamma({\rm He}^0)_{\rm AB}$& 0.861& 0.33& 0.011& 0.36& 0.16& 0.14\\
$\Gamma({\rm He}^0)_{\rm WD}$& 17.0& 0.99& 4.5e-3& 5.9e-3& 8.6e-4& 1.2e-6\\
$\Gamma_{He^+}({\rm He}^0)$& 0.916&\nodata&\nodata&\nodata&\nodata&\nodata\\
$\Gamma_{He^0}({\rm He}^0)$& 7.10&\nodata&\nodata&\nodata&\nodata&\nodata\\
$\zeta_{\rm CR}({\rm He}^0)$& 14.4&\nodata&\nodata&\nodata&\nodata&\nodata\\
Total $\Gamma({\rm He}^0)$& {\bf 
97.3}&\nodata&\nodata&\nodata&\nodata&\nodata\\[5pt]
$\Gamma({\rm He}^+)_{\rm ext.}$& 
16.9&\nodata&\nodata&\nodata&\nodata&\nodata\\
$\Gamma({\rm He}^+)_{\rm MS}$& 8.32&\nodata&\nodata&\nodata&\nodata&\nodata\\
$\Gamma({\rm He}^+)_{\rm AB}$& 
0.0978&\nodata&\nodata&\nodata&\nodata&\nodata\\
$\Gamma({\rm He}^+)_{\rm WD}$& 
0.224&\nodata&\nodata&\nodata&\nodata&\nodata\\
$\Gamma_{He^+}({\rm He}^+)$& 0.317&\nodata&\nodata&\nodata&\nodata&\nodata\\
Total $\Gamma({\rm He}^+)$\tablenotemark{d}& {\bf 
25.9}&\nodata&\nodata&\nodata&\nodata&\nodata\\[5pt]
$\Gamma({\rm Ar}^0)_{\rm ext.}$& 251.& 0.94& 0.018& 0.042& 1.1e-3& 3.9e-3\\
$\Gamma({\rm Ar}^0)_{\rm MS}$& 82.2& 0.77& 0.014& 0.18& 0.035& 3.9e-3\\
$\Gamma({\rm Ar}^0)_{\rm AB}$& 10.5& 0.77& 3.6e-3& 0.12& 0.053& 0.049\\
$\Gamma({\rm Ar}^0)_{\rm WD}$& 45.3& 0.99& 5.6e-3& 8.2e-3& 1.2e-3& 1.8e-6\\
$\Gamma_{He^+}({\rm Ar}^0)$& 1.35&\nodata&\nodata&\nodata&\nodata&\nodata\\
$\Gamma_{He^0}({\rm Ar}^0)$& 93.6&\nodata&\nodata&\nodata&\nodata&\nodata\\
$\zeta_{\rm CR}({\rm Ar}^0)$& 113.&\nodata&\nodata&\nodata&\nodata&\nodata\\
Total $\Gamma({\rm Ar}^0)$& {\bf 
597.}&\nodata&\nodata&\nodata&\nodata&\nodata\\[5pt]
$\Gamma({\rm Ar}^+)_{\rm ext.}$& 
123.&\nodata&\nodata&\nodata&\nodata&\nodata\\
$\Gamma({\rm Ar}^+)_{\rm MS}$& 47.2&\nodata&\nodata&\nodata&\nodata&\nodata\\
$\Gamma({\rm Ar}^+)_{\rm AB}$& 1.57&\nodata&\nodata&\nodata&\nodata&\nodata\\
$\Gamma({\rm Ar}^+)_{\rm WD}$& 49.0&\nodata&\nodata&\nodata&\nodata&\nodata\\
$\Gamma_{He^+}({\rm Ar}^+)$& 0.969&\nodata&\nodata&\nodata&\nodata&\nodata\\
Total $\Gamma({\rm Ar}^+)$\tablenotemark{d}& {\bf 
222.}&\nodata&\nodata&\nodata&\nodata&\nodata\\
\enddata
\tablenotetext{a}{The internal rates (with subscripts MS, AB and WD) are 
based on a volume density $n({\rm H}_{\rm tot})=0.5\,{\rm cm}^{-3}$.  Such 
rates scale inversely with density (although not exactly so, because the 
secondary ionization efficiencies change when $x_e$ changes).}
\tablenotetext{b}{Fractions of the values shown in column (2).  See 
Eqs.~\protect\ref{Gammas}, \protect\ref{Hphi}, \protect\ref{Hephi}, 
\protect\ref{auger_H}, and \protect\ref{auger_He}.}
\tablenotetext{c}{Meaning of the subscripts that follow the different forms 
of $\Gamma$: ext.~=~external X-ray background radiation 
(Section~\protect\ref{external_radiation}), MS~=~embedded main-sequence stars 
(Section~\protect\ref{ms}), AB~=~embedded active binary stars 
(Section~\protect\ref{ab}), and WD~=~embedded WD stars 
(Section~\protect\ref{wd}).}
\tablenotetext{d}{The total ionization rates for the ions He$^+$ and Ar$^+$
do not include the ionizations from secondary electrons or cosmic rays.
Hence these totals underestimate the true rates.}
\end{deluxetable}
Heiles \& Troland (2003) estimate 
that an overall average $\langle n({\rm H}^0)_{\rm WNM})\rangle=0.28\,{\rm 
cm}^{-3}$ translates into local densities of $0.56\,{\rm cm}^{-3}$ if the WNM 
has a volume filling factor of 0.5.  A second method for estimating $n({\rm 
H}^0)_{\rm WNM}$ is to note where the WNM branch of the theoretical thermal 
equilibrium curve of Wolfire et al. (2003) intersects 
the average thermal pressure $p/k=3800\,{\rm cm}^{-3}$K of Jenkins \& Tripp 
 (2011).  This exercise yields a value of 
$0.4\,{\rm cm}^{-3}$.

Table~\ref{known_source_rates} shows a detailed accounting of the ionization 
rates from the different radiation sources, under the condition that the 
value of $x_e$ corresponds to what the equilibrium equations yield for a WNM 
at $T=7000\,$K.  The entries in this table reveal that for neutral hydrogen 
the ionization caused by all of the sources of EUV and X-ray photons result 
in a combined total rate $\Gamma({\rm H}^0)$ that is about 1.6 times the rate of 
ionization by cosmic rays $\zeta_{\rm CR}$.  For neutral He and Ar, the 
photon ionization rates are several times higher than those from cosmic rays.  
The changing patterns in the distribution of different ionization mechanisms 
revealed in Columns (3) to (7) in the table reflect differences in the 
distribution of fluxes with energy shown in Fig.~\ref{flux_plot}.  For 
sources with hard spectra, about half of the hydrogen ionization arises from 
the ionization of He$^0$, which produces energetic electrons that can cause 
secondary collisional ionizations, i.e., the process associated with 
$\Gamma_{s,{\rm He}^0}({\rm H}^0)$.  It is only for the very soft spectrum of 
the collective radiation from WD stars that we find that 
$\Gamma_p({\rm H}^0)$ strongly dominates over the other ionization routes.

The ionization rates and concentrations of various primary constituents are 
coupled to each other by the network of reactions described in 
Section~\ref{fundamentals} and Appendices \ref{gamma_s} through 
\ref{cr_ioniz}.  The first column of Table~\ref{properties_table} lists 
various quantities of interest, and their derived values under the assumption 
that $n({\rm H_{tot}})=0.50\,{\rm cm}^{-3}$ and $T=7000\,$K are given in 
Column (2). (The remaining two columns of this table will be discussed later 
in Section~\ref{additional_photoionization}.)

\begin{figure}[h!]
\epsscale{1.0}
\plotone{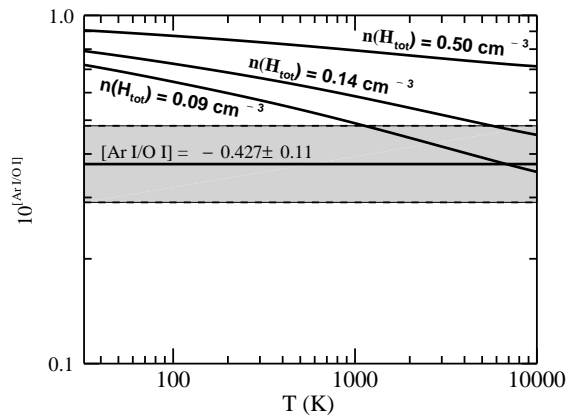}
\caption{Calculated values for $10^{\rm [Ar I/O I]}$ for three different 
densities $n({\rm H}_{\rm tot})$ and a range of different gas temperatures 
$T$ shown on the $x$ axis, computed using estimates for the ionization rates 
from known sources of radiation, both external and internal.  A general value 
for $10^{\rm [Ar I/O I]}$ that arose from the survey is indicated by the 
horizontal line, and the shaded band shows the range of possible 
systematic errors.\label{sdosdb_plot}}
\end{figure}

Figure~\ref{sdosdb_plot} shows the predicted outcomes for [Ar~I/O~I] after we 
perform the calculations, again using the equations and reaction rates given 
in Section~\ref{fundamentals} and the Ar ionization rates derived from the 
information in Sections~\ref{external_radiation} and 
\ref{internal_radiation}.  The upper curve in this figure that represents 
$n({\rm H}_{\rm tot})=0.50\,{\rm cm}^{-3}$ is clearly inconsistent with our 
findings for [Ar~I/O~I].  At $T=7000\,$K, a lower value $n({\rm H}_{\rm 
tot})=0.14\,{\rm cm}^{-3}$ is consistent with the upper error bound for 
[Ar~I/O~I].  At lower temperatures, however, the predictions for [Ar~I/O~I] 
once again are found to be considerably above the observed values. It is not 
until a density of $0.09\,{\rm cm}^{-3}$ is reached, a value that is 
unacceptably low, that the predicted ionization conditions for $T=7000\,$K 
fit comfortably with the nominal value from the observations.  Even here, 
however, this result is not fully satisfactory because a good fraction of the 
WNM is known to be at temperatures well below 7000$\,$K, where the gas is 
thermally unstable (Heiles \& Troland 2003).

Our overall conclusion is that the ionization rates arising from what we 
consider to be known sources of ionizing radiation are not able to maintain a 
level of ionization in the WNM that is consistent with our low observed 
values for [Ar~I/O~I].  Our quest to resolve this problem by exploring some 
possible supplemental means for ionizing the medium will be addressed later 
in Section~\ref{additional_photoionization}.

\begin{deluxetable}{ 
l   
c   
c   
c   
}
\tablewidth{0pt}
\tablecaption{Gas Properties\tablenotemark{a}\label{properties_table}}
\tablehead{
\colhead{} & \colhead{Without SNR} & \colhead{With SNR} & \colhead{Greater 
Low}\\
\colhead{Quantity} & \colhead{Contrib.\tablenotemark{b}} & 
\colhead{Contrib.\tablenotemark{c}} & \colhead{Energy 
Penetration\tablenotemark{d}}\\
\colhead{(1)} &\colhead{(2)} &\colhead{(3)} & \colhead{(4)}
}
\startdata
$P_{\rm Ar}$\tablenotemark{e}\dotfill&13.1&19.3&12.6\\
$P^\prime_{\rm Ar}$\tablenotemark{f}\dotfill&13.4&22.8&15.9\\
$n(e)\,({\rm cm}^{-3})$\dotfill&0.018&0.033&0.052\\
$n({\rm H}^0)\,({\rm cm}^{-3})$\dotfill&0.48&0.48&0.46\\
$n({\rm H}^+)\,({\rm cm}^{-3})$\dotfill&0.015&0.025&0.036\\
$n({\rm He}^0)\,({\rm cm}^{-3})$\dotfill&0.047&0.042&0.035\\
$n({\rm He}^+)\,({\rm cm}^{-3})$\dotfill&3.0e-03&8.1e-03&0.014\\
$n({\rm He}^{++})\,({\rm cm}^{-3})$\dotfill&2.1e-05&2.1e-04&1.0e-03\\
Total $\Gamma({\rm H}^0)\,({\rm s}^{-1})$\dotfill&3.2e-16&9.2e-16&2.1e-15\\
$\ell_c$ ($10^{-27}{\rm erg~s}^{-1}{\rm 
H~atom}^{-1})$\tablenotemark{g}\dotfill&4.3&6.2&8.5\\
\enddata
\tablenotetext{a}{For $n({\rm H}_{\rm tot})=0.50\,{\rm cm}^{-3}$ and 
$T=7000\,$K.}
\tablenotetext{b}{Applies to the photoionization rates expressed in 
Table~\protect\ref{known_source_rates} from the known sources only.}
\tablenotetext{c}{The radiation from known sources is supplemented by the 
average flux over 1$\,$Myr from 3 supernova remnants created by explosions 
with energies $E_{\rm SN}=3\times 10^{50}\,{\rm erg}$ in a medium with 
$n({\rm H})=1.0\,{\rm cm}^{-3}$ located at a distance of 100$\,$pc.  This 
supplemental radiation is shown by the dashed line in the upper panel of 
Fig.~\protect\ref{flux_plot}.}
\tablenotetext{d}{Allows for a low energy cutoff of 60$\,$eV instead of 
90$\,$eV for the external radiation background; see 
Section~\protect\ref{penetration}.}
\tablenotetext{e}{As defined in Eq.~\protect\ref{p_ar}.}
\tablenotetext{f}{As defined in Eq.~\protect\ref{Pprime_ar}.}
\tablenotetext{g}{Cooling rate from radiation by C$^+$ in its excited 
fine-structure state.}
\end{deluxetable}

\subsection{Dependence of Recombination with Time}\label{recomb_time}

An argument that helps to support the concept that the internal sources 
spread their ionization rather evenly throughout the medium is that they move 
at a rate that makes their local residence time short compared to a 
characteristic $e$-folding time $t_{\rm recomb.}$ for the decay of the proton 
density from an initial high value to some end equilibrium state $n({\rm 
H}^+)_{\rm eq.}$ as $t\rightarrow\infty$,
\begin{equation}\label{t_recomb}
t_{\rm recomb.}=-{n({\rm H}^+)-n({\rm H}^+)_{\rm eq.}\over dn({\rm H}^+)/dt}~,
\end{equation}
where
\begin{eqnarray}\label{dnpdt}
&&dn({\rm H}^+)/dt=-\Big[\alpha({\rm H}^0,T)n(e)\nonumber\\
&+&\alpha_g({\rm 
H}^0,n(e),G,T)n({\rm H_{tot}})\nonumber\\
&+&\Gamma({\rm H^0})\Big] n({\rm 
H}^+)+\Gamma({\rm H}^0)n({\rm H_{tot}})~.
\end{eqnarray}
This expression overlooks the complications arising from charge exchange 
reactions, and when we evaluate the trend of $n({\rm H}^+)$ with time we 
assume that $n(e)$ is always equal to $1.2n({\rm H}^+)$, as is the case when 
the gas is in a steady-state ionization condition.  Another simplification is 
that $T$, and hence $\alpha({\rm H}^0,T)$, remains constant.\footnote{A lack 
of variation in $T$ is probably a safe assumption for isobaric recombination, 
but in the isochoric case $T$ may deviate to lower values at intermediate 
times (Dong \& Draine 2011), and this would increase the value of 
$\alpha({\rm H}^0,T)$ and make the recombination more rapid.  This could be 
of importance for very large regions of space ionized by the radiation from 
old SNRs, which will be considered later in 
Section~\protect\ref{previous_SNR}.}  When the gas is highly ionized, 
$\Gamma({\rm H}^0)$ is not equal to the value that we computed for $n({\rm 
H}^+)_{\rm eq.}=0.015\,{\rm cm}^{-3}$ because the secondary ionization 
processes change with $n(e)$ and depend on the strength of the helium 
ionization.  However, we can ignore this complication that occurs at early 
times because the recombination terms in Eq.~\ref{dnpdt} are considerably 
larger than the terms involving $\Gamma({\rm H}^0)$.  The ionization rate 
influences the character of the decay only when the gas is weakly ionized.  
Hence there is no harm in declaring that at all times $\Gamma({\rm 
H}^0)=3.23\times 10^{-16}{\rm s}^{-1}$, as stated in 
Table~\ref{known_source_rates}. We can adopt for the final state 
(equilibrium) densities the values $n({\rm H}^+)_{\rm eq.}=0.015\,{\rm 
cm}^{-3}$ and $n(e)_{\rm eq.}=0.018\,{\rm cm}^{-3}$ listed in Column (2) of 
Table~\ref{properties_table}.  

\begin{figure}[h!]
\epsscale{1.0}
\plotone{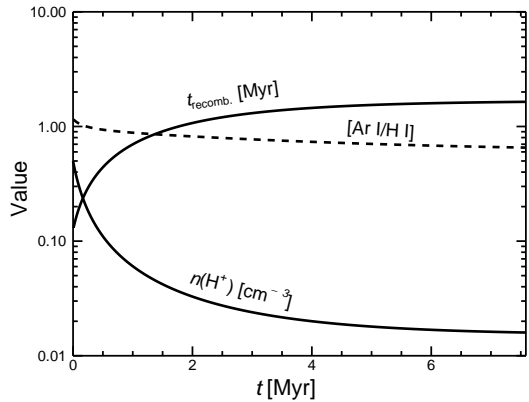}
\caption{The behavior of $t_{\rm recomb.}$ (upper solid curve) and $n({\rm 
H}+)$ (lower solid curve) with time $t$, starting from a fully ionized 
condition to an equilibrium state $n({\rm H}+)_{\rm eq.}$ for $n({\rm 
H_{tot}})=0.50\,{\rm cm}^{-3}$ and $T=7000\,$K. Values of $t_{\rm recomb.}$ 
are defined by Eq.~\protect\ref{t_recomb}, and $n({\rm H}+)$ as a function of 
$t$ can be found by solving the differential equation for $dn({\rm H}+)/dt$ 
given in Eq.~\protect\ref{dnpdt}.  By solving a similar equation for Ar$^+$ 
and comparing it to that for H$^+$, we obtain the trend for [Ar~I/H~I] shown 
by the dashed line.\label{ioniz_decay}}
\end{figure}

The change of $t_{\rm recomb.}$ with time and the relaxation of $n({\rm 
H}^+)$ from a fully ionized condition to $n({\rm H}^+)_{\rm eq.}$ for $n({\rm 
H_{tot}})=0.50\,{\rm cm}^{-3}$ and $T=7000\,$K is illustrated in 
Fig.~\ref{ioniz_decay}.  Values for $t_{\rm recomb.}$ start at 0.13$\,$Myr 
for an initial $n({\rm H}^+)=0.50\,{\rm cm}^{-3}$ and increase to 1.0$\,$Myr 
when $n({\rm H}^+)$ reaches $0.036\,{\rm cm}^{-3}$ at $t=1.77\,$Myr, beyond 
which $t_{\rm recomb.}$ very gradually climbs toward a steady value of 
1.68$\,$Myr and the subsequent decay in $n({\rm H}^+)$ toward $n({\rm 
H}^+)_{\rm eq.}=0.015\,{\rm cm}^{-3}$ is almost purely exponential. 

Of the three different classes of internal sources, WD stars emit 
the softest radiation, which means the influences of their ionizations can be 
more sharply bounded than those of the others.  While this consideration may 
present a challenge to our assumption about the uniformity of ionization in 
space, we note that these stars have observed radial velocity dispersions of 
about $25\,{\rm km~s}^{-1}$ (Falcon et al. 2010), or rms speeds of 
$43\,{\rm km~s}^{-1}$, which are consistent with the transverse velocity 
dispersion measurements reported by Wegg \& Phinney (2012).  This indicates
that WDs typically move about 76$\,$pc during the time interval that
$n({\rm H}^+)$ decays from 0.50 to $0.036\,{\rm cm}^{-3}$.  This amount of
travel is considerably larger than the radius of a Str\"omgren Sphere
\begin{equation}\label{r_S}
r_{\rm S}=\left( {3r^2_*\int_{13.6\, {\rm eV}}^\infty F(E)_{\rm WD}dE\over 
n({\rm H})^2\alpha({\rm H}^0,T)}\right)^{1/3}
\end{equation}
that would be established if a WD star were stationary.  (Recall 
that $F(E)_{\rm WD}$ is the flux emitted at the surface of the star with a 
radius $r_*$.)  A typical value for $r_{\rm S}$ in the WNM would be about 
1.8$\,$pc; this value applies to a WD star with $r_*=0.014r_\odot$
that has $T_{\rm eff}=50,000\,$K and is situated within a gas with a density 
$n({\rm H_{tot}})=0.50\,{\rm cm}^{-3}$.

One can view the dispersal of ionization caused by the star's motion in the 
context of the simplified description of ``cometary H~II regions'' described 
by Raga (1986); the dimensionless elongation parameter in this 
development $\beta=3v/[n({\rm H_{tot}})\alpha({\rm H}^0,T)r_{\rm S}]=14$ if 
$r_{\rm S}=1.8\,$pc and $v=43\,{\rm km~s}^{-1}$, which would result in a tube 
of ionization with a radius $\sim 0.5r_{\rm S}$ and a very long tail.  The 
fact that the ionizations produced by WD stars could be highly 
diluted because they are distributed over large volumes could explain why a 
sensitive survey of H$\alpha$ emission from regions around such stars carried 
out by Reynolds (1987) yielded only one detection out of 
nine targeted regions surrounding these stars.

\section{PROPOSALS FOR ADDITIONAL 
PHOTOIONIZATION}\label{additional_photoionization}

Given that the outcome for a reasonable value of $n({\rm H}_{\rm tot})$ and 
the ionization rate from known sources discussed in 
Section~\ref{known_photoionization} do not agree with our observations, we 
must explore some alternatives for boosting the average rate of 
photoionization in the WNM.  Two possibilities are discussed in the
following sections.

\subsection{Ionization from Previous Supernova Remnants}\label{previous_SNR}

A popular theme in the ISM literature of early 1970's was the consideration 
that the low density portions of the ISM were ionized impulsively by UV and 
X-ray radiation from supernovae and their remnants (SNRs), which represented 
sources of ionization that had limited durations and that materialized at 
random locations and times (Bottcher et al. 1970; Werner et al. 1970; Jura \&
Dalgarno 1972; Gerola et al. 1973; Schwarz 1973).  A particularly instructive
example that shows the wide variations in temperature and fractional
ionization can be seen in the results from a numerical simulation by Gerola
et al. (1974).  This topic has also been approached by Slavin et al. (2000),
who estimated the effects of irradiation of the neutral medium by SNRs, with
a special emphasis on its influence on gas in the Galactic halo. The spectral
character and strength of the emitted radiation from a remnant depends on a
combination of supernova energy and the density of the ambient medium
(Mansfield \& Salpeter 1974).  The emission of soft X-rays as a remnant
evolves can last up to a late phase when radiative cooling causes a dense
shell to form, which could be opaque to the lowest energy X-rays.  However
some of the radiation may continue to escape afterward if instabilities
cause the shell to break up and create gaps (Vishniac 1994; Blondin et al.
1998).  The magnitudes of such instabilities are uncertain, since they might
be diminished if either the ambient medium has a low density and high
temperature (Mac Low \& Norman 1993) or if the shock is partly stabilized by
an embedded transverse magnetic field
(T\'oth \& Draine 1993).

Chevalier (1974) has computed the strength and 
distribution over energy for the radiation emitted by various models for 
SNRs.  As a representative example, we can use the emission 
over a time interval of $2.5\times 10^5\,{\rm yr}$ from his model that had an 
energy $E_{\rm SN}=3\times 10^{50}\,{\rm erg}$ and was expanding into a 
medium with an average density of $1\,{\rm cm}^{-3}$.  If we assume that 
there was a succession of about 3 such remnants that occurred once every Myr 
(i.e., of order $t_{\rm recomb.}$ defined in Eq.~\ref{t_recomb} for $n({\rm 
H}^+)=0.036\,{\rm cm}^{-3}$; see Fig.~\ref{ioniz_decay}) and they were at a 
distance of 100$\,$pc from some location in the ISM, the resulting 
time-averaged flux shown by the dashed line in the upper panel of 
Fig.~\ref{flux_plot} would increase the total ionization of H from 
$\Gamma({\rm H}^0)=3.2\times 10^{-16}{\rm s}^{-1}$ for the steady ionization 
sources to the much higher value of $9.2\times 10^{-16}{\rm s}^{-1}$ and 
create conditions that would yield values for [Ar~I/O~I] that are virtually 
the same as those depicted by the curve labeled $n({\rm H}_{\rm 
tot})=0.14\,{\rm cm}^{-3}$ in Fig.~\ref{sdosdb_plot}, but in this case for a 
density $n({\rm H}_{\rm tot})=0.50\,{\rm cm}^{-3}$.  At $T=7000\,$K, the 
densities of various constituents are represented by the numbers that are 
shown in Column (3) of Table~\ref{properties_table}.  For this temperature, 
the calculation of [Ar~I/O~I] is just consistent with the upper error bound 
from the observations.

A stronger flux from SNRs would be needed to make the 
calculated value for [Ar~I/O~I] match nominal value for the observed ones.  
To obtain a curve that matches the lowest curve (labeled $n({\rm H}_{\rm 
tot})=0.09\,{\rm cm}^{-3}$) in Fig.~\ref{sdosdb_plot} and still maintain an 
assumed local density $n({\rm H}_{\rm tot})=0.50\,{\rm cm}^{-3}$, we would 
require either an increase in the average energy of the supernovae (SNe), or the 
rate would need to be increased from 3 to 8~SN~${\rm Myr}^{-1}$ at a distance 
of 100$\,$pc.  In this circumstance, $n(e)=0.047\,{\rm cm}^{-3}$.

While one might be tempted to ask whether some average SN rate in our 
location within the Galaxy (van den Bergh \& McClure 1994; McKee \& Williams
1997; Ferri\`ere 2001) is consistent with at least 3 SNe Myr$^{-1}$ at
separations of about 100$\,$pc from some representative location, it is
probably more realistic to draw upon actual estimates of the recent history
of explosions in the general neighborhood of the Sun, which have been
inspired by the evidence of how the ISM has been disrupted by the expansions
of the explosion remnants.  The Local Bubble (see footnote 2 in
Section~\ref{strategy}) and the neighboring radio emitting Loop~I
superbubble\footnote{Loop~I is coincident with the North Polar Spur of X-ray
radiation. Its center is estimated to be only 180 pc away from us  (Bingham
1967).}, which intersect one another (Egger \& Aschenbach 1995), are both
thought to have been created by a series of SN explosions that
occurred over the past 10-15$\,$Myr
(Ma\'iz-Apell\'aniz 2001; Bergh\"ofer \& Breitschwerdt 2002).  Fuchs
et al. (2006) analyzed the numbers and mass functions of stars within nearby
O-B associations and traced their motions back in time.  From this
information, they concluded that the Local Bubble was created by 14-20
SNe.  This estimate is consistent with the 19 SNe that
Breitschwerdt \& de~Avillez (2006) used in their hydrodynamic simulation of
the creation of the Local Bubble.

Within a shorter time frame, there is some evidence that atoms from a nearby 
SN were deposited on the Earth some 2.8$\,$Myr ago, as revealed by an 
enhancement of $^{56}$Fe in a thin layer within deep-sea ferromanganese crust 
 (Knie et al. 2004).  Fitoussi et al. (2008) 
attempted to replicate this result in a marine sediment, but the outcome did 
not agree with the earlier result at the same level of significance.  
However, there are alternative means for investigating past depositions of 
SN elements in terrestrial samples (Bishop \& Egli 2011; Feige et al.
2012), and new results may emerge soon.

If we add to the Local Bubble contribution the past SNe in the Sco-Cen 
association that created Loop~I (Iwan 1980; Egger 1998; Diehl et al. 2010),
one can imagine that, to within the uncertainties in the X-ray production
rates, the time-averaged level of ionizing radiation could conceivably be of
the same order of magnitude as that discussed in the preceding paragraphs.

A legitimate question to pose is whether or not, at some intermediate stage 
of recombination, the value of [Ar~I/H~I] makes a brief excursion to a level 
below the final equilibrium state.  If this were the case, we might be able 
to relax the requirement for a high frequency of SN explosions.  The 
recombination rates $\alpha({\rm H}^0,T)$ and $\alpha({\rm Ar}^0)$ are nearly 
equal to each other, but the dust grain recombination rate $\alpha_g$ for Ar 
should be much less than that for H because the thermal velocities of Ar are 
lower.  Thus, at early times where recombinations still dominate over the 
steady-state ionization rate but the electron density has decreased to a 
point that neutralization of Ar and H by collisions with dust grains become 
important, one could imagine that the difference in grain recombination rates 
might be influential in allowing the H$^+$ to recombine more quickly than 
Ar$^+$.  In order to investigate this possibility, we can evaluate the time 
history of the Ar recombination using Eqs.~\ref{t_recomb} and \ref{dnpdt} but 
with a substitution of ``Ar'' for each appearance of ``H.''  A resulting 
comparison of the Ar recombinations to those of H shows that [Ar~I/H~I] shows 
a steady decrease from an initial value slightly greater than 1.0 (when the 
gas is just starting to recombine) to the final equilibrium state for this 
quantity without any excursion to lower values.  This behavior is illustrated 
by the dashed line in Fig.~\ref{ioniz_decay}.

\subsection{Penetration of the WNM by Low-energy Photons}\label{penetration}

Until now, we have regarded the emission of X-rays from the hot ($T\sim 
10^6\,$K) gas in our Galaxy as an external source of ionizing radiation that 
must penetrate the bulk of the WNM regions under study.  With this condition 
in mind, we established a 90$\,$eV cutoff for the external radiation, below 
which the X-rays were assumed to be absorbed in the outer layers of the 
neutral gas.  However, some of the soft X-ray emitting gas may be threaded 
through the WNM, a picture that is reminiscent of the network of hot 
interstellar tunnels proposed by Cox \& Smith (1974).  
Hot, interspersed gas could allow some or all of the neutral medium to have 
access to lower energy radiation.  Observations indicate that O~VI, an 
indicator of gas that is collisionally ionized at $T\sim 3\times 10^5\,$K, 
can be seen in emission (Dixon et al. 2006; Otte \& Dixon 2006) and absorption
(Jenkins 1978a, b; Bowen et al. 2008) almost everywhere in the disk of the
Galaxy.  The simulations of SNe exploding at random in the Galactic
disk by Ferri\`ere  (1995) and de~Avillez \& Breitschwerdt (2005a, b) indicate 
that this hot gas should be sufficiently pervasive and frothy that, even 
though our measurements penetrate through column densities well in excess of 
$10^{19}\,{\rm cm}^{-3}$, individual parcels of gas may be exposed to 
adjacent sources of radiation through significantly lower column densities.  
Indeed, there could be unusually strong X-ray emission at the boundaries 
where the neutral regions come into contact with the hot gas, where charge 
exchange reactions between neutrals and highly ionized species could produce 
an enhancement of soft X-ray emission over that which is emitted by just the 
hot gas (Wang \& Liu 2012).

It is clear from the gray parts of the external radiation spectrum depicted 
in Fig.~\ref{flux_plot} that even a modest lowering of the cutoff below our 
adopted 90$\,$eV threshold will create a significant increase in the 
ionization of the WNM.  For instance, if the threshold were dropped to 
60$\,$eV, an energy just below the strong M-line complex of emission features 
from highly ionized Fe at 70$\,$eV, the level of ionization would increase to 
an extraordinarily large total hydrogen ionization rate $\Gamma({\rm 
H}^0)=2.1\times 10^{-15}{\rm s}^{-1}$.  For $n({\rm H}_{\rm tot})=0.50\,{\rm 
cm}^{-3}$ and $T=7000\,$K, this increase would create a value for [Ar~I/O~I] 
that is consistent with the upper error bound for the measured outcomes.  In 
this particular case, most of the ionization is caused by photons with 
energies near 70$\,$eV.  For this energy, $P_{\rm Ar}(E)$ is near its minimum 
value (see the lower panel of Fig.~\ref{p_ar_plot} and the entry in Column 
(4) of Table~\ref{properties_table}).  As a consequence, in order to match 
our observations, Eq.~\ref{ar_h} shows us that the required $n({\rm H}^+)$ 
(and thus $\Gamma({\rm H}^0)$) must be increased beyond the value needed from 
the time-averaged SNR radiation with its larger value of 
$P_{\rm Ar}$.

A consideration that disfavors the presence of a strong ionizing flux at 
energies in the vicinity of 70$\,$eV, at least at our location, is that it 
does not seem to be present at anywhere near the intensity level shown in 
Fig.~\ref{flux_plot}.  For instance, from observations by the {\it Cosmic Hot 
Interstellar Plasma Spectrometer\/} ({\it CHIPS\/}) Hurwitz et al. 
 (2005) stated an upper limit for the flux emitted by 
the Fe lines near 70$\,$eV.  This limit was consistent with an emission 
measure for a plasma with a solar abundance pattern at $T=10^6\,$K that is 
less than 5\% of the value ${\rm EM}=10^{16.37}{\rm cm}^{-5}$ that we used 
for reconstructing the emission spectrum from the soft component of the X-ray 
background described in Section~\ref{external_radiation}.  Jelinsky et al. 
 (1995) and Bloch et al. (2002) 
similarly found flux upper limits at low energies (but not as stringent as 
those from {\it CHIPS\/}) that were much lower than model predictions at this 
energy that we obtained from fits to the observed radiation at higher 
energies.  A marginal detection of the Fe emission complex toward high 
Galactic latitudes by McCammon et al. (2002), 
$F(E)=100\pm 50\,{\rm phot~cm}^{-2}{\rm s}^{-1}{\rm sr}^{-1}$ is also well 
below the peak at 70$\,$eV that appears in our reconstructed flux.  The 
weakness of these fluxes could be caused either by a deficiency of Fe below 
the solar abundance ratio in the hot emitting gases or by our inability to 
see hot gas regions whose low energy radiation is not absorbed by intervening 
neutral material.

\section{DISCUSSION}\label{discussion}

We have now reached a point where it is appropriate to investigate the 
consistency of the newly derived WNM ionization levels with other 
observational and theoretical findings, along with some consequences of our 
results on various relevant physical processes.

\subsection{Other UV Observations}\label{other_obs}

Along a number of sight lines, there have been detailed investigations of UV 
absorption lines observed at high enough velocity resolutions to identify 
which components came from either mostly neutral or mostly ionized clouds 
 (Spitzer \& Fitzpatrick 1993, 1995; Fitzpatrick \& Spitzer 1994, 1997; Wood
\& Linsky 1997; Holberg et al. 1999; Welty et al. 1999; Jenkins et al. 2000a;
Gry \& Jenkins 2001; Sonnentrucker et al. 2002).  These investigations relied
either on the relative populations of the excited fine-structure level of
${\rm C}^+$ or the ratios
of ions to neutrals for various elements.
The ion-to-neutral ratios of different elements gave very different outcomes 
for $n(e)$ (but systematically went up and down together from one velocity 
component to the next); these disparities probably arose as a result of a 
lack of a good understanding of the physical processes involved.  Welty et 
al. (2003) have presented a good summary of the results 
that exhibited this problem.

Values of $n(e)$ identified with the WNM by the investigators cited above 
generally ranged between 0.04 and $0.12\,{\rm cm}^{-3}$, which seem to be 
higher than our values given in Table~\ref{properties_table}.  A common theme 
in the discussions of these results was that such high partial ionizations 
could not be attributed to the known EUV, X-ray and cosmic ray ionization 
rates -- a conclusion that is stated once again from our [Ar~I/O~I] results.

Electron densities for clouds embedded in the Local Bubble have an average 
value of $0.11\,{\rm cm}^{-3}$ (Redfield \& Falcon 2008), a result 
that once again relied on comparisons of excited C$^+$ to other species 
(either unexcited C$^+$ or S$^+$).  The fractional ionization of these clouds 
appears to be much higher than what we found for the general WNM outside the 
Local Bubble, if one assumes that the characteristic total density $n({\rm 
H}_{\rm tot})\approx 0.2\,{\rm cm}^{-3}$ within the clouds (Redfield \&
Linsky 2008). By solving for the time-dependent ionization in Eq.~\ref{dnpdt}
for $n({\rm H_{tot}})=0.2\,{\rm cm}^{-3}$, we find that the decay from a
fully ionized condition to an approximately half ionized state takes only
0.40$\,$Myr.  It is likely that this higher level of ionization might be
explained by supplemental radiation from evaporative boundaries that surround
the clouds (Slavin \& Frisch 2002), if indeed they are embedded in a very
hot, low density gas, or by the infiltration of some ionizing radiation from
$\epsilon$~CMa, which strongly dominates over other radiation sources within
the Local Bubble at low energies (Vallerga \& Welsh 1995).

\subsection{Pulsar Dispersion Measures}\label{pulsar_DM}

Cordes \& Lazio (2002) have derived characteristic 
electron densities for three volumes within about 1$\,$kpc of the Sun.  
Inside the Local Bubble, their model seems to best fit an average density of 
only $\langle n(e)\rangle=0.005\,{\rm cm}^{-3}$, which is not surprising 
because much of the volume has probably been cleared of material by SN 
explosions, except for some isolated, warm clouds with an average filling 
factor in the range $5.5-19$\% (Redfield \& Linsky 2008).  Regions 
outside the Local Bubble have higher densities: two volumes that they studied 
yield $\langle n(e)\rangle=0.012$ and $0.016\,{\rm cm}^{-3}$.  However, these 
regions are identified with ellipsoidal volumes designated as either a 
``local superbubble'' (LSB) or a ``low density region'' (LDR), so they too 
could have significant voids that would dilute the apparent electron 
densities.  The values of $\langle n(e)\rangle$ given for these regions could 
be consistent with our measures of $n(e)$ given in the last two columns of 
Table~\ref{properties_table} for the WNM if this medium had a filling factor 
of about 1/3 in these regions and there were no significant contributions 
from the warm ionized medium (WIM).

An independent analysis of dispersion measures was carried out by de Avillez 
et al. (2012) for 24 pulsars with known distances 
between 0.2 and 8$\,$kpc from the Sun and with $\vert z\vert<0.2\,$kpc. They 
found a distribution of $n(e)$ outcomes that was consistent with a log-normal 
distribution centered on $\log n(e)=-1.47$ and a dispersion $\sigma[\log 
n(e)]=0.17$.  The fact that their representative values of $n(e)$ are higher 
than the determination of Cordes \& Lazio (2002) 
and close to our findings based on [Ar~I/O~I] may be accidental, since their 
results may be strongly influenced by sight-line interceptions of fully 
ionized regions.

\subsection{Emission Lines}\label{emission_lines}

Observations of diffuse line emissions in the sky most generally apply to 
probing the physical conditions in fully ionized gases, either the bright 
H~II regions around hot stars or the WIM
(Reynolds et al. 1977, 2002; Haffner et al. 1999; Madsen et al. 2006).  Most
of the emissions are dominated by contributions from species that are
expected to be abundant in such highly ionized media, and they should
overwhelm any contributions from the very dilute ionization in the WNM. 
Nevertheless, there are a few cases where emission line fluxes from some
neutral species in the WIM are expected to be very weak, and thus a
contribution from the WNM might in principle be identified.  We need to
explore whether or not the predictions for line strengths from a medium with
our enhanced electron densities are not in serious violation of detections or
the upper limits for the fluxes.  We explore three such cases in the
subsections that follow.  The first test involving line emission from He
recombinations is especially important, because we predict that the fraction
of singly-ionized He could be as high as 16$-$28\% of the total amount of
helium. 

\subsubsection{He $\lambda$5876 Recombination Radiation}\label{He_recomb}

Reynolds \& Tufte (1995) attempted to compare 
the strength of the recombination line of He$^0$ at 5876$\,$\AA\ to that of 
H$\alpha$ in parts of the sky away from well defined H~II regions. Their 
motive was to determine the hardness of the radiation that maintains the 
ionization in the WIM. While an explicit upper limit for the He recombination 
line flux was not stated by Reynolds \& Tufte (1995), we estimate that the
flux in the two directions that they sampled was found to be less than 0.1$\,
$R (R~=~Rayleigh~=~$10^6/(4\pi) {\rm phot~cm}^{-2}{\rm sr}^{-1}{\rm s}^{-
1}$).

Using the line emission rates given by Benjamin et al. (1999) we find that
for $n(e)=0.04\,{\rm cm}^{-3}$, $n({\rm He}^+)=0.01\,{\rm cm}^{-3}$, and $T=
7000\,$K (representative values for the enhanced ionization cases presented
in the last two columns of Table~\ref{properties_table}) the emission should
be $3.0\times 10^{-17}{\rm phot~cm}^{-3}{\rm s}^{-1}$.  If we were to
propose that the emission is seen over a path of 200$\,$pc with no
extinction (i.e., imagine a length of 400$\,$pc with a filling factor for the
WNM of 50\%), we could expect to find an emission of $1500\,{\rm phot~cm}^{-
2}{\rm sr}^{-1}{\rm s}^{-1} = 0.02\,$R.  This expectation is well below the
sensitivity of the observations by Reynolds \& Tufte (1995), so our predicted 
intensity from the enhanced electron density and ionization of He does not 
violate their upper limit.

\subsubsection{Emission of [O~I] $\lambda$6300 from Electron 
Collisions}\label{OI_excitation}

In order to determine the neutral fraction of H in the WIM, Reynolds et al. 
 (1998) measured the strength of the [O~I] 
$\lambda$6300 line emitted in parts of the sky that had a uniform, moderately 
strong H$\alpha$ emission (but away from obvious H~II regions excited by 
stars), much as they had done for the He recombination line discussed above.  
In three different directions, they detected intensities of 0.2, 0.09, and 
0.11$\,$R.

According to Federman \& Shipsey (1983), 
electron collisions dominate over those by hydrogen for the excitation of the 
$^1D_2$ state of neutral oxygen when the electron fraction exceeds 
$1.5\times 10^{-4}$.  Hence, we can ignore hydrogen impacts.  From the 
fitting formula of P\'equignot (1990) to the collision strengths computed 
by Berrington \& Burke (1981), we derive a 
collisional rate constant $C_e$ for electron excitations to be $8.0\times 
10^{-11}{\rm cm}^3{\rm s}^{-1}$ at $T=7000\,$K.  If we again make the 
conservative assumption expressed in Section~\ref{reference_abund} that at 
the low densities of the WNM the depletion of O is negligible, we expect that 
$n({\rm O}^0)=2.7\times 10^{-4}{\rm cm}^{-3}$ if $n({\rm H}^0)=0.47\,{\rm 
cm}^{-3}$.  For $n(e)=0.04\,{\rm cm}^{-3}$, we expect the emissivity to be 
equal to $n({\rm O}^0)n(e)C_e$ multiplied by a branching fraction 0.76 
 (Froese Fischer \& Tachiev 2010) for the proportion of decays from 
the $^1D_2$ level to the lower $^3P_2$ level.  This product 
equals $6.6\times 10^{-16}{\rm phot~cm}^{-3}{\rm s}^{-1}$, which should 
produce 0.41$\,$R over a path of 200$\,$pc.  This value is greater than the 
three measurements by Reynolds et al. (1998).  The 
magnitude of this violation is not large, considering the uncertainties of 
our assumptions about the lack of depletion of O and the adopted path length 
estimate.  Also, for temperatures less than 7000$\,$K, the expected strength 
of the emission will be considerably less: for instance, at $T=5000\,$K, the 
emission should be 4.5 times weaker than at 7000$\,$K.

\subsubsection{Emission of [N~I] $\lambda$5201 from Electron 
Collisions}\label{NI_excitation}

An upper limit of 0.13$\,$R for the [N~I] $\lambda$5201 line was determined 
for a single direction in the sky by Reynolds et al. (1977).  If we adopt
the same calculations as in the previous section for O~I, but make the
substitution that $n({\rm N}^0)=8.0\times 10^{-5}n({\rm H}^0)=3.8\times
10^{-5}{\rm cm}^{-3}$ (again, assuming no depletion), and $C_e=8.6\times
10^{-11}{\rm cm}^3{\rm s}^{-1}$ at $T=7000\,$K (Tayal 2006), we obtain an
emissivity equal to $1.3\times 10^{-16}{\rm phot~cm}^{-3}{\rm s}^{-1}$,
which yields an intensity of 0.08$\,$R over a 200$\,$pc path.  This value is
below the upper limit determined by Reynolds et al. (1977).

\subsection{Cooling Rates from Carbon Ions and Oxygen 
Atoms}\label{cooling_rate}

A major coolant for the neutral ISM is the singly charged carbon ion 
 (Dalgarno \& McCray 1972; Wolfire et al. 1995, 2003), whose $^2P_{3/2}$
excited fine-structure level in the ground electronic state can be excited
by collisions with electrons and hydrogen atoms.  The cross section for
excitation by electrons is substantially greater than that for neutral
hydrogen.  For this reason, the level of partial ionization is an important
factor in the excitation rate.  After excitation, an energy loss can occur
because the excited level can undergo a spontaneous radiative decay with a
transition probability $A_{21}=2.29\times 10^{-6}{\rm s}^{-1}$ to the lower
$^2P_{1/2}$ state (Nussbaumer \& Storey 1981), a process that liberates a
photon with a wavelength of $158\,\mu$m.  If we assume that the abundance
ratio $({\rm C}^+/{\rm H}^0)=9.5\times 10^{-5}$ (see footnote~1 in
Section~\ref{intro}), we can use the densities that we derived for $n({\rm
H}^0)$ and $n(e)$ along with the excitation cross section by H$^0$ impacts
computed by Barinovs et al.  (2005) and electron collision strengths of
Wilson \& Bell (2002) to compute the ${\rm C}^{+*}$ energy 
loss rates $\ell_c$ per H atom.  These rates are given in the last row of 
Table~\ref{properties_table}.  They do not change in direct proportion to 
$x_e$ because about half of the excitations come from collisions with H 
atoms, which are more numerous than the electrons.  In addition, the overall 
cooling rate for the WNM does not increase in direct proportion to $\ell_c$ 
because it accounts for only about one-third of the total cooling.

Most of the remaining cooling comes from the fine-structure excitation of O~I 
and the subsequent emissions at 44 and $63\,\mu$m. The O~I cooling is 
insensitive to changes in $x_e$ because the collision rate constants for 
electrons (Bell et al. 1998) are much less than those for neutral 
hydrogen.  Using an extrapolation of the atomic hydrogen collisional rate 
constants given by Abrahamsson et al. (2007) 
above a temperature of $10^3\,$K, the spontaneous decay rates of the two 
excited levels given by Galav\'is et al. (1997), and ${\rm O/H}=5.75\times 
10^{-4}$, we find that for $n({\rm H}^0)=0.47\,{\rm cm}^{-3}$ and $T=7000\,$K 
that the energy loss rate per H atom $\ell_o=1.13\times 10^{-26}{\rm 
erg~s}^{-1}$.

The results for $\ell_c$ shown in the table are significantly lower than the 
average value of $2\times 10^{-26}{\rm erg~s}^{-1}{\rm H~atom}^{-1}$ found 
for low-velocity clouds by Lehner et al. (2004), who 
measured the column densities of C~II$^*$ from spectra recorded by {\it 
FUSE}.  [Lehner et al. (2003) obtained similar results 
for sight lines toward WD stars within or just outside the Local 
Bubble.]  Lehner et al. (2004) compared their results 
with measurements of H$\alpha$ emission in the same directions, and they 
concluded that about half of the C~II$^*$ that they detected came from fully 
ionized gas (but with large variations from one sight line to the next).  
This inference was supported by the fact that their values of $\ell_c$ were 
slightly lower for sight lines with large values of $N$(H~I).  In principle, 
one can gain an insight on the relative importance of H~II regions by 
comparing the observed abundances of C$^{+*}$ to those of N$^{+*}$, since the 
former can come from both neutral and ionized regions, while the latter 
arises only from ionized regions.  Gry et al. (1992) 
compared these two species in a comprehensive study of both the absorption 
lines observed by the {\it Copernicus\/} satellite (Rogerson et al. 1973) and
the 158 and 205$\mu$m emission lines observed by the {\it COBE\/}
satellite (Wright et al. 1991).  They concluded that H~II regions were
responsible for a major portion of the C$^{+*}$ that was observed, but this
result is clearly dependent on the assumed ratio of atomic C to N in the H~II
regions.

An important advantage of the $\ell_c$ determinations synthesized from our 
results for [Ar~I/O~I] is that they apply {\it only\/} to the WNM; hence they 
give more accurate indications of the carbon cooling rates within this medium 
without any contamination from H~II regions.  They do, of course, rely on the 
value of assumed relative abundances of C and H in the gas phase.

\subsection{Heating Rates}\label{heating_rates}

\subsubsection{Thermal Time Constants}\label{thermal_time_constants}

The cooling time for a medium with $T=7000\,$K and thermal pressure 
$nT=3800\,{\rm cm}^{-3}$K (Jenkins \& Tripp 2011) is about 
4.1$\,$Myr [ (Wolfire et al. 2003); see their Eq.~(4)].  However, when 
the medium is impulsively heated and ionized by the EUV and X-ray 
illuminations from a SNR, the temperature can approach or exceed $10^4\,$K, 
and the onset of L$\alpha$ cooling creates a dramatic increase in the overall 
cooling rate (Dalgarno \& McCray 1972).  When this happens, the 
thermal relaxation timescale becomes much shorter than the mean interval 
between the bursts of radiation, each of which last only about one to a few 
times $10^5\,$yr.  Thus, while we can compute a time-averaged heating rate 
for the SNR illuminations that might explain our high levels of partial 
ionization, we have no reason to expect that this average should be balanced 
by the cooling rates that would apply for the medium at $T=7000\,$K.

\subsubsection{Secondary Electrons}\label{secondary_heating}

The same energetic electrons that are responsible for the secondary 
ionizations $\Gamma_s$ and $\Gamma_{s^\prime}$ can also heat the gas through 
collisions with other electrons.  As with the calculations of the efficiency 
of secondary ionizations described in Appendix~\ref{gamma_s}, we use the 
analytic approximations of Ricotti et al. (2002) for 
the heating efficiencies, based on the numerical results for various 
conditions obtained by Shull \& Van Steenberg (1985).  Aside from replacing
ionization efficiencies with heating efficiencies and multiplying by the
energies of the respective secondary electrons, the calculations here are
virtually the same as for the ionization rates.

Table~\ref{heating_rates_table} shows the outcomes for the evaluations of 
secondary electron heating rates.  It is no surprise that there is a 
substantial increase in the heating rates when we advance from the ionization 
created by known sources to either of the two hypothetical enhanced 
ionization examples that could explain the electron fractions indicated by 
the observations of [Ar~I/O~I].

Our high result ($1.2\times 10^{-25}{\rm erg~s}^{-1}{\rm H~atom}^{-1}$) for 
the heating in the regime where there is increased penetration of X-rays down 
to 60$\,$eV creates a serious problem for this model, since this steady-state 
rate is considerably larger than the corresponding cooling rate 
$\ell_c+\ell_o=1.98\times 10^{-26}\,{\rm erg~s}^{-1}{\rm H~atom}^{-1}$ given 
in Section~\ref{cooling_rate}.

\subsubsection{Cosmic Ray Heating of Electrons}\label{CRheating_elec}

In addition to ionizing the gas and creating secondary electrons that can 
heat the gas, cosmic rays can also interact with free electrons in the medium 
and heat them.  The heating rate per unit volume is approximately equal to 
$A\zeta_{\rm CR}({\rm H}^0)n(e)$, where $A\approx 4.6\times 10^{-10}{\rm 
erg}$ 
(Draine 2011, p. 338). With the electron densities listed in 
Table~\ref{properties_table} and $\zeta_{\rm CR}=1.25\times 10^{-16}{\rm 
s}^{-1}{\rm H~atom}^{-1}$, we find that for the three different electron 
densities in this table the heating rates are $1.0\times 10^{-27}$ (for no 
SNR contribution), $1.9\times 10^{-27}$ (with SNR contributions), and 
$3.0\times 10^{-27}{\rm erg~cm}^{-3}{\rm s}^{-1}$ (for the lower energy 
penetration example).  After dividing these numbers by values of $n({\rm 
H}^0)$ given in Table~\ref{properties_table}, we find that these rates are 
about 0.4 to 1.2 times the rate from secondary electrons liberated by the 
cosmic rays, and 0.05 to 0.5 times the respective heating rates from the 
secondary electrons generated by ionizations from the X-ray backgrounds.

\begin{deluxetable}{ 
l   
c   
}
\tablewidth{0pt}
\tablecaption{Secondary Electron Heating 
Rates\tablenotemark{a}\label{heating_rates_table}}
\tablehead{
\colhead{} & \colhead{Rate}\\
\colhead{Source\tablenotemark{b}} & \colhead{($10^{-27}{\rm erg~s}^{-1}\,{\rm 
H~atom}^{-1}$)}
}
\startdata
X-ray background (no SNR)\dotfill&4.1\\
X-ray background (plus SNR)\dotfill&50\\
X-ray background (down to 60$\,$eV)\dotfill&110\\
Main-sequence stars\dotfill&1.5\tablenotemark{c}\\
Active binaries\dotfill&0.063\tablenotemark{c}\\
WDs\dotfill&0.84\tablenotemark{c}\\
Cosmic rays\dotfill&4.9\tablenotemark{c}\\
He$^+$ recomb.\tablenotemark{d} (background with no SNR)\dotfill&0.059\\
He$^+$ recomb.\tablenotemark{d} (background plus SNR)\dotfill&1.2\\
He$^+$ recomb.\tablenotemark{d} (background down to 60$\,$eV)\dotfill&8.8\\
\enddata
\tablenotetext{a}{Heating by secondary electrons that are liberated by the 
ionizations of H$^0$, He$^0$ and He$^+$, expressed in terms of an energy 
dissipation rate per neutral H atom.  Direct interactions of cosmic rays with 
free electrons produce an additional heating which is discussed in 
Section~\protect\ref{CRheating_elec}.}
\tablenotetext{b}{The conditions for the top and bottom three rows correspond to those 
for the last three columns in Table~\protect\ref{properties_table}.}
\tablenotetext{c}{To first order, the heating rates from internal sources and 
cosmic rays should not change as the strength of the background radiation 
increases above the basic rate from known sources.  In 
reality, they increase by modest amounts ($\sim$30\%) when the values of 
$x_e$ (which drive the heating efficiency) increase.}
\tablenotetext{d}{From secondary electrons that are produced by 
$\Gamma_{He^+}({\rm He}^0)$ and $\Gamma_{He^+}({\rm H}^0)$.  Other 
ionizations arising from helium recombinations do not produce electrons with 
sufficient energy to cause any appreciable heating.}
\end{deluxetable}
\begin{deluxetable}{ 
l   
c   
c   
c   
c   
c   
c   
c   
c   
c   
c   
c   
}
\tablecolumns{12}
\tablewidth{0pt}
\tablecaption{Dust Grain Heating and Cooling 
Rates\tablenotemark{a}\label{grain_rates_table}}
\tablehead{
\colhead{} & \multicolumn{3}{c}{Without SNR} & \colhead{} & 
\multicolumn{3}{c}{With SNR} & \colhead{} & \multicolumn{3}{c}{Greater Low}\\
\colhead{$b_c$\tablenotemark{b}} & 
\multicolumn{3}{c}{Contrib.\tablenotemark{c}} & \colhead{} & 
\multicolumn{3}{c}{Contrib.\tablenotemark{c}} & \colhead{} & 
\multicolumn{3}{c}{Energy Penetration\tablenotemark{c}}\\
\cline{2-4} \cline{6-8} \cline{10-12}
\colhead{$\times 10^5$} & \colhead{H} & \colhead{C} & \colhead{$\Delta$} & 
\colhead{} & \colhead{H} & \colhead{C} & \colhead{$\Delta$} & \colhead{} & 
\colhead{H} & \colhead{C} & \colhead{$\Delta$}\\
\colhead{(1)} & \colhead{(2)} & \colhead{(3)} & \colhead{(4)} & \colhead{} & 
\colhead{(5)} & \colhead{(6)} & \colhead{(7)} & \colhead{} & \colhead{(8)} & 
\colhead{(9)} & \colhead{(10)}
}
\startdata
0.0&9.9&5.1&4.8&~&16.1&7.5&8.5&~&22.0&10.0&12.0\\
2.0&15.1&7.3&7.8&~&23.2&11.0&12.1&~&30.7&14.9&15.8\\
4.0&24.4&10.6&13.9&~&36.4&16.0&20.4&~&46.9&21.8&25.1\\
6.0&31.6&13.8&17.8&~&47.4&20.9&26.5&~&61.3&28.5&32.9\\
\enddata
\tablenotetext{a}{Expressed in units of $10^{-27}{\rm erg~s}^{-1}\,{\rm 
H~atom}^{-1}$.  H~=~heating, C~=~cooling, and $\Delta={\rm H-C}$ (i.e., net 
heating rate).  The values were computed for the average interstellar 
radiation field (ISRF), as defined by Eq.~31 and Table~1 of Weingartner \& 
Draine (2001b) (with a radiation intensity 
$G=1.13$).  The adopted value of $R_V\equiv A_V/E(B-V)=3.1$.  The values for 
the heating and cooling rates per H atom tabulated here were computed from 
Eq.~44 (for grain heating) and Eq.~45 (for grain cooling) of Weingartner \& 
Draine (2001b) (but without the $n({\rm H})$ 
factor) and the coefficients listed in their Tables~2 and 3.}
\tablenotetext{b}{The abundance of carbon, relative to hydrogen, in the 
grains.  A high value of $b_c$ implies a population of dust grains that is 
rich in polycyclic aromatic hydrocarbons (PAHs).}
\tablenotetext{c}{The same conditions as for Columns (2)-(4) in 
Table~\protect\ref{properties_table}.}
\end{deluxetable}

\subsection{Dust Grains: Photoelectric Heating and Recombination 
Cooling}\label{photoelectric_grain_heating}

When dust grains are illuminated by starlight, they emit photoelectrons, 
which can heat the medium (Watson 1972; Draine 1978; Pottasch et al. 1979;
Bakes \& Tielens 1994; Weingartner \& Draine 2001b).  This heating is
partly offset by collisional cooling via grain-ion recombination (Draine \&
Sutin 1987).  The efficiencies of these mechanisms are regulated by the
charge of the grains, which in turn depends on $n(e)$, $T$ and $G$ (the
density of starlight).  Table~\ref{grain_rates_table} shows our
calculations of grain heating and cooling for the three cases (no SNR, with
SNR, and lower energy X-ray penetration) using the fitting formulae and
coefficients given by Weingartner \& Draine (2001b); see note~$a$ of the
table for details.

\placetable{grain_rates_table}

It is generally regarded that the photoelectric heating from grains is 
greater than that from cosmic rays by about one order of magnitude
 (Draine 2011, p. 339) [or considerably more than this if the assumed value 
for $\zeta_{\rm CR}({\rm H}^0)$ is lower than that adopted here, e.g., 
Wolfire et al. (1995, 2003)].  
Any enhancement in $x_e$ tends to increase the heating rate from grains, 
since the grains will be more negatively charged.  However, this increase is 
not as strong as the effect of a greater electron fraction on the direct 
cosmic ray heating, so the disparity between the two rates is decreased.

\section{SUMMARY AND CONCLUSIONS}\label{summary}

We have developed a means for deriving the representative rates of 
ionization, and thus the resulting electron densities, along sight lines that 
penetrate the WNM and that extend out to several 
hundred pc from us, well beyond the edge of the Local Bubble. Our method 
makes use of the fact that when a mostly neutral medium is exposed to the 
ambient EUV and soft X-ray ionizing radiation, the argon atoms are far more 
susceptible to being ionized than hydrogen atoms.  Thus, by comparing the 
abundances of Ar~I to those of H~I, we gain an understanding of the strength 
of the photoionization and secondary processes related to it.  For a number 
of practical reasons, we find that it is desirable to use O~I as a proxy for 
H~I.  The partial ionization of oxygen is strongly coupled to that of 
hydrogen through a rapid charge exchange process.  Using this strategy, we 
compare the column densities of Ar~I and O~I derived from absorption lines 
seen in the {\it FUSE\/} spectra of 44 hot subdwarf stars.

Since the neutral forms of argon and oxygen are virtually absent in regions 
that are fully ionized (either the prominent H~II regions around hot stars or 
the much lower density but more pervasive WIM), our 
probes sample only regions that have appreciable concentrations of H~I.  This 
sets our measurements apart from conventional determinations of average 
electron densities (e.g., pulsar dispersion measures, H$\alpha$ intensities, 
C~II fine-structure excitation, etc.), which are strongly influenced by 
contributions from the fully ionized regions.

We find that, on average, the abundance of neutral argon, relative to that of 
neutral oxygen, is $[{\rm Ar~I/O~I}]=-0.427\pm 0.11\,$dex below what we would 
expect if both species had no partial ionization and the solar abundance 
ratio is a proper standard of comparison.  We interpret this deficiency in 
terms of the greater susceptibility of argon to photoionization.  After 
accounting for a broad range of processes that can modify the fractional 
ionizations, we conclude that with known sources of ionizing radiation, both 
external and internal, the only straightforward way to reconcile the large 
deficiency of Ar~I is to propose that the neutral medium has a characteristic 
density of only $0.09\,{\rm cm}^{-3}$, which is far below a generally 
accepted value of about $0.5\,{\rm cm}^{-3}$.  At this latter, higher volume 
density of hydrogen, known sources of ionization should produce an electron 
density $n(e)=0.018\,{\rm cm}^{-3}$ and create a result $[{\rm 
Ar~I/O~I}]=-0.14$, which is greater than all but a small fraction of the 
measurements and well above the overall average value.

In order to achieve a result for [Ar~I/O~I] that is consistent with both our 
observations and a density $n({\rm H~I})=0.5\,{\rm cm}^{-3}$, we must propose 
nonconventional sources of additional ionization.  We discuss two such 
possibilities; in reality we might be witnessing the outcome of some 
combination of both of them working together.

The first explanation is that the shielding of the external, low energy X-ray 
flux is less effective than expected: if we lower the cutoff energy from a 
nominally expected 90$\,$eV to only 60$\,$eV, we can gain sufficient 
additional ionizing photons from the external background to create a level of 
ionization that is consistent with an upper bound for the value implied by 
our observed [Ar~I/O~I]. To explain this lower shielding, one might envision 
a nearly sponge-like topology for the WNM, where the internal holes and 
channels are filled with hot, X-ray emitting gas.  Indeed, de Avillez et al. 
 (2012) have performed hydrodynamical simulations 
of the effects from random SN explosions and stellar winds in the 
plane of the Galaxy, and the outcome of their model exhibits a complex, 
turbulent entanglement of hot, warm and cold gas complexes.  Their 
simulation, which includes non-equilibrium ionization calculations, yields an 
average electron density $n(e)=0.04\,{\rm cm}^{-3}$, a value that is not far 
removed from our determination for the WNM.  This near match may be 
coincidental, however.  In the simulation, voids containing hot, 
collisionally ionized gas with very low $n(e)$, in combination with dense 
regions that are photoionized, may produce an outcome that is close to the 
value for the WNM.

A serious shortcoming of the interpretation that we are viewing a 
steady-state maintenance of a high level of ionization in the WNM is that the 
calculated heating rate of $1.2\times 10^{-25}\,{\rm erg~s}^{-1}\,{\rm 
H~atom}^{-1}$ from secondary electrons is unacceptably high -- much higher 
than the cooling rate from atomic fine-structure excitations or 
recombinations of ions onto dust grains. 

Our second explanation is a time-dependent solution that can sidestep the 
problem of overheating in the equilibrium case discussed above.  Over a 
period of one to a few times $10^5\,$yr, an SNR can create a 
burst of ionizing radiation in the X-ray region that can briefly elevate the 
ionization of the ISM to levels well above normal out to distances of order a 
few hundred pc away.  Initially, the recombination timescale is quite short, 
but then it advances to about 1$\,$Myr when the level of ionization 
approaches our observed average state.  For heating by secondary 
electrons, the temperature should exhibit a similar rapid adjustment that is 
followed by a much slower trend with time.  The radiation burst will indeed 
create a very high heating rate, but as the gas temperature approaches or 
exceeds $10^4\,$K, the abrupt onset of the very strong L$\alpha$ cooling will 
dump most of the heat over a time scale that is considerably shorter than the 
cooling time ($\sim 4\,$Myr) for the gas in its usual state.  A good 
exposition of this quick thermal recovery to $T<10^4\,$K and a comparatively 
more gentle relaxation in the heat loss and recombinations can be seen in 
plots of $T$ and $x_e$ vs. time in Fig.~2 of Gerola et al. (1974).

We propose that the ionization imprint of an SN explosion on the WNM 
can last well beyond the time that the remnant is recognizable in X-rays.  We 
have good evidence that a number of SNe have exploded in our general 
vicinity over the past $1-10\,$Myr.  As a consequence, if one considers that 
the activity in our neighborhood has recently been higher than normal for our 
Galaxy, the WNM out to several hundred pc from the Sun may be ionized to a 
level that is somewhat higher than the low-density neutral gas at similar 
galactocentric distances elsewhere in the Galactic plane.

\acknowledgments

This research was supported by a NASA Astrophysics Data Processing grant No. 
NNX10AD44G to Princeton University.  All of the data presented in this paper 
were obtained from the Mikulski Archive for Space Telescopes (MAST) at the 
Space Telescope Science Institute (STScI). STScI is operated by the 
Association of Universities for Research in Astronomy, Inc., under NASA 
contract NAS5-26555. Support for MAST for non-{\it HST\/} data is provided by the 
NASA Office of Space Science via grant NNX09AF08G and by other grants and 
contracts. The author acknowledges many useful discussions with B.~T.~Draine.  
Drs. Draine and C.~Gry supplied suggestions for improvement after reading a 
draft of this paper.  Some of the conclusions presented here relied on the 
use of the CHIANTI database and software, which is a collaborative project 
involving the NRL (USA), the Universities of Florence (Italy) and Cambridge 
(UK), and George Mason University (USA); other conclusions made use of the 
T\"ubingen Model Atmosphere Fluxes within the framework of the German 
Astrophysical Virtual Observatory (GAVO).  The coordinates and apparent 
magnitudes displayed in Table~\ref{aro_results} were provided by the 
SIMBAD database, operated at CDS, Strasbourg, France.

{\it Facility:\/} \facility{FUSE}

\appendix
\section{IONIZATION BY SECONDARY ELECTRONS \boldmath{$\Gamma_s$} AND 
\boldmath{$\Gamma_{s^\prime}$}}\label{gamma_s}

\subsection{Electrons from H and He}\label{electrons_H_He}

In order to determine $\Gamma_s({\rm H}^0)$, [and, later, $\Gamma_s({\rm 
He}^0)$ and $\Gamma_s({\rm Ar}^0)$], we start by considering the effects from 
the secondary electrons that are produced by the primary photoionizations of 
both H and He. To do so, we evaluate various forms of the quantity $\phi({\rm 
H}^0,E_e,x_e)$, which represent the number of additional ionizations created 
by the collisions from the photoejected electrons with energy $E_e=E-{\rm 
IP}({\rm H^0,~He^+,He^{++}})$ and any cascade of additional electrons that 
may follow.  ($E$ is the original photon energy, and IP represents the 
ionization potentials 13.6, 24.6 and 54.4~eV for H and the two stages of He, 
respectively.)  As implied by the notation for $\phi$, we must know not only 
the energy $E_e$ of the primary electron but also the ambient electron 
fraction $x_e=n(e)/[n({\rm H}^0+n({\rm H}^+)]$ since this 
quantity\footnote{Shull \& Van Steenberg (1985) defined their efficiencies
in terms of a parameter $x=n({\rm H}^+)/[n({\rm H}^0)+n({\rm H}^+)]$ and
then assumed that the ionization fraction of He is equal to that of H.  This
assumption is not valid for the conditions considered in the current work. 
Hence, we substitute our term $x_e$ for their $x$, which allows for a proper
accounting of a more generous contribution of free electrons from the
ionization of He.} influences the relative efficiency of collisions that can
create new ionizations, as opposed to collisional heating of free electrons
or the excitation of atomic states  (Shull 1979).

For our calculations of $\phi({\rm H}^0,E_e,x_e)$, we make use of the 
empirical fits by Ricotti et al. (2002) to the 
secondary ionization efficiencies calculated by Shull \& Van Steenberg 
 (1985).\footnote{For readers who might 
compare these two references, we provide some guidance about differences in 
meaning for the representations of the efficiencies for ionizations by 
secondary electrons that each employed:  Shull \& Van Steenberg used  $\phi$ 
to denote an energy fraction that goes into secondary ionizations, whereas 
Ricotti et al. used $\Phi$ to mean a number fraction of ionizations that take 
place for each primary ionization.  Our $\phi$ function in Eqs.~\ref{Hphi} 
and \ref{Hephi} is equivalent to $\Phi$ defined by Ricotti et al.}  A 
complete accounting for the secondary ionization of H that includes electrons 
from the primary ionizations of both H and He is given by the relation
\begin{eqnarray}\label{Hphi}
\Gamma_s({\rm H}^0)&=&\Gamma_{s,{\rm H}^0}({\rm H}^0)+ \Gamma_{s,{\rm 
He}^0}({\rm H}^0)+\Gamma_{s,{\rm He}^+}({\rm H}^0)\nonumber\\
&=&\int \Big[ \sigma({\rm H}^0,E)\phi({\rm H}^0,E-{\rm IP}({\rm 
H}^0),x_e)\nonumber\\
&+&{n({\rm He}^0)\over n({\rm H}^0)}\sigma({\rm He}^0,E)\phi({\rm H}^0,E-{\rm 
IP}({\rm He}^0),x_e)\nonumber\\
&+&{n({\rm He}^+)\over n({\rm H}^0)}\sigma({\rm He}^+,E)\phi({\rm H}^0,E-{\rm 
IP}({\rm He}^+),x_e)\Big] F(E)\,dE
\end{eqnarray}
Values of $\sigma({\rm He}^0)$ and $\sigma({\rm He}^+)$ are from Marr \& West 
 (1976) and 
Spitzer (1978, pp. 105-106), respectively.

Free electrons created by the primary and secondary ionizations of He can 
exceed 10\% of those from H and thus should not be neglected.  For this 
reason, we must also consider the helium counterpart to Eq.~\ref{Hphi}, which 
is given by
\begin{eqnarray}\label{Hephi}
\Gamma_s({\rm He}^0)&=&\Gamma_{s,{\rm H}^0}({\rm He}^0)+ \Gamma_{s,{\rm 
He}^0}({\rm He}^0)+\Gamma_{s,{\rm He}^+}({\rm He}^0)\nonumber\\
&=&10\int \Big[ \sigma({\rm H}^0,E)\phi({\rm He}^0,E-{\rm IP}({\rm 
H}^0),x_e)\nonumber\\
&+&{n({\rm He}^0)\over n({\rm H}^0)}\sigma({\rm He}^0,E)\phi({\rm 
He}^0,E-{\rm IP}({\rm He}^0),x_e)\nonumber\\
&+&{n({\rm He}^+)\over n({\rm H}^0)}\sigma({\rm He}^+,E)\phi({\rm 
He}^0,E-{\rm IP}({\rm He}^+),x_e)\Big] F(E)\,dE
\end{eqnarray}
Again, we use the empirical fits of Ricotti et al. (2002) to obtain
values of $\phi({\rm He}^0,E_e,x_e)$.

In order to compute $\Gamma_s({\rm Ar}^0)$, we can estimate from the data 
compiled by Lennon et al. (1988) and Bell et al. 
 (1983) that the cross section for collisional ionization 
of Ar by low energy electrons arising from H and He ionizations is about 3.85 
times that for H.  The effects of secondary ionizations of Ar$^0$ are small 
compared to the primary ones, so any moderate deviation from our 
approximation $\Gamma_s({\rm Ar}^0)=3.85\Gamma_s({\rm H}^0)$ should be 
inconsequential.

\subsection{Electrons from Inner Shell Ionizations of Heavy 
Elements}\label{heavy_elem}

In spite of the fact that heavy elements have abundances at least three 
orders of magnitude below those of hydrogen, X-ray ionizations of their inner 
shell electrons create a perceptible increase in opacity of the neutral ISM 
at energies above 300$\,$eV, and they strongly dominate the absorption of 
X-rays above about 1$\,$keV (Morrison \& McCammon 1983; Wilms et al. 2000). 
These ionizations create energetic electrons, which, like those from the
primary ionizations of H and He, can create additional collisional
ionizations $\Gamma_{s^\prime}$ of the species that we are considering.

Electrons from the ionization of the inner shells of atoms arise from two 
fundamental processes.  First, an electron is ejected by the primary 
ionization, and it has an energy equal to that of the ionizing photon minus 
the ionization potential of the inner shell electron.  To calculate the 
ionization cross section $\sigma(X_{i,j},E)$ for level $j$ of an element 
$X_i$ as a function of photon energy $E$, we used the coefficients and 
fitting formula given by Verner \& Yakovlev (1995).  The primary ionization
is followed by the ejection of one or more Auger electrons, which can also 
collisionally ionize H, He, and Ar.  The energies $E_{i,j}$ of the Auger 
electrons, or the sum of energies of two or more electrons, have been 
calculated by Kaastra \& Mewe (1993).  In our 
assessments of the strengths of the resulting collisional ionizations of H, 
He, and Ar, we use equations analogous to Eqs.\ref{Hphi} and \ref{Hephi},
\begin{equation}\label{auger_H}
\Gamma_{s^\prime}({\rm H}^0)=\sum_i\Big\{ A(X_i)\sum_j\int 
\sigma(X_{i,j},E)\Big[ \phi({\rm H}^0,E-{\rm IP}(X_{i,j}),x_e) + \phi({\rm 
H}^0,E_{i,j},x_e) \Big] F(E)\,dE\Big\}
\end{equation}
and
\begin{equation}\label{auger_He}
\Gamma_{s^\prime}({\rm He}^0)=10\sum_i\Big\{ A(X_i)\sum_j\int 
\sigma(X_{i,j},E)\Big[ \phi({\rm He}^0,E-{\rm IP}(X_{i,j}),x_e) + \phi({\rm 
He}^0,E_{i,j},x_e) \Big] F(E)\,dE\Big\}~,
\end{equation}
where $A(X_i)$ is the assumed abundance of element $X$ relative to that of H 
and the $\phi$ functions are identical to those described in 
Section~\ref{electrons_H_He}.  As we did for the secondary ionizations of Ar, 
we assumed a rate equal to that of H multiplied by a factor of 3.85.  For the 
Auger electrons, we have no information on the distribution function of 
electrons with energy.  Thus, we fall back on the simple assumption that the 
overall effect is the same as that for a single Auger electron with an energy 
equal to the average sum of energies of multiple electrons (if more than 
one).  In doing so, we overlook the shortcoming that when two or more 
electrons have almost equal energies, the efficiency for secondary 
ionizations should be slightly lower than that for a single electron with the 
same energy.

For the element abundances $A_X$, we assume that only the gas-phase 
(depleted) abundances are relevant.  While it is true that high energy X-rays 
can penetrate dust grains and ionize the inner shells of electrons for atoms 
inside the grains, the escape probabilities for these electrons are low.  For 
instance, Weingartner et al. (2006) calculated 
that dust grains with radii $a=0.1\,\mu$m  have ultimate electron yields less 
than 30\% per photoionization for $E<1\,$keV.  In various models for the 
grain size distributions, about half of the mass of material is in grains 
with $a$ either above or below about $0.1\,\mu$m (Draine 2011, p. 
281).
For the gas-phase depletions, we adopt values given by Jenkins 
(2009) for low density gas, i.e., strengths characterized 
by his parameter $F_*=0$.  The elements considered were, in decreasing order 
of importance, O, C, Ne, N, S, Si, and Mg.  These elements accounted for more 
than 99\% of the free heavy atoms in the ISM.

Finally, we follow the strategy of Adamkovics et al. (2011) in neglecting
ionizations caused by fluorescence photons from heavy elements that are
exposed to X-rays.  They showed that this is an effect that is small compared
to those from primary and Auger electrons.

\section{IONIZATION BY PHOTONS FROM THE RECOMBINATION OF HELIUM 
IONS}\label{gamma_he}

When a free electron with an energy of order $kT$ combines with a He ion and 
creates a lower ionization stage of He (with an ionization potential IP) in 
either a level with an energy $E_0$ above the ground electronic state or in 
the ground state itself ($E_0=0$), a photon having an energy $E={\rm 
IP}-E_0+kT$ is generated.  A transition between bound levels also emits a 
photon with an energy $E$ equal to the energy difference of the levels.  
Since the mean free path of a photon that can ionize H or He is short 
compared to the size of the neutral gas region, it is safe to assume that any 
recombination radiation photon capable of creating an additional ionization 
will do so and not escape.  (We neglect the absorption of photons by dust 
grains.)

In the following subsections, we consider the various ways that the photons from 
He recombinations can be produced and how they add to the overall ionization 
rates.  In most instances, they can ionize more than one kind of atom or ion.  
Thus, as with the treatment of primary ionizations by internal sources 
discussed in Section~\ref{predicted_level}, we need to consider the relative 
probability $y(X,E)$ defined earlier by Eq.~\ref{y(X,E)} that any photon of 
energy $E$ will ionize a given species $X$.  This sharing function will 
appear in the detailed treatments that follow.

While the only the rates $\Gamma_{He^+}({\rm H}^0)$ and $\Gamma_{He^0}({\rm 
H}^0)$ (and their counterparts for Ar) are of direct interest in the 
development of the relative ionizations of H and Ar, the influence of helium 
recombination photons in ionizing He$^0$ and He$^+$ will affect the 
ionization balance equations given in Sections~\ref{equilibrium} and 
\ref{electron_density}.  Except for the relatively rare recombinations to the 
He$^+$ ground $1s$ level that can then photoionize H$^0$ (see 
Section~\ref{he+(h0)} below), ionizations of H and He arising from He 
recombinations do not create free electrons with enough energy to contribute 
to $\Gamma_s$ for either H or He.

\subsection{\boldmath{$\Gamma_{He^+}({\rm He}^+)$} from the Reaction 
\boldmath{${\rm He}^{++}+e\rightarrow{\rm He}^++\gamma$}}\label{he+(he+)}

The only recombination channel that creates a photon with an energy that is 
capable of reionizing He$^+$ is that which goes directly from the free 
electron continuum to the $1s$ ground state of He$^+$.  This recombination 
results in a photoionization rate
\begin{equation}\label{gamma_he+(he+)}
\Gamma_{He^+}({\rm He}^+)={n({\rm He}^{++})n(e)\over n({\rm 
He}^+)}\alpha_{1s}({\rm He}^+,T)y({\rm He}^+,54.4\,{\rm eV}+kT)~{\rm 
s}^{-1}~,
\end{equation}
where an approximate fit for the recombination coefficient to the $1s$ state 
is given by the formula
\begin{equation}\label{alpha_1s(he+)}
\alpha_{1s}({\rm He}^+,T)=3.16\times 
10^{-13}(T_4/4)^{-0.540-0.017\ln(T_4/4)}~{\rm cm}^3{\rm s}^{-1}~,
\end{equation}
with $T_4=T/10^4\,$K (Draine 2011, p. 138; all future equations of 
this form are from the same source).
Were it not for the fact that some fraction of the emitted photons ultimately 
ionize He$^0$ and H$^0$, as indicated by the fact that the $y$ term in 
Eq.~\ref{gamma_he+(he+)} is less than 1, the effect of $\Gamma_{He^+}({\rm 
He}^+)$ would be simply to convert the familiar Case~A recombination rate (to 
all electronic levels) to that of a Case~B rate (to levels with $n\geq 2$) 
 (Baker \& Menzel 1938).

\subsection{\boldmath{$\Gamma_{He^+}({\rm He}^0)$} from the Reaction 
\boldmath{${\rm He}^{++}+e\rightarrow{\rm He}^++\gamma$}}\label{he+(he0)}

As with the ionization of He$^+$, some of the photons that arise from 
recombinations directly to the $1s$ state will ionize He$^0$, giving a rate 
equal to that expressed in Eqs.~\ref{gamma_he+(he+)} and \ref{alpha_1s(he+)}, 
except that the term $y({\rm He}^0,54.4\,{\rm eV}+kT)$ replaces $y({\rm 
He}^+,54.4\,{\rm eV}+kT)$.  These photons are supplemented by direct 
recombinations to the $2p$ level, which then decays to the ground state by 
emitting a 40.8$\,$eV photon.  Finally, recombinations to levels above $n=2$ 
eventually emit photons with an average energy of about 50$\,$eV.  (This 
average energy can degrade toward 40.8$\,$eV if $n({\rm He}^+)/n({\rm H}^0)$ 
is large enough to create resonant absorptions of the He$^+$ Lyman series 
photons before they can ionize H$^0$.)  These three processes yield an 
ionization rate
\begin{eqnarray}\label{gamma_he+(he0)}
\Gamma_{He^+}({\rm He}^0)&=&{n({\rm He}^{++})n(e)\over n({\rm He}^0)}\{ 
\alpha_{1s}({\rm He}^+,T)y({\rm He}^0,54.4\,{\rm eV}+kT) + \alpha_{2p}({\rm 
He}^+,T)y({\rm He}^0,40.8\,{\rm eV})\nonumber \\ 
&+&[ \alpha_{\rm B}({\rm He}^+,T)-\alpha_{2p}({\rm He}^+,T)-\alpha_{{\rm 
eff,}\,2s}({\rm He}^+,T) ] y({\rm He}^0,50\,{\rm eV}) \}~{\rm s}^{-1}~,
\end{eqnarray} 
where the direct rate to just the $2p$ state is given by 
\begin{equation}\label{alpha_2p(he+)}
\alpha_{2p}({\rm He}^+,T)=1.07\times 
10^{-13}(T_4/4)^{-0.681-0.061\ln(T_4/4)}~{\rm cm}^3{\rm s}^{-1}~,
\end{equation}
recombinations to all levels except the ground $1s$ state have a combined 
rate constant
\begin{equation}\label{alpha_B(he+)}
\alpha_{\rm B}({\rm He}^+,T)=5.18\times 
10^{-13}(T_4/4)^{-0.833-0.035\ln(T_4/4)}~{\rm cm}^3{\rm s}^{-1}~,
\end{equation}
and all recombinations that pass through the $2s$ state have a combined rate
\begin{equation}\label{alpha_eff,2s(he+)}
\alpha_{{\rm eff,}\,2s}({\rm He}^+,T)=1.68\times 
10^{-13}(T_4/4)^{-0.7205-0.0081\ln(T_4/4)}~{\rm cm}^3{\rm s}^{-1}.
\end{equation}
[The last equation is a Draine-style fit to the three numbers in Table~14.2 
of Draine (2011).]

\subsection{\boldmath{$\Gamma_{He^+}({\rm H}^0)$} from the Reaction 
\boldmath{${\rm He}^{++}+e\rightarrow{\rm He}^++(1~{\rm 
or}~2)\gamma$}}\label{he+(h0)}

There are multiple processes that generate photons capable of ionizing H$^0$.  
First, there are direct recombinations to the $1s$ state.  Direct 
recombinations to the $2p$ state, followed by decays to the $1s$ level, first 
generate a photon with an energy $E=13.6\,{\rm eV}+kT$ that can ionize 
hydrogen, and then produce a 40.8$\,$eV photon that can ionize either H$^0$ 
or He$^0$.  Recombinations to the metastable $2s$ state again start with a 
creation of a hydrogen-ionizing photon, and later the decay of this level 
creates 2 photons, which have a combined average hydrogen ionization 
efficiency of 1.42 photons (Osterbrock 1989, p. 29).
As discussed earlier in connection with the derivation of $\Gamma_{He^+}({\rm 
He}^0)$, photons with $E\approx 50\,$eV are emitted by decays from levels 
with $n>2$.  Finally, there are cascades from upper levels to the $2s$ state 
that generate only 1.42 photons without the added effect of the initial 
$E=13.6\,{\rm eV}+kT$ photon.  For the processes that have just been 
discussed, we have
\begin{eqnarray}\label{gamma_he+(h0)}
\Gamma_{He^+}({\rm H}^0)&=&{n({\rm He}^{++})n(e)\over n({\rm H}^0)}\{ 
\alpha_{1s}({\rm He}^+,T)y({\rm H}^0,54.4\,{\rm eV}+kT) + \alpha_{2p}({\rm 
He}^+,T)[1+y({\rm H}^0,40.8\,{\rm eV})]\nonumber \\ 
&+&2.42\alpha_{2s}({\rm He}^+,T) + [ \alpha_{\rm B}({\rm 
He}^+,T)-\alpha_{2p}({\rm He}^+,T)-\alpha_{eff,\,2s}({\rm He}^+,T) ] y({\rm 
H}^0,50\,{\rm eV})\nonumber\\
&+&1.42[\alpha_{eff,\,2s}({\rm He}^+,T)-\alpha_{2s}({\rm He}^+,T)] \}~{\rm 
s}^{-1}~,
\end{eqnarray}
where
\begin{equation}
\alpha_{2s}({\rm He}^+,T)=4.68\times 10^{-14}(T_4/4)^{-0.537-0.019\ln(T_4/4)}
\end{equation}

\subsection{\boldmath{$\Gamma_{He^+}({\rm Ar}^+)$} from the Reaction 
\boldmath{${\rm He}^{++}+e\rightarrow{\rm He}^++\gamma$}}\label{he+(ar+)}
The ionization potential of Ar$^+$ is 27.62$\,$eV, which is only about 
3$\,$eV higher than that of He$^0$.  There are no photons for cascades of 
levels within recently formed He$^+$ that have energies that fall between the 
two ionization potentials, hence Eq.~\ref{gamma_he+(he0)} can be used for the 
ionization of Ar$^+$, after substituting every occurrence of ``He$^0$'' with 
``Ar$^+$.''

\subsection{\boldmath{$\Gamma_{He^+}({\rm Ar}^0)$} from the Reaction 
\boldmath{${\rm He}^{++}+e\rightarrow{\rm He}^++(1~{\rm 
or}~2)\gamma$}}\label{he+(ar0)}

The processes that ionize Ar are very similar to those expressed for H in 
Section~\ref{he+(h0)}, except that direct decays from the continuum to the 
$2p$ or $2s$ level do not have enough energy to ionize Ar$^0$.  The absence 
of these two sources of photons for ionizing Ar$^0$ eliminates several terms 
that we had in Eq.~\ref{gamma_he+(h0)}, which now leads to
\begin{eqnarray}\label{gamma_he+(ar0)}
\Gamma_{He^+}({\rm Ar}^0)&=&{n({\rm He}^{++})n(e)\over n({\rm Ar}^0)}\{ 
\alpha_{1s}({\rm He}^+,T)y({\rm Ar}^0,54.4\,{\rm eV}+kT) + \alpha_{2p}({\rm 
He}^+,T)y({\rm Ar}^0,40.8\,{\rm eV})\nonumber \\ 
&+&[ \alpha_{\rm B}({\rm He}^+,T)-\alpha_{2p}({\rm 
He}^+,T)-\alpha_{eff,\,2s}({\rm He}^+,T) ] y({\rm Ar}^0,50\,{\rm 
eV})\nonumber\\
&+&1.29\alpha_{eff,\,2s}({\rm He}^+,T) [n({\rm Ar}^0)/n({\rm H}^0)] [\langle 
\sigma({\rm Ar}^0)\rangle_{f(E)}/\langle\sigma({\rm H}^0)\rangle_{f(E)} ] 
\}~{\rm s}^{-1}~,
\end{eqnarray}
For Ar$^0$, the factor 1.42 for the effective number of H ionizing photons 
arising from the 2-photon decay of the $2s$ level must be reduced because the 
ionization potential of Ar$^0$ (15.76$\,$eV) is slightly higher than that of 
H$^0$ (13.60$\,$eV).  After integrating the photon energy distribution $f(E)$ 
given by Spitzer \& Greenstein (1951) from 
the ionization potential of Ar$^0$ to 40.8$\,$eV, we find that the photon 
yield is reduced to a value equal to 1.29.  Since these photons are spread 
over a broad range of energy, we must calculate a flux-weighted ratio of the 
photoionization cross sections of Ar and H that appears in 
Eq.~\ref{gamma_he+(ar0)}, given by
\begin{equation}\label{two_phot_sigma_ar/sigma_h}
\langle \sigma({\rm Ar}^0)\rangle_{f(E)}/\langle\sigma({\rm 
H}^0)\rangle_{f(E)} =\int_{\rm IP(Ar^0)}^{40.8\,{\rm eV}}f(E)\sigma({\rm 
Ar}^0)\,dE\Bigg/\int_{\rm IP(H^0)}^{40.8\,{\rm eV}}f(E)\sigma({\rm 
H}^0)\,dE~,
\end{equation}
which has a numerical value of 13.8.  

\subsection{\boldmath{$\Gamma_{He^0}({\rm He}^0)$} from the Reaction 
\boldmath{${\rm He}^++e\rightarrow{\rm He}^0+\gamma$}}\label{he0(he0)}

As with the case for He$^+$ reionizing itself discussed in 
Section~\ref{he+(he+)}, we have for He$^0$
\begin{equation}\label{gamma_he0(he0)}
\Gamma_{He^0}({\rm He}^0)={n({\rm He}^+)n(e)\over n({\rm 
He}^0)}\alpha_{1s}({\rm He}^0,T)y({\rm He}^0,24.6\,{\rm eV}+kT)~{\rm 
s}^{-1}~,
\end{equation}
where an approximate fit for the recombination coefficient to the $1s$ state 
is given by the formula
\begin{equation}\label{alpha_1s(he0)}
\alpha_{1s}({\rm He}^0,T)=1.54\times 10^{-13}T_4^{-0.486}~{\rm cm}^3{\rm 
s}^{-1}
\end{equation}
 (Benjamin et al. 1999).

\subsection{\boldmath{$\Gamma_{He^0}({\rm H}^0)$} from the Reaction 
\boldmath{${\rm He}^++e\rightarrow{\rm He}^0+\gamma$}}\label{he0(h0)}

There are two channels that contribute to $\Gamma_{He^0}({\rm H}^0)$.  The 
first is direct decays to the $1s$ state.  The second is all possible decays 
to levels with $n\geq 2$, of which 96\% of them generate photons which can 
ionize H$^0$ in the low density limit (Osterbrock 1989, p. 26). 
Hence, we find that
\begin{equation}\label{gamma_he0(h0)}
\Gamma_{He^0}({\rm H}^0)={n({\rm He}^+)n(e)\over n({\rm 
H}^0)}[\alpha_{1s}({\rm He}^0,T)y({\rm H}^0,24.6\,{\rm 
eV}+kT)+0.96\alpha_{\rm B}({\rm He}^0,T)]~{\rm s}^{-1}
\end{equation}
where
\begin{equation}\label{alpha_B(he0)}
\alpha_{\rm B}({\rm He}^0,T)=2.72\times 10^{-13}T_4^{-0.789}~{\rm cm}^3{\rm 
s}^{-1}
\end{equation}
 (Benjamin et al. 1999).
 
\subsection{\boldmath{$\Gamma_{He^0}({\rm Ar}^0)$} from the Reaction 
\boldmath{${\rm He}^++e\rightarrow{\rm He}^0+\gamma$}}\label{he0(ar0)}

The ionization of Ar$^0$ from recombinations to He$^0$ is similar to that for 
H$^0$ given in Eq.~\ref{gamma_he0(h0)}:
\begin{eqnarray}\label{gamma_he0(ar0)}
\Gamma_{He^0}({\rm Ar}^0)&=&{n({\rm He}^+)n(e)\over n({\rm Ar}^0)}\{ 
\alpha_{1s}({\rm He}^0,T)y({\rm Ar}^0,24.6\,{\rm eV}+kT)\nonumber\\
&+&0.94\alpha_{\rm B}({\rm He}^0,T)n({\rm Ar}^0)\sigma({\rm Ar}^0,\,20\,{\rm 
eV})/[ n({\rm H}^0)\sigma({\rm H}^0,\,20\,{\rm eV})]\} ~{\rm s}^{-1}
\end{eqnarray}
Drake et al. (1969) have computed the energy 
distribution of the two-photon decay from the singlet 2s level of He$^0$.  
From this distribution, one finds that the average yield of H-ionizing 
photons from each decay is 0.56.  With the slightly higher ionization 
potential of Ar$^0$, this yield drops to 0.30 photons, which lowers the 
coefficient in front of $\alpha_{\rm B}({\rm He}^0)$ from 0.96 to 0.94.  The 
average energy of all these decays is 20$\,$eV, which must be used in the 
calculation of the ratio of the Ar primary photoionization cross section to that of 
H.

\section{COSMIC RAY IONIZATION}\label{cr_ioniz}

We consider that a cosmic ray ionization rate for molecular hydrogen has a 
representative rate $\zeta_{\rm CR}({\rm H}_2)=3.5\times 10^{-16}{\rm 
s}^{-1}$ (Indriolo \& McCall 2012).  (There are large variations 
found from one region to the next however.)  This rate includes both primary 
cosmic ray ionizations and the additional ionizations from the resulting 
secondary electrons, and it should be twice as large as that for atomic 
hydrogen 
(Nagy \& V\'egh 1992). The measurements of Indriolo \& McCall (2012) applied
to cool, diffuse cloud regions\footnote{These regions are well shielded from
low energy X-rays, and hence virtually all of the ionization is caused by
cosmic rays.  Of the 50 sight lines covered in the survey by Indriolo \&
McCall (2012), only one ($\epsilon$~Per) had $N({\rm H~I+2H}_2)<10^{21}\,{\rm
cm}^{-2}$.  Under these circumstances the X-ray ionization rate should be
less than $10^{-18}\,{\rm s}^{-1}$, i.e., far less than the rate from cosmic
rays.} where some molecules are present.  Since these regions have
$x_e\approx 3\times 10^{-4}$
(Draine 2011, p. 185), we must adjust $\zeta_{\rm CR}$ to reflect the 
fact that the efficiency of secondary ionizations is reduced when larger 
values of $x_e$ are present (see Appendix~\ref{gamma_s}).  According to 
Spitzer
(1978, p. 144) the average energy of such secondary 
electrons is 35~eV.  Neglecting the fact that there is some spread in the 
actual energies, we derive an ionization rate for primary ionizations alone 
for H in atomic form,
\begin{equation}\label{zetaH_pCR}
\zeta_{p,{\rm CR}}({\rm H}^0)={3.5\times 10^{-16}{\rm s}^{-1}\over 
2[1+\phi({\rm H}^0,35\, {\rm eV},3\times 10^{-4})]}~.
\end{equation}
(See Section~\ref{electrons_H_He} for an explanation of the term $\phi({\rm 
H}^0,E_e,x_e)$.)  Within the context of the far more diffuse regions that are 
studied here, which have much higher electron fractions $x_e$, we must allow 
for a reduction in the ionization efficiency for the secondary electrons to 
obtain a corrected total rate for both primary and secondary ionizations,
\begin{equation}\label{zetaH_CR}
\zeta_{\rm CR}({\rm H}^0)=[1+\phi({\rm H}^0,35\, {\rm eV},x_e)]\zeta_{p,{\rm 
CR}}({\rm H}^0)~.
\end{equation}
As with their susceptibility to ionization by photons, argon atoms are about 
10 times more easily ionized by protons than hydrogen (for proton energies 
$E\sim 1\,$MeV) (Hooper et al. 1962; Kingston 1965).  In the 
description of secondary ionizations in Section~\ref{electrons_H_He}, it was 
noted that at low electron energies the cross section for collisional 
ionization for Ar is about 3.85 times that of hydrogen.  Accordingly, we find 
that
\begin{equation}\label{zetaAr_CR}
\zeta_{\rm CR}({\rm Ar}^0)=[10+3.85\phi({\rm H}^0,35\, {\rm 
eV},x_e)]\zeta_{p,{\rm CR}}({\rm H}^0)~.
\end{equation}
Finally, as one of the ingredients for calculating $P^\prime({\rm Ar})$ 
(Section~\ref{electron_density}) and the He recombination photons 
(Section~\ref{gamma_he}) involves the ionization equilibrium of He 
(Eqs.~\ref{n0}$-$\ref{n(e)}), we must know in principle the cosmic-ray 
ionization rate of He,
\begin{equation}\label{zetaHe_CR}
\zeta_{\rm CR}({\rm He}^0)= 1.36\zeta_{p,{\rm CR}}({\rm H}^0)
\end{equation}
Again, we used the results of Hooper et al. (1962) and Kingston (1965) 
to find that $\zeta_{p,{\rm CR}}({\rm He}^0)/\zeta_{p,{\rm CR}}({\rm 
H}^0)=1.36$.  Most of the secondary electrons are not energetic enough to 
ionize He, so Eq.~\ref{zetaHe_CR} does not have a $\phi$ factor included.  In 
practice, $\zeta_{\rm CR}({\rm He}^0)\ll\Gamma({\rm He}^0)$.

\end{document}